\documentclass[nofootinbib,
reprint,
floatfix
]{revtex4-1}

\usepackage{isomath}
\usepackage{upgreek} % to have roman greek letter
\usepackage{siunitx}

\usepackage{epsfig, amsmath,amsfonts, amssymb,graphicx,amsthm,color}
\usepackage{mathtools}% For \MoveEqLeft
\usepackage{dcolumn}% Align table columns on decimal point
\usepackage{bm}% bold math
\usepackage{siunitx}
\usepackage{amssymb}
\usepackage[latin1]{inputenc}
\usepackage{graphicx}
\usepackage{epstopdf}
\usepackage{comment}

\newcolumntype{M}[1]{>{\raggedright}m{#1}}

\begin{document}

\title{Cesium atoms in cryogenic argon matrix}
\author{T.  Battard$^{(1)}$}
\author{S.  Lahs$^{(1)}$}
\author{C.  Cr\'epin$^{(2)}$}
\author{D. Comparat$^{(1)}$}
\email[Corresponding author:]{daniel.comparat@universite-paris-saclay.fr}

\affiliation{$^{(1)}$ Universit\'e Paris-Saclay, CNRS,  Laboratoire Aim\'e Cotton, 91405, Orsay, France. }

\affiliation{$^{(2)}$
Universit\'e Paris-Saclay, CNRS, Institut des Sciences Mol\'eculaires d'Orsay, 91405, Orsay, France}

\date{\today}

\begin{abstract}

This paper presents both experimental and theoretical investigations into the spectroscopy of dilute cesium (Cs) atoms within a solid argon (Ar) matrix at cryogenic temperatures. This system is relevant for matrix isolation spectroscopy and in particular for recently proposed methods for investigating phenomena that extend beyond the standard model of particle physics. We record absorption spectra at various deposition temperatures and examine the evolution of these spectra post-deposition with respect to temperature changes.

Taking advantage of Cs-Ar and Ar-Ar pairwise interaction potentials, we conduct a stability study of trapping sites, which indicates a preference for T$_{\rm d}$ (tetrahedral, 4 vacancies) and O$_{\rm h}$ (cubic, 6 vacancies) symmetries. By implementing a mean-field analysis of the long-range Cs(6s,6p)-Ar-Ar triple dipole interaction, combined with a temperature-dependent shift in zero point energy, we propose effective Cs(6s,6p)-Ar pairwise potentials.
Upon integrating these pairwise potentials with spin-orbit coupling, we achieve a satisfactory agreement between the observed and simulated absorption line positions. 
The observed line broadening is reasonably well reproduced by a semi-classical thermal Monte Carlo approach based on Mulliken-type differences between excited and ground potential curves.
Additionally, we develop a simple, first-order crystal field theory featuring only 6 interaction mode coordinates. It uses the reflection approximation and incorporates quantized (phonon) normal modes. This produces a narrow triplet structure but not the observed amount of splitting. % Further investigation is required, as we cannot definitively assert that the two structures observed result from T${\rm d}$ (tetrahedral, 4 vacancies) and O${\rm h}$ (cubic, 6 vacancies) symmetries.

\end{abstract}

\maketitle

\section{Introduction}

Despite matrix isolation spectroscopy having been studied since the 1950s \cite{barnes81matrix,almond89,bondybey1996new}, our understanding of the shape and behavior of trapping environments within the matrix remains limited. This is particularly true for even the simplest systems comprising single valence electron atoms (alkali atoms) in rare gas matrix environments, where no consensus has yet been reached regarding the trapping site with the host atom, especially for heavier atoms \cite{crepin1999photophysics,davis2018investigation,ozerov2021generic}. 

Gaining a more accurate understanding of the trapping site and the matrix's effect on the dopant would prove invaluable for precision spectroscopy experiments, such as magnetometry, spin-dependent interactions, and investigations into physics beyond the Standard Model. These investigations may include searches for violations of fundamental symmetries (parity, time-reversal) like electric dipole moment (EDM) particles or axion-like dark matter candidates. Precision spectroscopy experiments utilize spin-induced transitions, which are not directly affected by the matrix's pure electrostatic interaction at first order. However, a detailed study is necessary to assess the precision that these experiments can attain \cite{pryor1987artificial,arndt1993can,kozlov2006proposal,atoms1018Hessel,upadhyay2019spin, braggio2022spectroscopy,budker2022quantum}. 

In this article, our focus is on studying cesium atoms trapped in an argon matrix. Understanding this simple system will help to further characterize more complex experiments. Among the (stable) alkali atoms, Cs is the heaviest, making it highly sensitive to the effects of the electron EDM \cite{safronova2018search}. We have chosen Argon because it is the only rare gas whose naturally occurring isotopes do not possess a nuclear spin that would interact with the cesium spin. 

Absorption (or transmission) spectra form the foundation for all optical manipulations necessary for the above-proposed experiments. Intriguingly, only a few experiments have studied Cs embedded in Argon \cite{ien1962electron,goldsborough1964electron}, and only two optical transmission spectra have been published \cite{weyhmann1965optical,Kanda1971} (\cite{balling1983laser} gives only absorption and emission frequencies). For Cs in Ar, two structures were observed. They could arise from $6s\rightarrow 6p$ transition in two different trapping site environments: the $6p$ triplet degeneracy being lifted either by a low symmetry site \cite{weyhmann1965optical}  %(experiment likely performed at \SI{4.2}{K})% 
or by a dynamical Jahn-Teller effect in a cubic symmetry site \cite{Kanda1971}.
Ref. \cite{balling1983laser} observed, in addition, $6s\rightarrow 5d$ and $6s\rightarrow 7p$ transitions, along with relaxation and fluorescence from $6p$ and $5d$ levels.

In this article, we conduct new experiments with the aim of reconciling the discrepancies between previous measurements. We begin by presenting the experimental setup, which enables precise Cs density measurements in highly polycrystalline Ar samples. Then, using Cs-Ar and Ar-Ar pairwise interaction potentials, we first conduct a trapping site stability study. By correcting the pairwise interaction with effective third-order effects, we predict absorption line positions for the different trapping sites found. We then validate, using a simple crystal field model, whether the found linewidth agrees with the proposed trapping sites. Finally, we provide conclusions and potential improvements for further exploration.

\section{Experimental setup}

\begin{figure}[ht]
    \centering
    \includegraphics[width=1\linewidth]{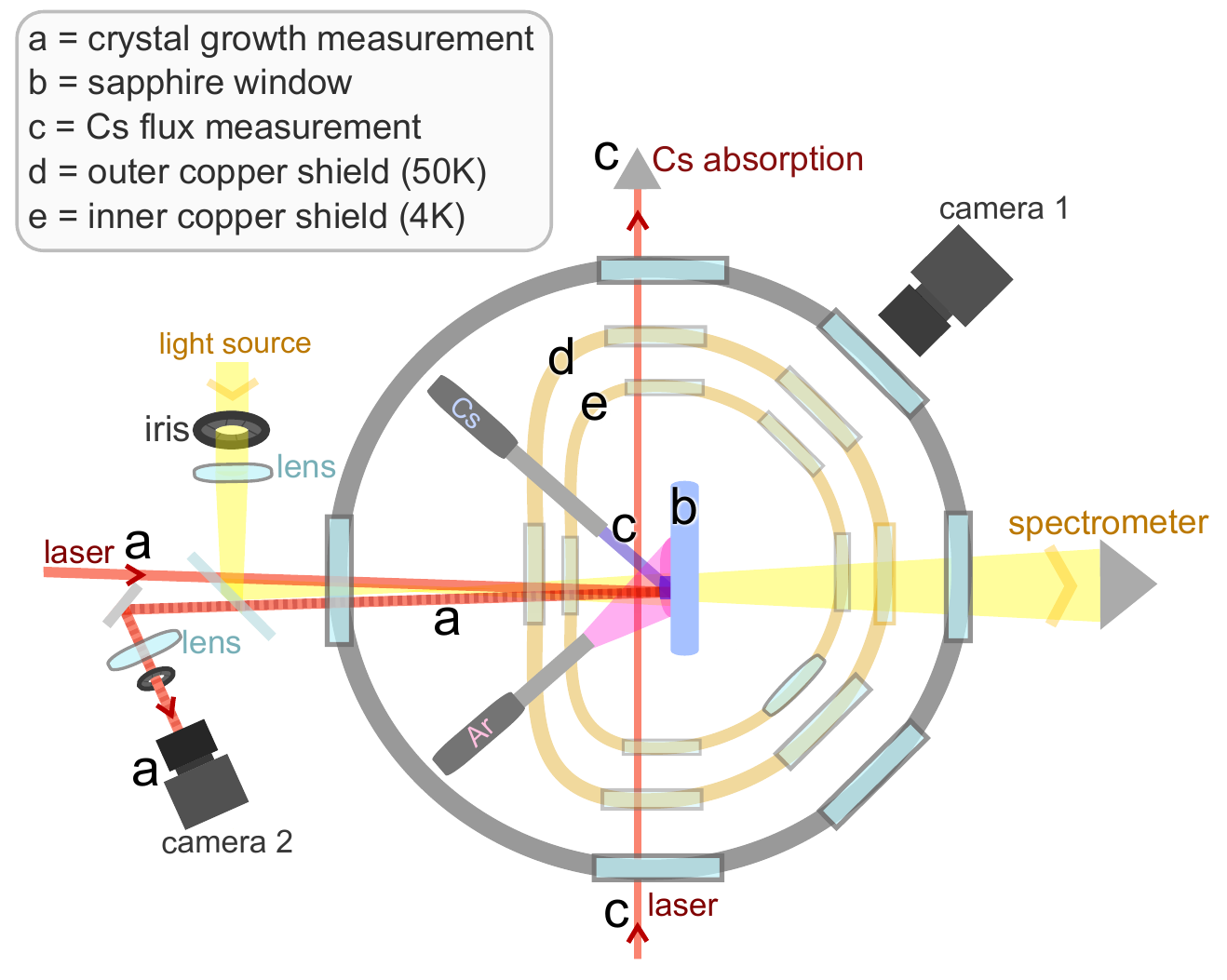}
    \caption{(Color available online) Schematic representation of the cryostat and the optical systems utilized to monitor the matrix growth rate and to measure the absorption and fluorescence spectra. The arrows indicate the propagation directions of the beams. The laser illumination of the matrix is imaged on CCD camera 2 to display the interference pattern.}
    \label{fig:Optical_elements}
\end{figure}

Figure \ref{fig:Optical_elements} illustrates the experimental setup. The experiment is performed under a residual gas pressure of $\sim10^{-7}\,$mbar in a two-stage cryostat manufactured by Mycryofirm using a pulsed tube cryocooler (SHI RP-082E2S). The cryogenic matrix is grown by depositing Ar gas onto a C-cut sapphire (Al$_2$O$_3$) plate, which is \SI{20}{mm} in diameter and \SI{1}{mm} in thickness, and embedded in a copper frame. The cryogenic temperature of the sample holder, ranging from \SI{3}{\kelvin} to \SI{30}{\kelvin}, is regulated via resistive heating. In order to minimize thermal disturbances caused by the kinetic energy transfer of the Ar beam, the sample holder's temperature is maintained at a constant level via a PID loop.

\subsection{Sample growth}

The thickness $L$ of the argon layer is continuously monitored via 2D reflection Fizeau interferometry. A pattern formed by the interference between the light reflected from the front and back sides of the matrix is captured by a lens onto a CCD camera. The 2D imaging offers numerous advantages compared to previous 1D transmission or reflection laser interferometry studies \cite{hollenberg1962absolute,groner1973measurement,jiang1975absolute}: it features a background-free environment ensuring excellent contrast, it is less restricted by beam size and spatial interference narrowing, and it enables the monitoring of the whole sample thereby ensuring the quality of deposition.

For this Ar growth measurement, a DFB diode laser operating at $\lambda=\SI{852}{nm}$ was employed, providing the advantage of also allowing the monitoring of the Cs flux through saturated absorption.

The interference fringes at a specific point evolve as per $ \frac{2n_\text{Ar}}{\lambda}\frac{dL}{dt}$, where $n_\text{Ar} = 1.3$ is the refractive index of an argon bulk crystal at 852 nm \cite{sinnock1969refractive}.

A growth of approximately 3 fringes per minute is observed, corresponding to a growth rate for an ideal argon bulk crystal of $\frac{dL}{dt} = 1 \mu\mathrm{m/min}$. After about one hour of argon deposition, a crystal is formed with a thickness of $\sim$\SI{50}{\micro \metre}.

\subsection{Cs density}

To embed cesium into the matrix, we utilize Alfasource 3S dispensers from Alfavakuo e.U., loaded with a CsBi$_{25}$ alloy. The alkali purity of the cesium is 99.980\%. The dispenser is situated outside the \SI{50}{K} and \SI{4}{K} thermal shields and is activated by resistive current heating. The Cs vapor is funneled to the substrate through a steel tube with an inner diameter $d$ of \SI{4}{mm}, which pierces both thermal shields. Independently, the tube is heated to approximately $\SI{40}{\degreeCelsius}$ via resistors to minimize blockage from condensing Cs. We have extensively studied in Ref. \cite{hahn2022comparative} the behavior of cesium atom effusion through a collimating tube and established that, irrespective of the tube material, Cs adheres to the wall. This results in a well-collimated and homogeneous Cs atomic beam.

We monitor the Cs flux by recording a laser absorption spectrum while adjusting the frequency of the low-intensity diode laser intersecting the cesium beam. We assume that the Cs beam diameter aligns with the inner tube diameter $d$ and its density $n$ is uniform. Confirmation of this is obtained by scanning the laser beam through the atomic beam or by a simple 2D absorption image of the Cs in the Ar matrix, where the cesium-doped zone creates a visible spot that is about \SI{5}{mm} wide. The velocity distribution is effusive \cite{hahn2022comparative}:
$
f_z(v_z) = \left( \frac{m}{\sqrt{2} k_B T_z}\right)^2 v_z^3 e^{ -\frac{m v_z^2}{2 k_B T_z} }
$
along $z$. The Cs beam is intersected by the laser at an angle of $\alpha =\SI{22}{\degree}$ relative to the beam axis, as shown in Fig. \ref{fig:Optical_elements}. Hence, the Beer-Lambert-Bouguer absorption is given by
$
n d \sigma_0 
 \int_{0}^{\infty} 
 \frac{1}{1+ \left( 
 \frac{\omega_L - \omega_{\rm Cs}   - k v_z \sin \alpha }{\Gamma/2}
 \right)^2 }
 f_z(v_z) d v_z 
$
where $\sigma_0 = \SI{1.4e-9}{cm^{2}}$ is the absorption cross section for an isotropic light polarization at the resonant $6s_{1/2}(F = 4) \rightarrow  6p_{3/2}( F = 5)$ frequency $\hbar \omega_{\rm Cs}$, which has a natural spontaneous decay rate of $\Gamma = 1/(\SI{30.4}{ns})$ \cite{steck2019cesium}. We fit the experimental absorption profile to this formula while scanning the laser angular frequency $\omega_L$. The results align with and are not highly sensitive to $T_z \approx \SI{400}{\kelvin}$. We then determine the beam density $n$ and consequently the Cs flux $n \pi (d/2)^2 \bar v$, where $\bar v = \sqrt{8 k_B T_z/ \pi m}$ is the average velocity. We only account for the population in the hyperfine level $F=4$ (which represents only 9/16 of the total Cs(6s) population in the beam, assuming Boltzmann-equidistribution between the 9 times degenerate $F=4$ and 7 times $F=3$ levels). Ultimately, we obtain a Cs(6s) flux of a few $10^{12}\ $at/s reaching the Ar matrix. We thus anticipate a $\sim 10^{-4}$ atomic ratio of Cs in the argon matrix (density on the order of $10^{18}$ cm$^{-3}$) and a typical average internuclear distance between Cs atoms on the order of $\SI{10}{nm}$.

To prevent Cs contamination of the window and to generate well-defined trapping environments for all the Cs atoms, we always initiate Cs deposition only after a few micrometers of Ar have already been deposited.

\subsection{Quality of the samples}

Undoubtedly, the epitaxy of an fcc Ar crystal is unfeasible on the hexagonal lattice of sapphire due to the incompatibility of their structural lattices \cite{nepijko2005morphology}, leading to the formation of several crystal defects, including holes or cavities, as observed in Ref \cite{nepijko2005morphology}.

In theory, Cs can be trapped in various defect types such as point defects surrounded by vacancies or line-dislocation defects (including screw, edge, etc.), surface defects (like stacking fault, tilt and twist grain boundaries), or volume defects (e.g., pores, cracks, or alternate phases like hcp inclusions) \cite{patterson2018solid}.

It is known from literature that the grain size is inversely proportional to the growth rate \cite{smith1970inert}, and the deposition temperature significantly influences the crystal quality. The crystals contain twins, stacking faults, and dislocations \cite{venables1977rare}. At $\SI{20}{K}$ and $\SI{10}{K}$, the inter-dislocation distances are roughly $\SI{8}{nm}$ and $\SI{3}{nm}$, respectively \cite{kovalenko1970structure,shakhsemampour1984magnetic}. This suggests the likely presence of small crystallites comprised of thousands of atoms, separated by dislocations. Additionally, if a film is grown at a condensation temperature greater than $2/3$ of the sublimation temperature (that is $\SI{30}{K}$ for Ar), the fcc grains are about $\SI{100}{nm}$ in size. Conversely, at 1/3 of the sublimation temperature (that is $\SI{10}{K}$ for Ar), the grain size reduces to around $\SI{10}{nm}$, with indications of a minority hcp phase \cite{smith1970inert,rudman1978rare,song2013self}. Furthermore, below a critical temperature (18 K for Ar), atomic-scale cavities appear in the lattice \cite{schulze1973density}.

These studies align well with our experimental observations regarding the quality of the Fizeau interference image during crystal growth, and the reduction in light transmission intensity due to scattering on grain boundaries. Specifically, as described in \cite{pollack1964solid}, a sample turns almost opaque when the Ar crystal grain diameter reaches around $0.1 \lambda$. We attribute this to exactly this increased scattering.

Although we attempted annealing up to $\SI{30}{\kelvin}$, the Ar crystal quality showed no significant improvement. This outcome aligns with expectations, as higher temperatures (closer to the triple point) would be necessary \cite{tovbin2015temperature,tovbin2017issues} to promote vacancy formation or atom-vacancy exchange through diffusion \cite{tishchenko1982self,tishchenko1984kinetics,leibin2021modeling}. However, achieving such temperatures in a free-standing sample in vacuum is not possible as sublimation would occur prematurely due to the high vapor pressure of solid argon.

\section{Absorption spectra}

\subsection{Experimental Spectra}

In our absorption (or transmission) spectroscopy, we use an Avantes AvaLight-HAL tungsten-halogen light source and an Ocean Optics QE65000 spectrometer. The spectrometer was calibrated by measuring the spectra of Hg and Kr lamps.

To measure the absorption in the doped Ar crystal, we shine light from the lamp through the sample. The position and shape of the light on the plate can be adjusted with irises and mirrors and monitored via a camera.

We then compute the spectral absorbance of the sample due to Cs as $A= \log_{10} \frac{I_0}{I_{\rm t}} $, where $I_t$ denotes the spectral irradiance density recorded in the presence of the Cs-Ar sample, and $I_0$ is obtained prior to any Cs deposition.

\begin{figure}[ht]
\centering
\includegraphics[width=0.95\linewidth]{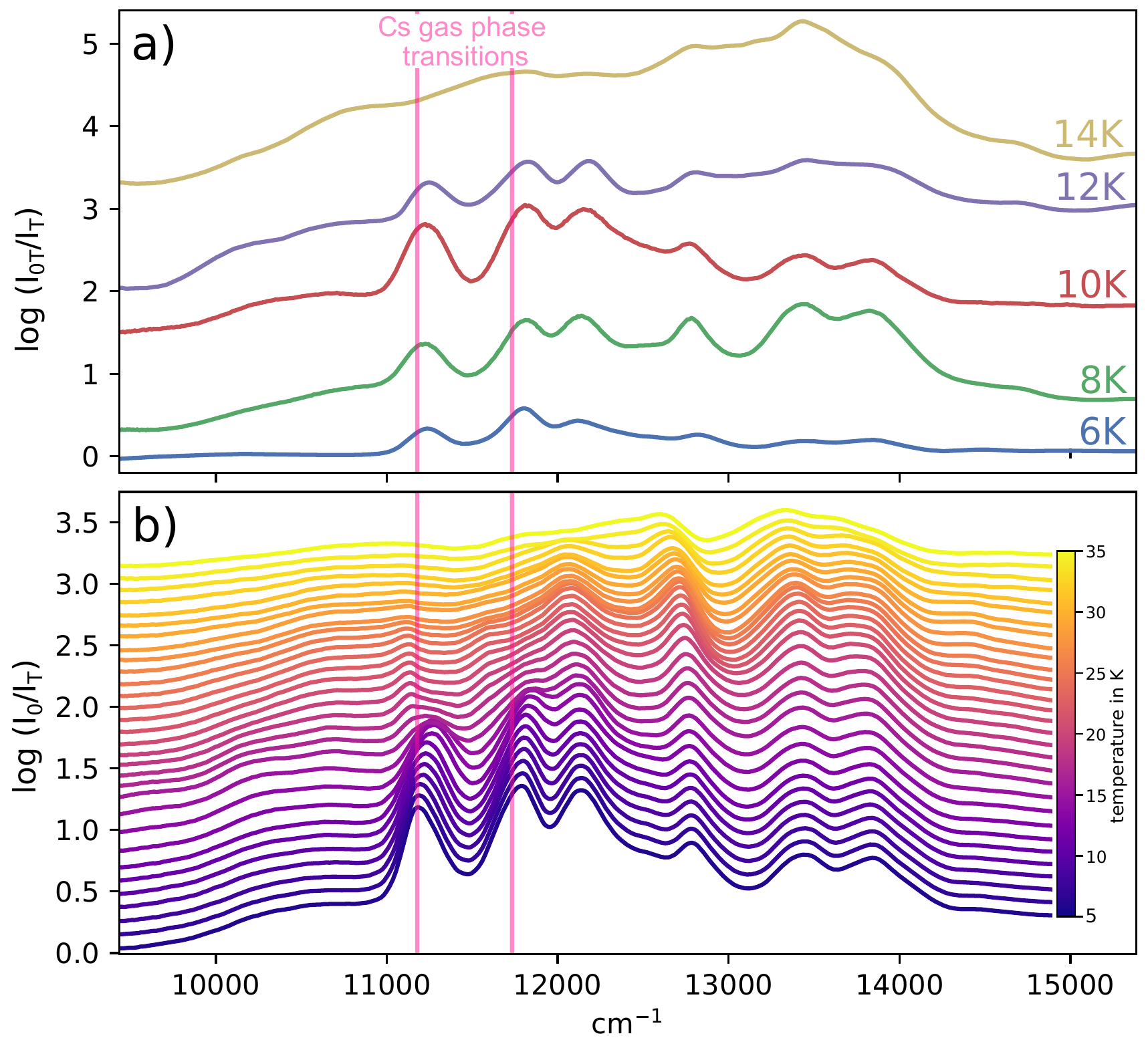}
\caption{a) Absorbance spectra of Cs embedded in Ar for deposits conducted at different temperatures. b) Evolution with temperature for a spectrum initially created at \qty{8}{K}. The spectra are offset in ordinate for better visual clarity. The positions of the 6s$_{1/2},\rightarrow$6p$_{1/2}$ (left) and 6s$_{1/2},\rightarrow$6p$_{3/2}$ (right) Cs gas-phase transitions are indicated for reference.}
\label{fig:abs_spectra}
\end{figure}

The captured spectra from varying deposition temperatures are presented in Fig. \ref{fig:abs_spectra} a). Above \qty{14}{K}, the spectra become significantly broader, and we encountered difficulties producing any spectra for deposition temperatures above this.

Conversely, if Cs and Ar are deposited at a low temperature (\qty{8}{K} in the case of Fig. \ref{fig:abs_spectra} b), the spectra remain visible even when the sample is subsequently heated up to approximately \qty{35}{K}, a temperature at which argon begins to sublimate under vacuum.

The spectra show two triplet structures similar to those observed by \cite{weyhmann1965optical,Kanda1971}, attributed to two types of trapping sites. For comparison,  spectra of previous experiments were obtained at \SI{1.8}{K} in Ref. \cite{Kanda1971}, likely performed at \SI{4.2}{K} in \cite{weyhmann1965optical} and performed at \SI{10}{K} in \cite{balling1983laser}. Typical positions and widths of the two triplets structures at 8K are presented in Table \ref{tableValues}. The differing relative intensities of the two triplets in the various samples (Fig. \ref{fig:abs_spectra} a) imply that depending on the deposition conditions (deposit rate, sample density, temperature, argon purity, etc.), one or the other trapping site is favored. It is worth noting that we used slightly different Ar and Cs fluxes for different samples. The flux estimates given in the previous section are  measured for the sample deposited at \SI{8}{K}, which will be the one we focus on in the following. For other deposits the rates used can easily vary by a factor of 3.

The temperature evolution shown in Fig. \ref{fig:abs_spectra} b) shows a more pronounced spectral deformation in the red site (the one between \qty{11000} and \qty{12500}{cm^{-1}}) than in the blue site (the one between \qty{12500} and \qty{14000}{cm^{-1}}). The temperature changes below 20K are mainly reversible. In the 6-14K temperature range corresponding to a well-defined red triplet, there is a slight blue shift ($\sim 50 $ cm$^{-1}$) of the two lower energy components from 6K to 14K, while the position of the highest energy component remains constant. The bandwidths of this triplet structure do not depend on the temperature. Conversely, a slight broadening of the components attributed to the blue site can be inferred from the analysis in Gaussian bands (less than 50 cm$^{-1}$ from 6 to 20K), while their positions do not shift with temperature. Heating beyond 20K typically leads to a drop in the baseline (likely due to stronger light scattering in the Ar), along with irreversible changes in the red triplet. The blue site, however, remains well-defined until 28K. Due to the lack of reproducibility, it is not possible to draw more precise conclusions about the evolution of temperature. These informations should be handled with care.

\begin{table}[]
\begin{tabular}{|l|l|l|}
\hline
line center (cm$^{-1}$) & line center (nm) & FWHM (cm$^{-1}$) \\ \hline
11 220 & 891 & 310 \\ \hline
11 780 & 849 & 360 \\ \hline
12 170 & 822 & 350 \\ \hline
12 770 & 783 & 400 \\ \hline
13 400 & 746 & 380 \\ \hline
13 850 & 722 & 400 \\ \hline
\end{tabular}
\caption{Line center and width extracted from Fig. \ref{fig:abs_spectra} b) at \SI{8}{K}. Both triplets were fitted by a sum of 3 Gaussians.}
\label{tableValues}
\end{table}

More broadly, these types of triplet structures are well-documented in alkali atoms trapped in rare gas matrices. They are not due to resolved phonon lines but are created by homogeneous broadening and result from the lifting of the degeneracy of the $p$ level by the crystal environment \cite{crepin1999photophysics}. However, due to the finite temperature, it is not straightforward to differentiate between the splitting created by a low symmetry crystalline field, as in dislocations or surface defects, and one created by a (temperature-dependent) dynamical Jahn-Teller effect in a more symmetric (cubic, tetrahedral, etc.) crystalline field. From the shape of the lines, however, it already seems evident that the main splitting is on the order of the fine structure splitting between $6p_{1/2}$ and $6p_{3/2}$ (of nearly $\SI{554}{cm^{-1}}$ in the gas phase). Then the static or dynamical electric crystal field will lift the degeneracy between the $|m|=3/2$ and $|m|=1/2$ of the $6p_{3/2}$ level.

\subsection{Trapping sites}

\subsubsection{Potential curves}

To explore potential trapping sites within an fcc lattice, we should conduct a stability analysis.

Utilizing a Lennard-Jones potential with an equivalent depth of $\sim \SI{45}{cm^{-1}}$ and an equilibrium distance of $\sim \SI{0.55}{nm}$ as the X$^2 \Sigma^+_{1/2}$, Cs-Ar potential \cite{blank2012m+,kobayashi2016ab,medvedev2018ab}, generic calculations for the accommodation of an atom in fcc and hcp rare-gas solids \cite{ozerov2019computational,ozerov2021generic} suggest that the likely trapping sites for Cs in an Ar fcc matrix could have 6, 8, or potentially 10 vacancies with respective symmetries of O$_h$, C$_{2v}$, C$_{4v}$. Nonetheless, a 7-vacancy fcc lattice (C$_{3v}$ symmetry), a 5-vacancy trapping site located in an hcp environment, or a grain boundary akin to a stacking fault accommodation (D$_{3h}$ symmetry) could also emerge \cite{davis2018investigation,kleshchina2019stable}.

\begin{figure}[ht]
    \centering    
    \includegraphics[width=1\linewidth]{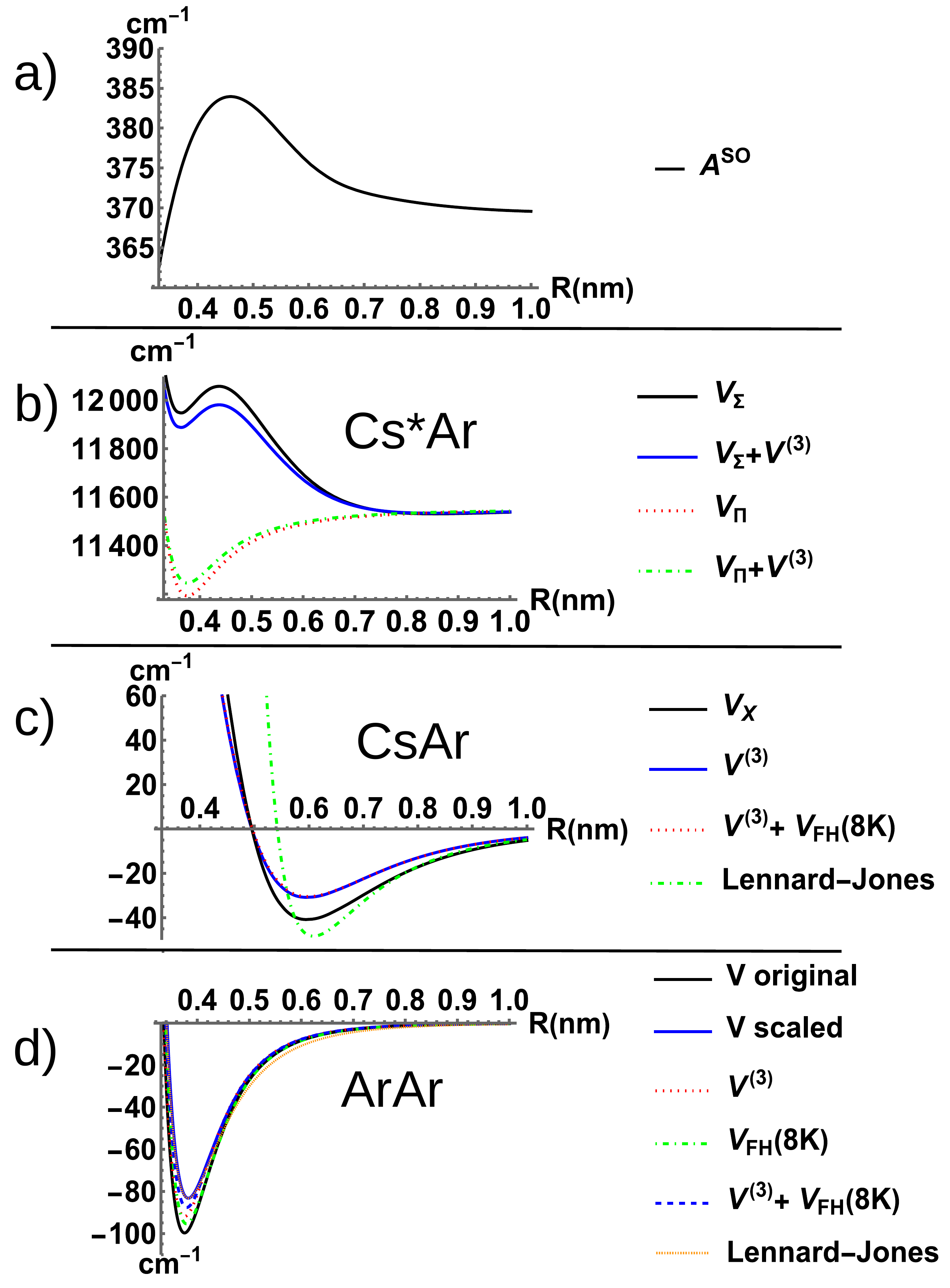}
    \caption{a) Spin-orbit constant for Cs(6p)-Ar interaction. b) Cs(6p)-Ar potential curves. c) Cs(6s)-Ar potential curves. d) Ar-Ar ground state potential curves. The potential curves $V$ used in this study are compared to the zero-point energy $V_{FH}$, and a third-order correction $V^{(3)}$. The original potentials are taken from \cite{kobayashi2016ab}. For the excited state, we chose to plot only a third-order case with no cut-off (in analogy with the ground state case, see text); i.e., the two-body potential $V$ (offset by $E(6p)$) multiplied by $\frac{1}{3} \frac{8\pi \rho}{9} \frac{C_9^*}{ C_6^* }$.
    }
    \label{fig:potentialCurves}
\end{figure}

However, as depicted in Fig. \ref{fig:potentialCurves} c), the Lennard-Jones potential fails to accurately reproduce the Cs-Ar ground state interaction. Consequently, we conducted our study using more precise potentials.

For the $V_{\rm Ar-Ar}$ potentials, we use the simple analytical formula provided by \cite{deiters2019two}. For the Cs-Ar ground state $V_{\rm Ar-Cs} = V_{\rm X\, ^2\Sigma_{1/2}^+}$, we employed the potential curve from Ref. \cite{kobayashi2016ab} with a third-order interpolation between points. Due to a lack of data points in the long-range, we incorporated parts of the potential from \cite{blank2012m+} for inter-nuclear distances exceeding $\SI{2}{nm}$. These potentials are presented in Fig. \ref{fig:potentialCurves}. Although the Lennard-Jones approximation accurately represents the Ar-Ar potential, it is considerably inaccurate for the Cs-Ar potential. 

For comprehensive Ar-Ar and Cs-Ar interactions, we use a simple two-body (pairwise potential) approximation. Here, the full interaction energy of a Cs-doped crystal with $(N-n)$ Ar atoms and $n$ vacancies is given by the following equation, using obvious notations:
\begin{equation}
 V = \sum_{i=1}^{N-n} V_{\rm Ar-Cs}(\bm R_{Cs-Ar_i})  + \sum_{1\leq i < j \leq N-n } V_{\rm Ar-Ar}(\bm R_{ij}) 
\label{eq_energy_tot}
\end{equation}

We began by using a simple cubic grid with $N= 4\times 7^3 = 1372$ atoms (the factor 4 arises from the 4 atoms in the fcc primitive cell). Convergence is achieved within one  percent of the energy for this configuration. This number of atoms enables us to simulate a small homogeneous chrystallite that would sit between dislocations. 

For a central Ar atom located at $\bm R_0$, the sum over all other atoms $\sum_i V_{\rm Ar-Ar}(\bm R_0 - \bm R_i)$ reaches $\approx \SI{773}{cm^{-1}}$ (here, a  larger grid is necessary for convergence). This significantly overestimates the experimental value for cohesive energy of $\SI{645}{cm^{-1}}$ \cite{lotrich1997three,schwerdtfeger2016towards}. Similarly, the equilibrium distance is established for a lattice constant of $\SI{0.521}{nm}$, compared to the correct value of $\SI{0.531}{nm}$ \cite{schwerdtfeger2016towards}. To reproduce the correct cohesive (also called atomization $-E_{\rm at}$) energy and the equilibrium lattice constant, we scaled the Ar-Ar potential by $V=\alpha V_{\rm Ar-Ar}(\beta R)$ with $\alpha= 645/773$ and $\beta = 0.521/0.531$. Other variants of empirically modified potentials, such as $\alpha V_{\rm Ar-Ar}(\beta + R)$ used in Ref. \cite{bezrukov2019empirically}, yielded similar outcomes. This method is a straightforward way to retain the simple two-body pairwise model while ensuring the appropriate cohesive energy and equilibrium lattice constant. This is achieved by incorporating many-body corrections and zero-point energy effectively \cite{schwerdtfeger2016towards}.

\subsubsection{Site Stability}

To evaluate the thermodynamic stability of a structure with $n$ vacancies, we calculate a variant of the Gibbs (or enthalpy) free accommodation energy for the structure \cite{lara2019effects,kleshchina2017modeling,ozerov2019computational}: $\Delta E_N(n) = E_{\rm Cs}(n,N) -E_{\rm Ar}(N) (N-n)/N$. Here,
$E_{\rm Cs}(n,N)$ signifies the total energy of the system when a Cs atom is introduced into the crystal. Conversely, $E_{\rm Ar}(N)$ represents the total energy of a pure Ar crystal comprising N atoms, excluding the vacancies.

For the system to exhibit stability, it is crucial that the dissociation of $M$ systems, each containing $n$ vacancies, into $M'$ subsystems with $n'$ vacancies and $M''$ subsystems with $n''$ vacancies be energetically unfavorable. This requirement can be represented as: $ M \Delta E_N(n) < M' \Delta E_N(n') + M'' \Delta E_N(n'')$, with the condition $n'' M'' + n' M' = n M$ ensuring the conservation of atom numbers. For instance, two systems each with 5 vacancies can evolve into a more stable configuration of one system with 4 vacancies and another with 6 vacancies. Thus, for $n$ vacancies to form a stable trapping site, $\Delta E_N(n)$ must fall within the convex hull of the free energy curve as it varies with the number of vacancies \cite{tao2015heat,kleshchina2017modeling}.

Despite its significance, $\Delta E_N(n)$ is just one of several vital factors, such as migration and activation energies, needed for understanding the trapping sites. The $\Delta E_N(n)$ formula is primarily beneficial for small crystals, or in our case, a small single crystallite in a poly-crystal. Alternatively, for $N\rightarrow \infty$, the thermodynamic limit is often applied with $E_{\rm Ar}(N) /N \approx E_{\rm at}$, resulting in $\Delta E^\infty_N(n) = E_{\rm Cs}(n,N) -E_{\rm Ar}(N) + n E_{\rm at}$ \cite{lam1985multiple,lam1986calculations,sabochick1988atomistic,zenia2016stability,lounis2016stability}.

\begin{figure}[ht]
\centering
\includegraphics[width=1\linewidth]{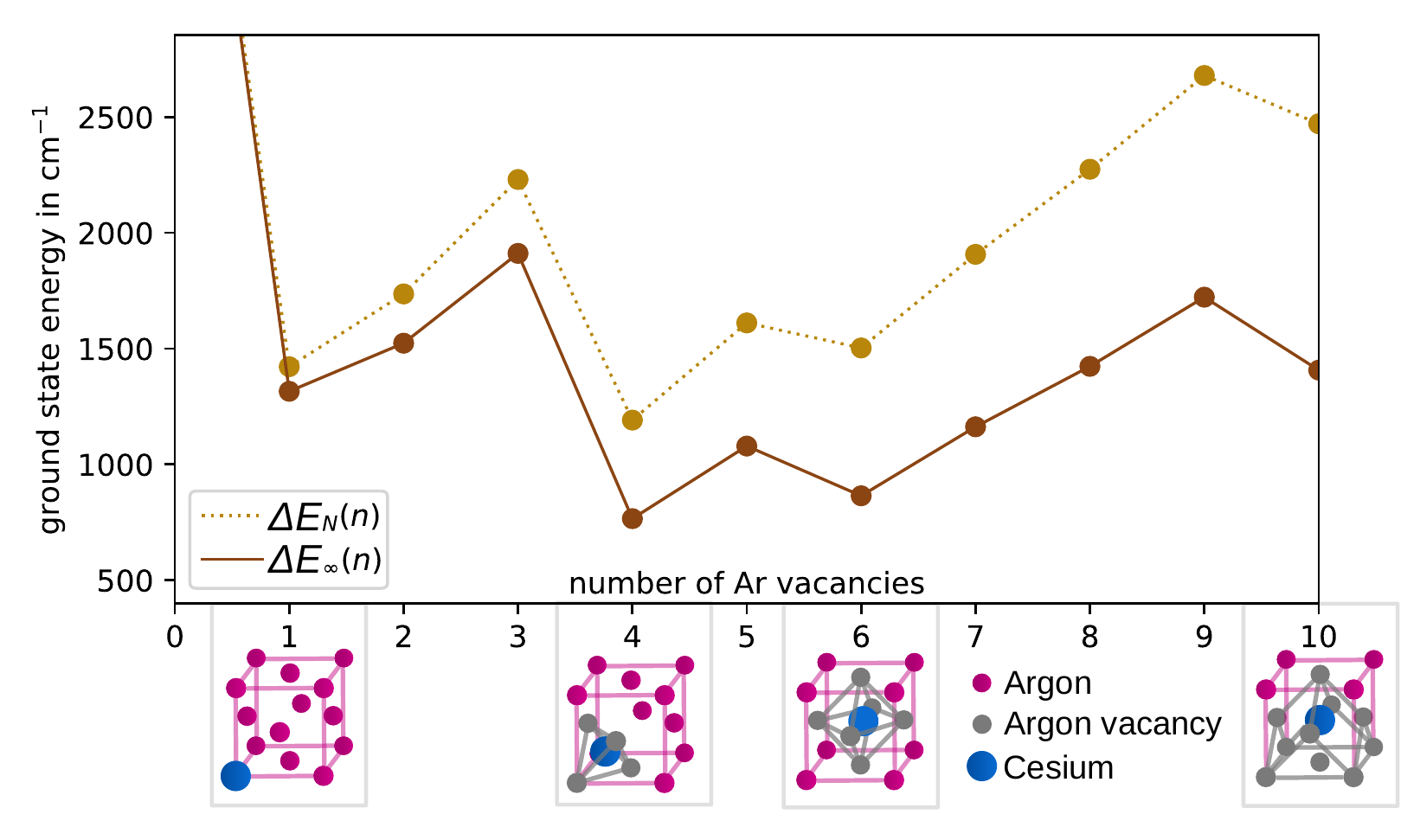}
\caption{For a given number of vacancies $n$, the optimized configuration found to lower $\Delta E(n)$ is displayed (calculation performed with $N=1372$ atoms). The figure illustrates both $\Delta E(n)$ and $\Delta E^\infty (n)$, alongside a depiction of the final structure realized for the Cs atom and a few of the nearest-neighbor Ar atoms.}
\label{fig:stabilityVacancies}
\end{figure}

In order to find the most stable (minimum energy) configuration at a fixed $n$ value, we do not explore the entire position space for the Cs atom and the vacancies through varying random initial configurations. Instead, we initialize with pre-established defect structures in fcc crystals, using positions sourced from existing literature \cite{cotterill1966morse,damask1968defects,johnson1973empirical,lam1985multiple,lam1986calculations,sabochick1988atomistic,nemirovich2007vacancies,wang2013defect,zenia2016stability,lounis2016stability,lara2019effects,peng2020solute}, specifically \cite{footnoteTrapping}.

In anticipation of the necessity for further studies, such as the Jahn-Teller dynamical effect, we permit the Cs atoms and the initial few (typically $\sim 60$) layers of nearest neighboring Ar atoms to move and optimize their position to minimize $\Delta E_N(n)$ as much as possible. We utilize the Adaptive Moment Estimation (Adam) \cite{kingma2014adam} and its modified version, Nesterov-accelerated Adaptive Moment Estimation (Nadam) \cite{dozat2016incorporating}, collectively considered as "the best overall choice" among gradient descent optimization algorithms \cite{ruder2016overview,haji2021comparison}. Upon nearing the optimum, employing the second-order derivative offers a faster method to adjust the equilibrium position of the normal coordinates \cite{simons2003introduction}. This procedure is iteratively applied, optimizing positions until a minimum is attained. Figure \ref{fig:stabilityVacancies} showcases results from such optimized geometry. However, it should be noted that using initial fixed (pure) fcc Ar and vacancy positions suffices to reproduce the results with satisfactory accuracy for stable site predictions, eliminating the need for atomic position optimization.

This study's findings contrast from the previous one carried out using a Lennard-Jones potential \cite{ozerov2019computational,ozerov2021generic}. Our study indicates that the most stable trapping sites probably are T$_d$ with 4 vacancies and O$_h$ with 6 vacancies. This seems consistent with our observation of two triplet lines in absorption.

\subsection{Simulation of the position of the absorption lines}

\subsubsection{Line Positions in the Pairwise Approach}\label{sec:rotation-matrices}

Contrary to the observations reported by Ref.~\cite{balling1983laser}, our data, as shown in Fig.~\ref{fig:abs_spectra}, does not show evidence of the Cs$(6s\rightarrow 5d)$ transition. Only Cs$(6s\rightarrow 6p)$ triplet structures appear visible, with minimal evidence of any other Cs levels involved. Also Cs dimers seem to be absent at the low density and temperature we are operating at. We are therefore confident that a simple pairwise approximation of the diatomic-in-molecule (DIM) type of approach, involving only the Ar ground state and Cs$(6s)$, Cs$(6p)$ levels, should yield accurate results to simulate the Cs$(6s\rightarrow 6p)$ absorption spectra.

The detailed calculation is outlined in Appendix~\ref{Energy_calculation}. Briefly, the energy shift of the $Cs(6s)$ is calculated in a similar manner as was done for the stability study. Now, we include the 6p manifold, where the energies are the eigenvalues of the Hamiltonian. These are calculated by summing up the interaction 
$\langle L' M' | \hat V_{\text{Cs,Ar}} (\bm R_{\text{Cs,Ar}} = \{X,Y,Z\}) | L M \rangle$ between a Cs atom (with $| 6p  M \rangle$ levels quantized along an arbitrary but fixed axis $z$) and each Ar atom. This interaction is defined by the matrix obtained using Wigner rotations:

\begin{widetext}
\begin{equation}
  \frac{1}{3} (2 V_\Uppi(R) + V_\Upsigma(R) ) \bm I_3 + \frac{1}{6} \frac{V_\Upsigma(R) - V_\Uppi(R)}{R^2} 
  \bordermatrix{
 & | M= - 1 \rangle  &  |M= 0 \rangle & | M=  1 \rangle \cr
 & X^2 + Y^2 - 2 Z^2  &  3 \sqrt{2}  (X + i Y) Z & -3   (X +  i Y)^2\cr
 & 3 \sqrt{2}   (X -  i Y) Z & -2   (X^2 + Y^2 - 2 Z^2) & -3 \sqrt{2}  (X +  i Y) Z \cr
 & -3   (X -  i Y)^2 & -3 \sqrt{2}   (X -  i Y) Z &  X^2 + Y^2 - 2 Z^2
   }
   \label{rot_matrix}
\end{equation}
\end{widetext}

Here, we also include the spin-orbit $A^{\rm SO}(R_{\text{Cs,Ar}}) \bm L.\bm S$ term, where $R = \sqrt{X^2+Y^2+Z^2}$.

The Hund's case (a) potential curves $V_\Uppi(R)$, $V_\Upsigma(R)$, and the spin-orbit coefficient $A^{\rm SO}(R)$, where $R$ denotes the Cs-Ar separation, are constructed as shown in Fig.~\ref{fig:potentialCurves}. They are based on the most recent $V_{\rm X\, ^2\Upsigma_{1/2}^+}$, $ V_{\rm A\, \Uppi_{1/2}}$, $V_{\rm A \,\Uppi_{3/2}}$, $V_{\rm B\, ^2\Upsigma_{1/2}^+}$ potentials \cite{kobayashi2016ab}. The long-range part is taken from the potentials of Ref.~\cite{blank2012m+}, which provide more points, but are calculated from a smaller basis sets and hence, are less accurate at short range.

Given that the significant variation of the spin-orbit coefficient $A^{\rm SO}(R)$ is limited to $R$ values smaller than the distance between the closest neighbours \cite{heaven2010potential}, as seen in Fig.~\ref{fig:potentialCurves} a), the use of a constant spin-orbit coefficient $A^{\rm SO}$ yields minimal impact (a maximum shift difference of $\SI{10}{cm^{-1}}$ in the final Cs(6s)-Cs(6p) transition lines). Hence, in the following analysis, we simply use the $V_\Uppi , V_\Upsigma$ potential curves and a constant spin-orbit coefficient $A^{\rm SO}$ based on the experimental value.

\begin{figure}[ht]
    \centering
    \includegraphics[width=0.9\linewidth]{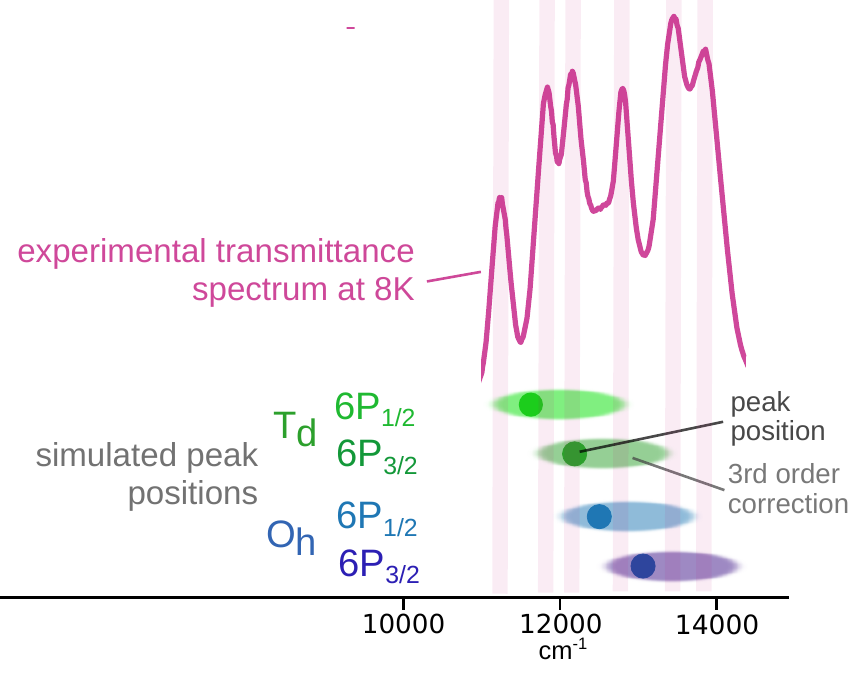}
    \caption{This figure presents a comparison between an experimental transmittance spectrum (represented by the solid pink lines), for the two trapping sites, O$_h$ with 6 vacancies, and T$_d$ with 4 vacancies. The lower section indicates the position of the $\SI{0}{K}$ lines employing pairwise potential, where dots correspond to the two-body cases with potentials sourced from \cite{kobayashi2016ab}. The influence of zero point energy on these lines is marginal (at most $\SI{150}{cm^{-1}}$). However, the incorporation of the third-order effect (refer to the main text) results in an additional shift that we predict to fall within the elliptical regions.}
    \label{fig:abs_spectra_th}
\end{figure}

The big dots on Figure \ref{fig:abs_spectra_th} display the results for the Cs(6s)-Cs(6p) transition line positions. In this $\SI{0}{K}$ theoretical computation, only doublets emerge, not triplets as would be anticipated in highly symmetric trapping sites, due to the lack of thermal or dynamical broadening. Nevertheless, the experimental and calculated positions show a reasonable degree of agreement.

It may seem surprising initially that the most shifted lines, relative to the gas phase case, are those with 6 vacancies. This happens while the more constrained structure with 4 vacancies appears to be hardly affected. The observation that the red triplet is T$_d$ and the blue one is O$_h$ is contrary to what has been noted (or simulated) for other systems such as Na in Ar \cite{ryan2010investigations,jacquet2011spectroscopic}. This discrepancy can be qualitatively explained by a simplified model where Cs is surrounded by Ar atoms in a symmetric spherical environment. % \footnote{Indeed, this scenario occurs in solid He, with Cs encapsulated in a He bubble \cite{moroshkin2008atomic}. }

\begin{figure}[ht]
    \centering
    \includegraphics[width=1\linewidth]{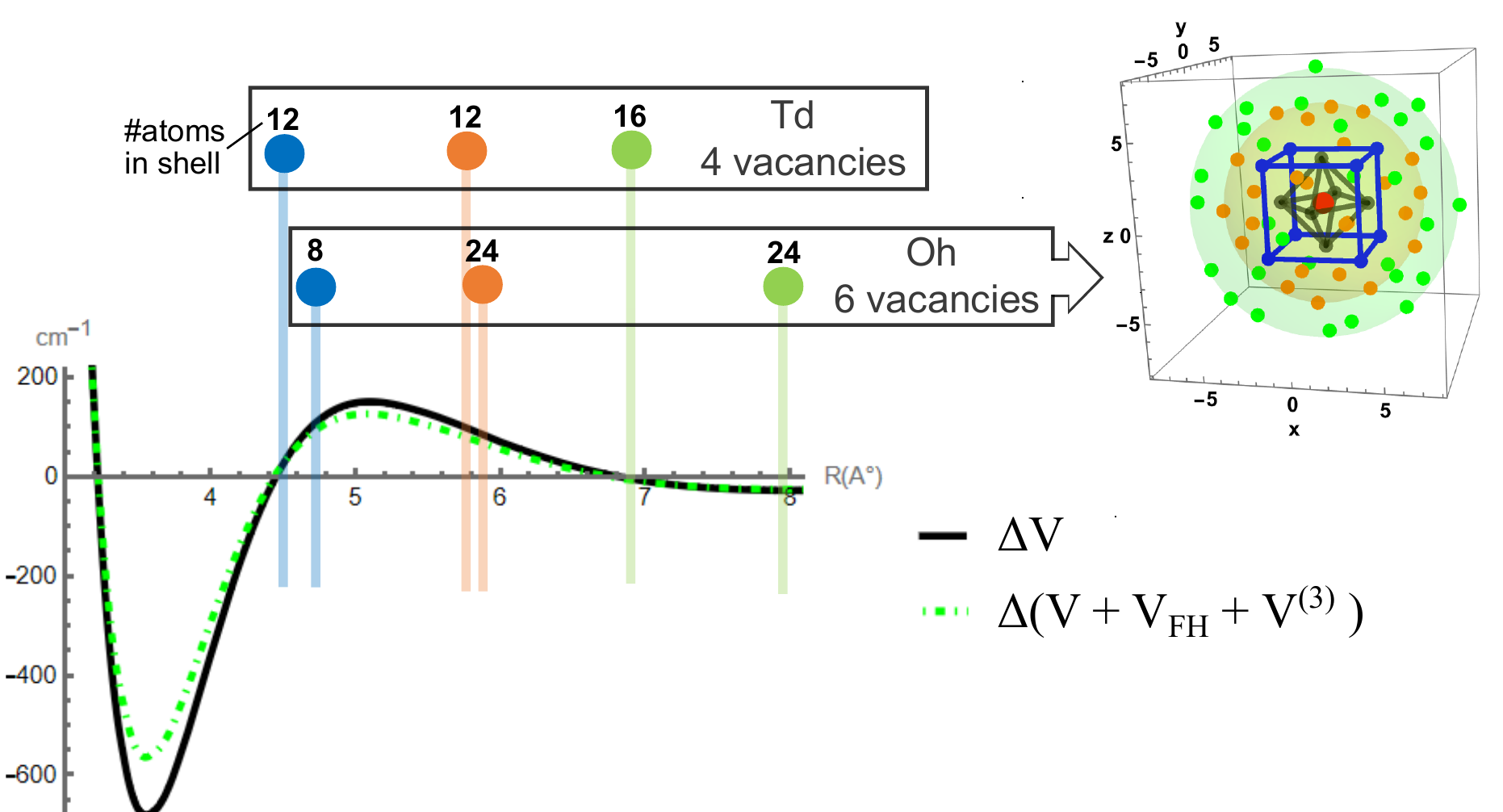}
    \caption{This figure illustrates $\Delta V = 2 V_\Uppi +  V_\Upsigma-V_{\rm X}$ Cs-Ar potentials with an offset of the asymptotic values. The first $R_1$, second $R_2$, and third $R_3$ nearest neighbor shells in the T$_d$ and O$_h$ scenarios are provided (for the O$_h$ case, an example of the Cs and Ar positions is also given).}
    \label{fig:potentialCurves_diff}
\end{figure}

In this spherical context, the average interaction term becomes $2 V_\Uppi + V_\Upsigma$ (refer to matrix (\ref{rot_matrix})). Consequently, a valuable approach to study the shift of the $6s-6p$ line position involves examining $2 V_\Uppi + V_\Upsigma-V_{\rm X}$. This is demonstrated in Fig. \ref{fig:potentialCurves_diff}, along with the positions of the nearest neighbors in both the T$_d$ and the O$_h$ structures as found in our stability study. 

Fig. \ref{fig:potentialCurves_diff} highlights that the blue shift, resulting from positive $\Delta V = 2 V_\Uppi + V_\Upsigma-V_{\rm X}$ values, is more pronounced in the O$_h$ case than in the T$_d$ scenario. This difference mainly arises because, in the O$_h$ situation, the first 8 and second 24 nearest neighbors (the solvation-type shell) each contribute to a blue shift of approximately $\SI{100}{cm^{-1}}$. Meanwhile, in the T$_d$ case, the first shell almost causes no shift due to the cancellation of $2 V_\Uppi + V_\Upsigma$ and $V_{\rm X}$ shifts, and the second shell only contains 12 atoms. The cumulative effects of these shifts, coupled with the red shifts produced by shells at greater distances, offer a qualitative explanation for the $\sim \SI{2000}{cm^{-1}}$ blue shift in the O$_h$ scenario. Also, the cancellation effect results in T$_d$ lines appearing near the gas phase location. 

In line with our approach for Ar-Ar potentials, to enhance the comparison of the line positions, we should include the zero-point energy and Third-Order Many-Body interactions. However, as these factors merely provide corrections, we didn't use them to re-optimize the atomic positions.

\subsubsection{Third order (dipole-dipole-dipole type) correction} \label{sec:third-order}

To the best of our knowledge, the third-order correction has not been previously utilized for simulating absorption spectra \cite{crepin1999photophysics,jacquet2011spectroscopic}. However, this correction is non-negligible, particularly for ground states and even more so for excited states.

We currently lack accurate three-body Cs-Ar-Ar potential curves, leading us to approximate their contribution \cite{diaz1980combination,dridi2022development}. As elaborated in Appendix \ref{third_order_appendix}, we employ two and three-body expressions for the pure long dipolar range part to extrapolate an appropriate three-body potential also applicable for shorter distances. Following that, we aggregate it over all atoms. Unfortunately, the nearest neighbors play a significant role \cite{jacquet2011spectroscopic}, and it is in the area where our formula proves to be the least accurate. Consequently, we opt for a simpler approximation, akin to the approach used by Mut\^{o}-Stenschke-Marcelli-Wang-Sadus \cite{stenschke1994effective,dridi2022development,akhouri2022thermodynamic}. We average the formula, employing a mean-field approach for the position of the third Ar atom, resulting in an effective correction to the two-body Cs-Ar potential. 

These are straightforward density-dependent expressions that connect the averaged three-body interactions to the two-body ones. This connection facilitates merging both types of interactions into effective Cs-Ar two-body potential curves. With this methodology, the final Cs-Ar potential curve is a corrected version of the original one, offset by the effective three-body contribution $\bar E_{12}^{(3)}$. Ultimately, we obtain an effective two-body potential as described by Eqs. (\ref{correction_potential}) and (\ref{coeff_C6_C9}).

The influence of this adjustment can be qualitatively estimated from Eq. (\ref{averaged_pot}). By utilizing $2 V_\Uppi +  V_\Upsigma-V_{\rm X}$ as a measure of the shift, we observe that the third-order effect $\bar E_{12}^{(3)}$ displays a $\frac{8\pi \rho}{3 R^6} (C_9^*-C_9) >0$ dependence. This implies a probable net blue shift, where $\rho$ is the Ar density and $C_9,C_9^*$ represent the triple dipole coefficient for the ground and excited states, respectively. This effect can also be interpreted by examining the potential curves provided in Fig. \ref{fig:potentialCurves}. In this figure, the depicted third-order excited state potential curve is computed similarly to the ground state case. Hence, we have $V_\Upsigma^{(3)} =  (V_\Upsigma -E(6p))  \frac{1}{3} \frac{8\pi \rho}{9 } \frac{C_9^*}{ C_6^* }$ and $V_\Uppi^{(3)} = (V_\Uppi -E(6p)) \frac{1}{3} \frac{8\pi \rho}{9 } \frac{C_9^*}{ C_6^* }$.

\subsubsection{Zero Point Energy Correction}

In order to incorporate the quantum effects attributable to the zero-point energy, we utilize the Feynman-Hibbs approach, equivalent to the Wigner-Kirkwood method. This approach introduces a temperature-dependent effective correction to a pair potential between ground state atoms, given as: $\Delta V_{FH}(R)=\sigma_0^2 (V''(R)+2V'(R)/R)$, where $\mu$ is the reduced mass, and $\sigma_0 = \sqrt{\frac{\hbar^2}{24 \mu k_B T}}$ represents the Gaussian width of quantum particles, as derived from the uncertainty principle \cite{berendsen2007simulating,stroker2022thermodynamic}.

\subsubsection{Absorption Line Positions}

Upon integrating the zero-point energy and third-order correction, we find novel line positions for the absorption spectra depicted in Fig. \ref{fig:abs_spectra_th}. In this computation, for the excited states, we applied the effective two-body potential (two-body corrected with three-body), and for the ground state, we additionally incorporated the Feynman-Hibbs zero-point energy correction, $\Delta V_{FH}(R)$, assuming an experimental temperature of $\SI{8}{K}$. This procedure was not used for the Ar-Ar interaction, given that we already have a well-scaled potential. However, it is intriguing to validate this method against the known Ar-Ar case. As Fig. \ref{fig:potentialCurves} suggests, our method of amending a two-body potential with $\Delta V_{FH}(R)+ \bar E^{(3)}$ accurately replicates the scaled potential we used.

A key insight is that the zero-point energy is quite negligible (on the order of $\SI{100}{cm^{-1}}$), whereas the third-order correction significantly impacts the outcome and thus should not be disregarded. Indeed, the shift generated by the third-order effect is extremely sensitive to the selected cut-off parameters, and it can be either attractive or repulsive, contingent on their values. As such, the induced shift should be viewed more as an indicator of the effect rather than a precise quantitative result, which is why we chose to represent its effect as an uncertainty in our findings. The ellipses in Fig. \ref{fig:abs_spectra_th}, acting as effective error bars, have been roughly assessed using different cut-off values and exponents in the power of the cut-off function (see Appendix \ref{third_order_appendix}), along with the potential curves $V_\Upsigma^{(3)}$ and $V_\Uppi^{(3)}$ shown in Fig. \ref{fig:potentialCurves}.

Despite the substantial uncertainty of the third order, we can observe a qualitative agreement between theoretical line positions and experimental ones, such as the O$_h$ lines being blue-shifted compared to the T$_d$ ones. However, the correct line separation between O$_h$ and T$_d$ remains elusive. Using theoretical potential curves from Ref. \cite{blank2012m+} yielded similar results but they were blue-shifted by approximately $\SI{1000}{cm^{-1}}$ compared to results obtained using the more precise and recent potentials from Ref. \cite{kobayashi2016ab}). Still, this does not resolve the issue of the line separation between O$_h$ and T$_d$, which consistently remains around $\SI{1000}{cm^{-1}}$, half of the value observed in the experiment. This discrepancy indicates the need for further investigations to reconcile the differences between experimental and simulated positions. High-accuracy Cs-Ar potential curves are required, and better Cs-Ar-Ar third-order calculations, particularly for the nearest neighbor distances, should be developed.

The line position alone is a good indicator of the trapping site but is not always sufficient to confirm it. Even though the calculated positions for the T$_d$ 4-vacancies and O$_h$ 6-vacancies cases align with the observation, it is prudent to also consider other stable trapping sites on the convex hull such as the single vacancy case, where one Ar atom is replaced by a Cs atom, and the 10-vacancies case. The simulation of the absorption lines for the single-vacancy substitution site results in very red-shifted theoretical lines below $\SI{10000}{cm^{-1}}$, where no experimental lines are observed, ruling out this case. However, the 10-vacancies case predicts line positions at $\SI{13760}{cm^{-1}}$ and $\SI{13210}{cm^{-1}}$.

To study the lineshape and not just the central position, we now need to incorporate thermal broadening of the lines due to electron-lattice interaction.

\begin{figure}[ht]
    \centering
   \includegraphics[width=0.92\linewidth]{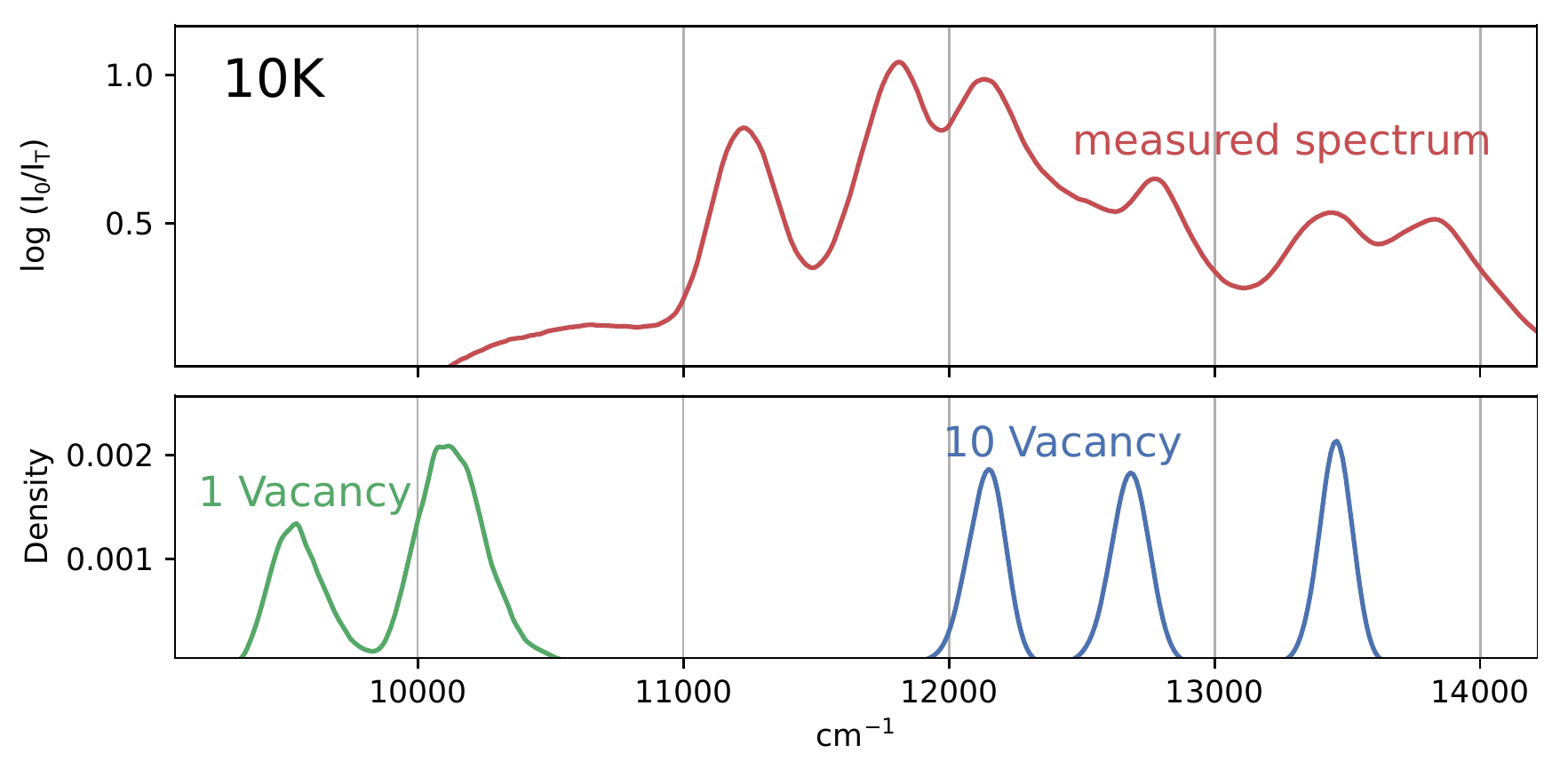}
    \caption{Comparison between the experimental absorption spectrum and theoretical semi-classical Mulliken formula (\ref{eq_Mulliken}) for the 1 and 10 vacancy sites at \SI{10}{K}.}
    \label{fig:1 and 10 vacancy}
\end{figure}

We have already provided the results for the 1 and 10 vacancies cases in Fig. \ref{fig:1 and 10 vacancy},  confirming that they can likely be dismissed due to the large shift for 1 vacancy and the overly large and inverted triplet splitting for 10 vacancies. % In addition to the fact that the stability diagram show them less probable than the 4 and 6 vacancies cases. 
Henceforth, we will focus primarily on the 4 and 6 vacancies cases.

\subsection{Lattice interaction}

We must also account for the fact that the nuclei are not stationary, due to zero-point energy and temperature effects.

\subsubsection{Electron-lattice Interaction for Non-equilibrium Positions}

We employ pair-wise potentials to compute the electronic ground state energy $V(\bm R)$ for the positions of the Cs and $(N-n)$ Ar nuclei $\bm R = \{\bm R_1, \ldots, \bm R_{N-n+1}\}$ around the equilibrium point $(\bm R_0)$, as determined by the stability study.

Subsequently, as detailed in Appendix \ref{JahnTeller}, we diagonalize the mass-scaled Hessian matrix, computed via finite difference, to identify the normal mode coordinates $\bm Q^{\rm n}$. The Hessian matrix provides all the oscillation frequencies; this method is a more accurate way of including the zero-point effect than the previously used Feynman-Hibbs formula.

\subsubsection{Semi-classical Approximation for the Transitions}

To study the laser excitation of the $(6s)$ cesium atom toward the $6p$ manifold under the dipolar operator $\hat d$, we start by noting (using the Beer-Lambert-Bouguer's law) that the spectral density optical absorption coefficient $A(E)$ for a photon of energy $E$ is given by the sum over all initial vibronic levels $|i\rangle$, populated with probability $P_i$, with state $|\Psi_i\rangle$ of energy $E_i$, towards all possible final ones $|\Psi_f\rangle$ of energy $E_f$:
\begin{equation}
A(E) \propto \sum_{f,i} P_i  \left| \langle \Psi_f | \hat d | \Psi_i \rangle \right|^2
\delta( E -(E_f-E_i) ) 
\end{equation}
$A(E) dE$ is the absorption coefficient for a photon in the energy band $(E, E+dE)$. Several approximations are derived in Appendix \ref{Semi_classical_approximations}. For instance, we neglect the variation of the dipole $d_{i f}(\bm Q)$ with the internuclear distances and assume it to be constant (it is simply the 6s to 6p dipole transition). We will primarily use two approximations: 
\begin{eqnarray}
A(E) &\propto &\int  P_{\rm g}(\bm Q)  \delta[  E -(V_{\rm e}(\bm Q) - V_{\rm g}(\bm Q)) ]  {\rm d} \bm Q
\label{eq_Mulliken} \\
A(E) & \propto & \sum_{i } P_i \int  \left|  \Psi_i(\bm Q) \right|^2  \delta[  E -(V_{\rm e}(\bm Q) - E_i) ]  {\rm d} \bm Q
\label{Reflection_approximation} 
\end{eqnarray}
The first one (equation (\ref{eq_Mulliken})) is called the Mulliken approximation. The second one (equation (\ref{Reflection_approximation})) is referred to as the Reflection approximation. $V_{\rm e}(\bm Q)$ is the excited electron (6p) potential energy curve, and $V_{\rm g}(\bm Q)$ is the ground state one.

$P_{\rm g}(\bm Q)$ can be chosen by classical statistics $P_{\rm g}(\bm Q) \propto e^{- V_{\rm g}(\bm Q)/k_B T}$, and the formulae are thus called \textit{classical}. However, it is clearly better to choose the quantum-statistical mechanical probability distribution $P_{\rm g}(\bm Q)  =  \sum_{i } P_i \left|  \Psi_i(\bm Q) \right|^2 $, in which case we call them \textit{semi-classical}.

Depending on the derivation chosen, the classical Mulliken approximation can be derived from the Reflection one with an additional approximation, as in Ref. \cite{lax1952franck}. Conversely, the reflection approximation can be derived from the Mulliken one, as in Ref. \cite{heller1978quantum}. Hence, it is not clear which formula is the most accurate. However, proper derivation (see Appendix \ref{Semi_classical_approximations}) tends to favor the (semi-)classical Mulliken approximation. Indeed, the Mulliken difference potential $E -(V_{\rm e}(\bm Q) - V_{\rm g}(\bm Q))$ is known to be a quite good approximation for a Franck-Condon factor \cite{mulliken1971role,tellinghuisen1987franck} as it favors transitions where the kinetic energy term is identical between ground and excited state. This is where the phase of the ground and excited wavefunctions matches best, thereby favoring their overlap, as also highlighted by the one-dimensional WKB approximation.

Thus, we start with the semi-classical Mulliken approximation.

We will use Monte Carlo integration to estimate the integrals numerically. Thanks to the spin-orbit interaction and the interaction matrix we have, in the Cs(6p$_{1/2}$,6p$_{3/2}$) manifold, six eigenvalues $V_{i,\rm exc}(\bm Q)$ (the degeneracy of which depends on the symmetry of the trapping sites). The spectra are simply given by summing the histogram of all eigenvalue cases.

\subsubsection{Phonons thermal effects}

At temperature $T$, the average number of phonons with frequency $\omega$ is given by the Bose-Einstein distribution $\bar n_{\omega}=\frac{1}{e^{\hbar \omega/k_B T}-1}$. The probability of exciting this phonon mode with $n$ quanta (of energy $E_n = (n+1/2) \hbar \omega$) is given by the geometric distribution $P_n = e^{- n  \hbar \omega /k_B T} \left[ 1-e^{-\hbar \omega/k_B T} \right]$. In the semi-classical approximation, we use the true (quantum) probability distribution function of a harmonic phonon mode of normal mode coordinate $Q_\omega$ and angular oscillation frequency $\omega$, as given by \cite{barragan2018harmonic}: $P_\omega(T)  = \frac{1}{\sqrt{2\pi \hbar [\bar n_{\omega}+1/2]/\omega }} e^{-\frac{\omega^2 Q_\omega^2}{2 \hbar \omega [\bar n_{\omega}+1/2]}}$. Thus, if we perform the classical Mulliken calculation with a Boltzmann distribution at a temperature $T'=T'(\omega)$ such that

\begin{equation}
    \frac{T'(\omega)}{T}=  \frac{\hbar \omega}{k_B T}  [\bar n_{\omega}+1/2] = \frac{\hbar \omega}{2 k_B T} \left[ 
\tanh{ \frac{\hbar \omega}{2 k_B T}}
\right]^{-1}, \label{scaled_temp} 
\end{equation} 

we will get the correct quantum spectral band contour in the approximation of harmonic motion under the $\frac{1}{2} \omega^2 Q_\omega^2$ potential, as if we had performed the full real quantum calculation at $T$. Thus, a classical Mulliken simulation at $T'$ is equivalent to a semi-classical Mulliken simulation at $T$. This scaled temperature is a way to include phenomenologically the position oscillation due to the zero point energy in classical simulations \cite{bergsma1984electronic}. The scaled temperature was used, for instance, to study Na in Ar using Molecular Dynamic simulation in Ref. \cite{ryan2010investigations}, where $T=\SI{12}{K}$ and so $T'=\SI{45}{K}$ was used for the pure argon subsystem that thermalizes the Na deposition on other Ar layers.

In argon, the Debye frequency is $ \omega_D = k_B \times \SI{93.3}{K}/\hbar$, leading to an effective temperature of $T'\approx \SI{44}{K}$ at low temperatures. Because we have several frequencies (one per mode), we used the scaled temperature $T'(\omega)$ for each of them and chose the mode coordinate $Q_\omega$ according to the distribution $P_\omega(T'(\omega)) \propto e^{-\frac{\omega^2 Q_\omega^2}{2 k_B T'(\omega) }}$.
The energy levels are then calculated for the chosen nucleus position, as done previously for the equilibrium positions. A histogram for the energy difference between excited levels and the ground state one finally gives the absorption coefficient at each energy according to Eq. (\ref{eq_Mulliken}). For this simulation, only Ar atoms within a specific cut-off distance $R_c$ from the Cs atom are considered. For all simulations presented here, convergence is obtained for $R_c=\qty{3}{nm}$. This corresponds to approximately 3000 Ar atoms, the positions of which have been optimized to relax the ground state energy. However, already $ R_c=\qty{1}{nm}$ leads to reasonable convergence.

To benchmark our method, we compared it with Na in Ar results from \cite{ryan2010investigations} and indeed found very similar absorption spectra for the $T_d$ site.

For our Cs-Ar system, calculations are done with a bare two-body Cs-Ar potential. We found no effect on the shape when adding third order corrections or when using other potentials such as from \cite{blank2012m+}. 

\begin{figure}[ht]
    \centering
    \includegraphics[width=1\linewidth]{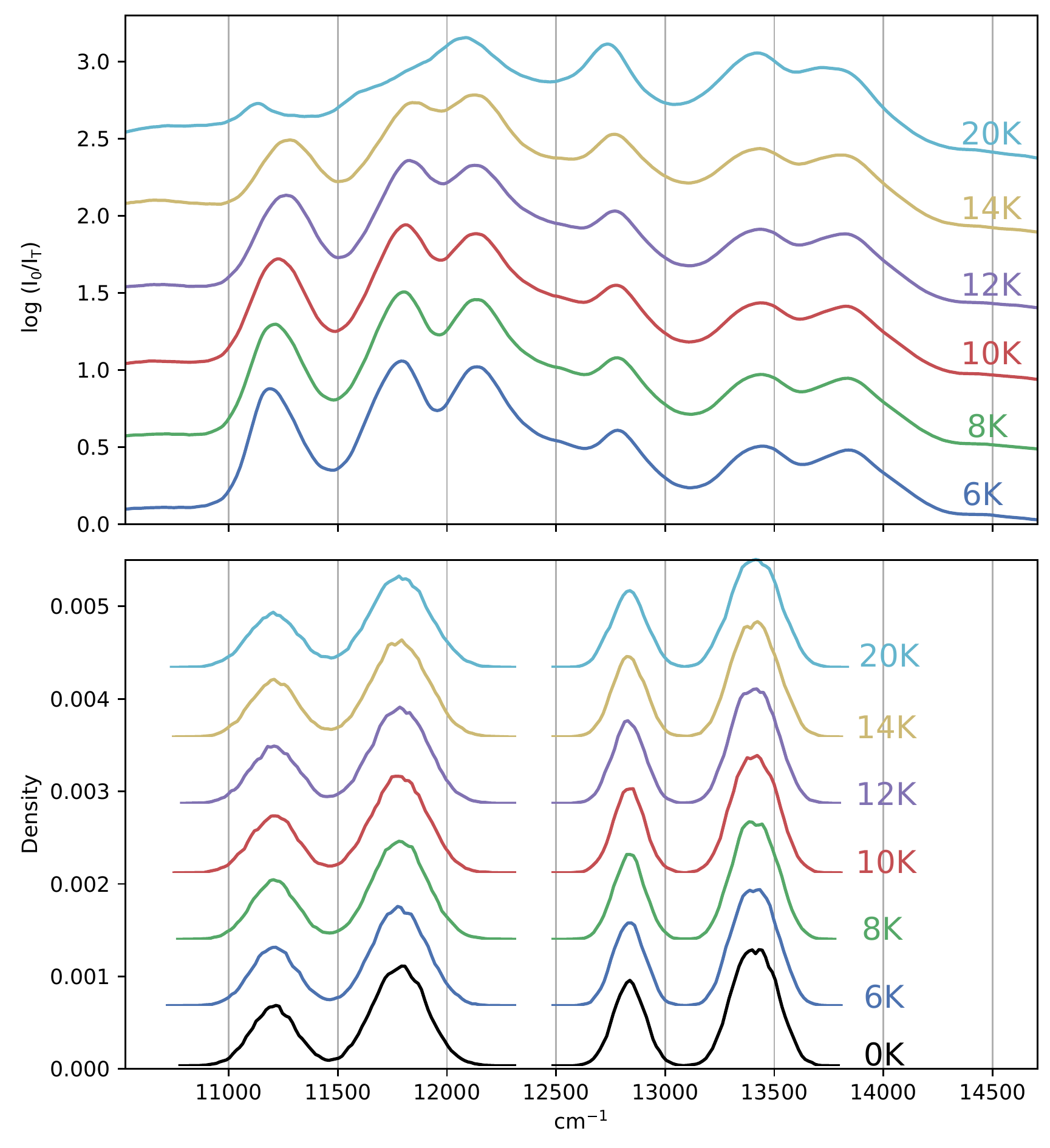}
    \caption{Comparison between experimental absorption spectra and theoretical semi-classical Mulliken formula (eq.(\ref{eq_Mulliken})) at different temperatures $T = \qtylist{6;8;10;12;14;20}{K}$ and also for $T = \qty{0}{K}$ for the theory. The theoretical spectra are created from a kernel density estimation of the Monte Carlo simulation results. The results for the Td and Oh site have been shifted in energy by a global offset (that are  $\SI{500}{cm^{-1}}$ and $\SI{-380}{cm^{-1}}$ for O$_h$ and T$_d$ respectively) to enable easier comparison with the experimental data. For better visual clarity, vertical offsets have been applied to both the experimental and the theoretical spectra.}
    \label{fig:pure_spectra_th}
\end{figure}

The results for 10000 simulated samples of the semi-classical Mulliken simulations are given in Fig. \ref{fig:pure_spectra_th} for the O$_h$ and T$_d$ cases. However, as discussed previously, the exact position cannot be perfectly reproduced for all presented simulations. Therefore, we shift the theoretical spectra for each O$_h$ and T$_d$ cases.
The experimentally observed line broadening is well reproduced, but the triplets are not visible because the splitting of the 6p$_{3/2}$ doublet is too low to be resolved.
The line positions barely evolve with temperature and the widths only increase by less than $\qty{20}{cm^{-1}}$ for the considered temperature range. 
The reason that almost no temperature effect arises is that we have a phonon angular frequencies spectral density very similar to the one for a pure fcc matrix, so ranging from  $\sim \SI{0}{K}$  to the Debye frequency $\sim \SI{94}{K}$. Thus, whatever the experimental temperature  $T = \SIrange{6}{14}{K}$, the energy spanned $V_{\rm g}(\bm{Q})$  by the ground state motion does not vary much with the temperature because the scaled temperature $T'(\omega)$ is always $T'(\omega)  \approx \frac{\hbar \omega}{2 k_B}$ and that way it is  not very sensitive to the actual experimental temperature.

It is interesting to note that a purely classical model using $T$, not $T'(\omega)$, as done for Na in Ar in Ref.  \cite{boatz1994monte}, is able to resolve the peak degeneracy and to show temperature effects when $T$ varies. This is an example of a faulty model that gives erroneous results but that might, at first glance, look reasonable.

Another intriguing aspect concerns the fact that a simple Crystal Field model used in Ref. \cite{Kanda1971} was able to reproduce the triplet structure quite well. Thus, we now study this Crystal Field model based only on first-order electron-lattice interaction in order to understand why a simplification of the model might, strangely, give better results than the more complete one we just used.

\subsubsection{Crystal Field: First order electron-lattice interaction}

The crystal field model is based on the natural expectation that, for better physical insight, it is worth linking the motional modes to the symmetry group of a given trapping site using the so-called crystal field theory \cite{pavarini2012crystal}.

The normal modes $\bm Q^{\rm n}$ coordinates have been chosen to obtain a diagonal Hamiltonian with ground state potential energy given by
$
    \langle 6s   |\hat V | 6s  \rangle =   V_0 +  \sum_{k} 
 \frac{1}{2} \omega_k^2 {Q^{\rm n} _k}^2$.
However, within the Cs(6p) manifold, these coordinates $\bm Q^{\rm n}$ are not necessarily the most appropriate anymore.
Indeed, if we restrict ourselves to the first order series in nuclear coordinates, each of the 9 coefficients $V_{m',m}=\langle 6p m'  |\hat V | 6pm  \rangle$ of the interaction matrix (\ref{rot_matrix}), when summed over all Ar positions, contains only linear combinations of the $\bm Q^{\rm n}$. Thus, at most 9 different so-called interaction modes matter, each of them being a linear combination of the normal modes. 
These interaction modes are much more appropriate than the normal modes to give a physical insight into the Cs(6p) interaction within the Ar matrix. 

They can be calculated when the interactions, so the potential curves, are known. But, in the crystal field theory, the $Q_{ \Gamma  \gamma}$ interaction mode coordinates are simply predicted using the symmetries of the trapping site. A given trapping site is an element of a symmetry group with irreducible representations $\Gamma$ (and its row $\gamma$). From it, we find the $Q_{ \Gamma  \gamma}$, which are the linear combination of the normal modes invariant under the $\Gamma $ symmetries. The symmetries are also reflected on the electronic wavefunction that should have an adapted basis set $|  \Gamma  \gamma \rangle $. Finally, the electron-lattice interaction can be written at first order (so-called crystal-field approximation) as
$
H_{\rm CF} =  V^{(0)}({\bm r}) + \sum_{ \Gamma  \gamma} V_{ \Gamma  \gamma}^{(1)}({\bm r}) Q_{ \Gamma  \gamma}.
$

Focusing on the $6p$ manifold, the symmetries of the zero-order $V^{(0)}({\bm r})$ potential in the cubic O$_h$ case suggest $\Gamma = T_{\rm 1u}$. The adapted electronic basis $| \Gamma \gamma \rangle $ is the $6p$ real (tesseral) spherical harmonics $|x\rangle, |y\rangle, |z\rangle$. Similarly, $\Gamma = \rm T_{\rm 1}$ applies in the $\rm T_d$ case.

To evaluate the first-order term $\sum_{ \Gamma \gamma} V_{ \Gamma \gamma}^{(1)}({\bm r}) Q_{ \Gamma \gamma} $ within this basis, we use the Wigner-Eckart theorem \cite{tsukerblat2006group,dresselhaus2007group,gtpack2}:
\begin{equation}
\langle \Gamma_1 \gamma_1 | V_{ \Gamma \gamma}^{(1)} | \Gamma_2 \gamma_2 \rangle  = \langle \Gamma_1 || \hat V_\Gamma^{(1)} || \Gamma_2 \rangle  \langle \Gamma_1 \gamma_1 | \Gamma_2 \gamma_2,\Gamma \gamma \rangle
\end{equation}
This theorem delivers the selection rules and predicts the non-zero coefficients $\langle \Gamma_1 \gamma_1 | V_{ \Gamma \gamma}^{(1)} | \Gamma_2 \gamma_2 \rangle $ that only arise for some $Q_{ \Gamma \gamma}$ \cite{van1939jahn,toyozawa1966dynamical,cho1968optical}. The calculations' specifics are presented in appendix \ref{crystal_fiel_matrix} for multiple symmetry groups in the $|x\rangle, |y\rangle, |z\rangle$ basis. To better align with prior calculations, we also offer the results in the $|m\rangle =|-1,0,+1\rangle$ basis. Here, the Hamiltonian matrix $M_{\rm CF}$ for the T$_{\rm d}$ case (refer to Eq. (\ref{MCF_Td_Oh})) is given by:
\begin{widetext}
\begin{equation}
    \begin{pmatrix}
 V_{A_1} Q_{A_1} -\frac{  V_{E} Q_{E,2}}{\sqrt{3}}  & V_{T_2} \frac{ i Q_{T_2,1} + Q_{T_2,2} }{\sqrt{2}}  & - V_{E} Q_{E,1} - i V_{T_2} Q_{T_2,3} \cr
V_{T_2} \frac{ -i Q_{T_2,1} + Q_{T_2,2} }{\sqrt{2}}  & V_{A_1} Q_{A_1}+  2\frac{V_{E} Q_{E,2}}{\sqrt{3}}  & V_{T_2} \frac{ -i Q_{T_2,1} - Q_{T_2,2} }{\sqrt{2}} \cr
 - V_{E} Q_{E,1} + i V_{T_2} Q_{T_2,3} & V_{T_2} \frac{ i Q_{T_2,1} - Q_{T_2,2} }{\sqrt{2}} &   V_{A_1} Q_{A_1} - \frac{ V_{E}  Q_{E,2}  }{\sqrt{3}} 
\end{pmatrix}
\label{matrice_CF} 
\end{equation}
\end{widetext}
For O$_h$, the results are analogous, with the symmetries' names slightly altered. For simplicity, we maintain these notations for both T$_d$ and O$_h$ in the following discussions. Comparing with the matrix (\ref{rot_matrix}) provides physical insight.

The $A_1$ mode, denoted as $Q_{A_1}$, which varies as $X^2 + Y^2 + Z^2$, is a symmetrical radial mode that preserves symmetry. The $E_g$ mode is a tetragonal distortion, represented by $Q_{E,1}$ varying as $X^2 - Y^2$ and $Q_{E,2}$ as $2Z^2 - X^2 - Y^2$. It transforms O$_h$ symmetry into D$_{4h}$. The $T_{2}$ mode bends the crystal to D$_{3d}$ or C$_{3v}$, with $Q_{T,1}$, $Q_{T,2}$, and $Q_{T,3}$ varying as $X Z$, $Y Z$, and $X Y$ respectively.

Note that even if the matrix elements are linear in the interaction mode coordinates, the eigenvalues, which are the potential curves, are not linear. Hence, to calculate the Cs(6s-6p) absorption spectra, a numerical integration method such as a Monte Carlo simulation is necessary. In the Mulliken formula (\ref{eq_Mulliken}), we have
$
V_{\rm g}(\bm Q) = V_{\rm Cs(6s)-Ar} (\bm Q) + V_{\rm Ar-Ar} (\bm Q)$ and 
$
V_{\rm e}(\bm Q) = V_{\rm Cs(6p)-Ar} (\bm Q) + V_{\rm Ar-Ar} (\bm Q)$
leading to

\begin{equation}
V_{\rm e} (\bm Q) - V_{\rm g}(\bm Q) = V_{\rm Cs(6p)-Ar} (\bm Q) -  V_{\rm Cs(6s)-Ar} (\bm Q). \label{eq_pot_diff}
\end{equation}

In the linear crystal field approximation, $V_{\rm Cs(6s)-Ar} (\bm Q)$ is null (or a constant that we offset), hence Eq. (\ref{eq_Mulliken}) simplifies to

\begin{equation}
A(E) \propto \int \sum_{i=1}^6 \delta[E - X_i (\bm Q^{\Upgamma})] e^{- \sum_{ \Gamma  \gamma} \omega_\Gamma^2 Q_{ \Gamma  \gamma}^2/ 2 k_B T}  {\rm d} \bm Q^{\Upgamma},
\label{line_shape}
\end{equation}

where $X_i(\bm Q^{\Upgamma})$ are the eigenvalues of $M_{\rm CF}$ plus the spin-orbit matrix (\ref{SO_matrix_cho}), $A^{\rm SO}  {\bm L} \cdot {\bm S}$. This can be estimated using a Monte Carlo method where each $\frac{\omega_\Gamma Q_{ \Gamma  \gamma}}{ \sqrt{k_B T}}$ follows a standard unit normal Gaussian distribution \cite{cho1968optical}.

Assuming in the Boltzmann distribution that the ground state energy can be written as $V_{\rm g}(\bm Q) = \frac{1}{2}\sum_{ \Gamma  \gamma} \omega_\Gamma^2 Q_{ \Gamma  \gamma}^2$, which isn't usually the case as $Q_{ \Gamma  \gamma}$ are not the normal mode coordinates, we found that in our study of the $6s-6p$ transition, if projected to the sole "active" coordinates $Q_{A_1},Q_{E,1},Q_{E,2},Q_{T,1},Q_{T,2},Q_{T,3}$, the ground state Hamiltonian indeed takes the desired quadratic diagonal form.

In this case, as used in Ref. \cite{toyozawa1966dynamical,cho1968optical} (see discussion in Appendix \ref{frequency_scaled}), the oscillation frequencies $\omega^\Upgamma$  are not needed for the Mulliken semi-classical case, because we can include them in the definition of new scaled coordinates $ \tilde Q_{ \Gamma  \gamma} = \frac{\omega_\Upgamma}{\sqrt{2}} Q$ with new interaction coefficients  $\tilde   V_{\Gamma \gamma} = \frac{\sqrt{2}}{\omega_\Upgamma} V$. This leads to $V_{\Gamma \gamma} Q_{\Gamma \gamma} = \tilde   V_{\Gamma \gamma}   \tilde Q_{\Gamma \gamma}$, thus Eq. (\ref{line_shape}) becomes

\begin{equation}
A(E) \propto \int \sum_{i=1}^6 \delta[E - X_i ({\tilde{\bm Q}}^{\Upgamma})]  e^{- \sum_{ \Gamma  \gamma} 
 {\tilde Q}_{ \Gamma  \gamma}^2/  k_B T}  {\rm d} {\tilde{\bm Q}}^{\Upgamma}.
 \label{line_shape_normalized}
\end{equation}

Usually, crystal field theory is used as an effective theory; the parameters $\tilde   V_{\Gamma \gamma}$ are adjusted to fit the data according to the trapping site symmetry. This fitting process, along with the 6s offset for the line position and potential fine structure splitting, typically yields a strong agreement between experiment and theory due to the large number of free parameters.

Compared to the Monte Carlo or molecular dynamics methods, the crystal field method is significantly faster. It only requires a few parameters (in our case, $\tilde V_{A_1}, \tilde V_{E},\tilde V_{T_2}$) and Gaussian random distributions ($\tilde Q_{A_1},\tilde Q_{E,1},\tilde Q_{E,2},\tilde Q_{T,1},\tilde Q_{T,2},\tilde Q_{T,3}$ in our case). Furthermore, the interaction is given by the analytical matrix (\ref{matrice_CF}). It is therefore a very attractive method. As previously mentioned in Ref. \cite{Kanda1971}, the observed triplets, similar to our experimental data, were attributed to cubic symmetry sites and were well-fitted using the matrix (\ref{matrice_CF}) with parameters listed in table \ref{table_crystal_field}. The fit was performed at a temperature of $T=\qty{1.8}{K}$ (classically, without any temperature scaling). Using these crystal field parameters, the calculated line shapes $A(E)$ from Eq. (\ref{line_shape_normalized}) well match the observed shape, and the triplet structure can be resolved. By varying $\tilde V_{A_1},\tilde V_{E}, \tilde V_{T_2}$ for the two sites, an almost perfect match can be achieved.

\begin{table}[ht]
    \label{potential_values}
        \begin{tabular}{r | c | c  | c | c | c | c}
           Site & \multicolumn{3}{c|}{$O_h$} & \multicolumn{3}{c}{$T_d$}\\
            \hline
            modes & $A_{1g}$ & $E_g$ & $T_{2g}$ & $A_1$ & $E$ & $T_2$ \\
            \hline
            %   $V_\Upgamma$ & 266.695 & 518.872 & 490.502 & 415.804 & 485.709 & 552.941 \\
            %  $\omega_\Upgamma^2$ & 2052.22 & 3177.52  & 2391.09 & 2336.74 & 2623.91 & 2977.1 \\
            %$\omega_\Upgamma$ (K) & 59.9958 & 74.6540 & 64.7600 & 64.0198 & 67.8396 & 72.261 \\
            %($V_\Upgamma/\omega_\Upgamma)^2$ (K) & 49.8654 & 121.906 & 144.770 & 106.454 & 129.350 & 147.76 \\
            $\hbar \omega_\Gamma/k_B$  & 60 & 75 & 65 & 64 & 68 & 72 \\
            $\frac{(V_\Gamma/\omega_\Gamma)^2}{k_B}$  & 50 & 120 & 140 & 110 & 130 & 150 \\
            from \cite{Kanda1971}  $\frac{(V_\Gamma/\omega_\Gamma)^2}{k_B}$ & 13600 & 84700 & 12600 & 13600 & 84700 & 12600 \\
    \end{tabular}
   \caption{Crystal field parameters, given in Kelvin, for the two types of sites. %The values are given for potential energy in cm$^{-1}$ and coordinate position in \AA.
Oscillation frequencies are such that the ground state potential is given by  $\langle 6s   |\hat V | 6s  \rangle = 
 V_0 +  \sum_{ \Gamma  \gamma} \frac{1}{2} \omega_\Gamma^2 Q_{ \Gamma  \gamma}^2$, while the parameters $V_{A_1}, V_E,V_{T_2}$ serve as coupling parameters for the excited state matrix
(\ref{matrice_CF}) incorporating linear terms of the $V_\Gamma Q_{\Gamma \gamma}$ type.
For comparison, values from the experimental spectrum of Ref. \cite{Kanda1971} are included. }
\label{table_crystal_field}
\end{table}

However, the coefficients $V_{A_1},V_{E},V_{T_2}$ can also be computed by comparing the matrix (\ref{matrice_CF}) to the gradient of the interaction matrix calculated via the summation of Cs-Ar potentials (refer to Appendix \ref{JahnTeller} and formulas (\ref{crystal_field_coeff}) and (\ref{ground_state_pot_int_coord}) for details).
The parameter $ (V_\Gamma/\omega_\Gamma)^2$ represents a Jahn-Teller energy shift for the excited potential, evident from the formula
$ V_\Gamma Q_{\Gamma \gamma} + \frac{1}{2} \omega_\Gamma^2 Q_{ \Gamma  \gamma}^2  =  \frac{1}{2} \omega_\Gamma^2 \left(  Q_{ \Gamma  \gamma} + 
\frac{V_\Gamma}{\omega_\Gamma^2} \right)^2  - \frac{1}{2} \frac{V_\Gamma^2}{\omega_\Gamma^2}$.
Results are presented in Table \ref{table_crystal_field}, where frequencies $\omega_\Gamma$ and interaction coefficients $V_\Gamma$ are displayed in temperature units using $\hbar \omega_\Gamma/k_B$ and  $(V_\Gamma/ \omega_\Gamma)^2/k_B$. 

It is noteworthy that the $V_\Gamma$ parameters derived from realistic interaction potentials markedly differ from those fitted in Ref. \cite{Kanda1971}. This shows the risk of using solely crystal field parameters for fitting, which may not align with physical reality, leading to potential inaccuracies and misleading predictions for subsequent studies \cite{lund1984magnetic}.
It is therefore crucial to validate them with an independent study, such as magnetic circular dichroism (MCD) \cite{lund1984magnetic}, temperature dependence of absorption lines, or theoretical predictions \cite{ammeter1973electronic,bersuker1975jahn,forstmann1977analysis,kolb1978analysis}.

Despite these findings, it is important to emphasize that using our derived $V_{A_1},V_{E},V_{T_2}$ coefficients, and limiting our absorption band shape Mulliken semi-classical simulation to the first order in the $\bm Q^{\Upgamma}$ coordinates for the excited state interaction matrix, we achieve crystal field results that are nearly identical to those obtained from the full Monte Carlo classical simulation (shown in Fig. \ref{fig:pure_spectra_th}). Here, excited states are calculated from the full potentials. Indeed, for our temperature range, the first-order matrix element values in the Monte Carlo simulation deviate less than 10\% from the actual ones.

In conclusion, both the full potential approach and the crystal field first-order approximation for the excited state matrix coefficients generate almost identical results. However, they fail to reproduce the observed triplet structure and its temperature evolution. The problem might originate from the inadequacy of the classical model for ground state motion at low temperatures where the vibrational energies are quantized.

 \subsubsection{Reflection approximation and quantization of the energy}

Classical simulations, despite temperature scaling via the appropriate spatial Gaussian distribution $\Psi_i (\bm Q)$, inherently fall short as they use a continuous energy $P_{\rm g} (\bm Q)$ in Eq. (\ref{eq_Mulliken}). In contrast, real systems possess quantized ground state energy $E_i$. For instance, at low temperatures with only a single zero phonon mode occupation, the ground state exhibits a single energy value, whereas classical simulation permits all energies from $0$ to this zero-point energy. 

Such a discrepancy suggests that classical simulations can induce artificial line broadening on the order of half the Debye oscillation frequency. In argon, this equates to roughly $\sim \SI{30}{cm^{-1}}$ per vibrational (phonon) mode. While this constitutes an extreme case, given that the majority of lattice vibrations between ground and excited states are similar. % it may provide a qualitative explanation for some failure of a classical simulation. In our case it broadens the triplet structure, resulting instead in a broad doublet. 

This indicates that the semi-classical Mulliken approximation, often employed to simulate atomic spectra in rare gas matrices may not always be perfectly adequate  \cite{o1971jahn,boatz1994monte,ryan2010investigations,jacquet2011spectroscopic,tao2015heat,kleshchina2019stable}. Previous studies have shown that this approximation cannot adequately account for observed line shapes, especially at low temperatures, for the $A \rightarrow T$ spectral band shape ($6s\rightarrow 6p$ is $A_{1_g}\rightarrow T_{\rm 1u}$ for $O_h$ or $A_1 \rightarrow T_{\rm 1}$ for $T_d$) \cite{o1971jahn,stockmann1984influence,bersuker2013jahn}.

In most cryogenic rare gas matrix experiments, a majority of phonon modes exhibit energies significantly larger than the thermal energy, $\hbar \omega \gg k_B T$. As a result, vibronic transitions often occur from zero-phonon occupation, a phenomenon observed in heavy atoms like Eu, Sm, or Yb embedded in Ar \cite{jakob1976zero,jakob1977observation,luchner1979phonon,lambo2012electronic,lambo2021high}. Such cases, confined to a single $A_1$ mode, lead to vibrational quantization that produces well-resolved Huang-Rhys-Pekar type peaks in the excited states due to the displacement of the harmonic potential in the ground state \cite{wagner1964structural,keil1965shapes,shi2019huang,zhang2019applications}. These peaks can only be elucidated via a full quantum treatment of kinetic energy, as opposed to a semi-classical Franck-Condon-Mulliken approximation.

We are dealing with a particularly complex case: the so-called $T\otimes (a+e+t_2)$ coupling, where a 6p triplet $T$ couples with the lattice through several modes plus spin-orbit interaction. This situation bears a strong similarity to the extensively studied Jahn-Teller $T\otimes (e+t_2)$ coupling, which has been investigated using a variety of approximation techniques in full quantum mechanical treatments \cite{STURGE1968a,bersuker1975jahn,bates1978jahn,bates1987analysis,bersuker2006jahn,bersuker2013jahn}. Even with some justifiable approximations for our case (see Table \ref{table_crystal_field}), of a $p$ state equally coupled to $E$ and $T$ vibrations \cite{o1971jahn}, the problem remains intricate. Nevertheless, full quantum treatments seem to agree with classical approaches \cite{o1980calculation,cho1968optical}.

The optimal path forward to derive a definitive answer would be to perform a full quantum calculation of nuclear motion in the calculation of probability transfers. However, this is outside the scope of this article, as our aim is to use semi-classical expressions to calculate electronic absorption spectra without the need to compute molecular vibrational eigenstates.

In our scenario, most phonon energies $\hbar \omega$ and Jahn-Teller energy shifts $\hbar (V_\Gamma/\omega_\Gamma)^2$ exceed the kinetic energy temperature $k_B T$. Therefore, the reflection approximation in Eq. (\ref{Reflection_approximation}) should provide greater accuracy than the Franck-Condon Mulliken approximation, as it treats the excited-state oscillator classically while the ground state is quantized \cite{lax1952franck,keil1965shapes}.

Regrettably, utilizing the reflection approximation gives rise to numerous complexities. This is particularly true when employing the harmonic approximation, as shown by:
$
V_{\rm g} (\bm Q) = V_{\rm Cs(6s)-Ar} (\bm Q) + V_{\rm Ar-Ar} (\bm Q) =  \sum_{k} \frac{1}{2} \omega_k^2 {Q^{\rm n} _k}^2
$.
From this, we must calculate the following:
$
V_{\rm e} (\bm Q) - E_i = V_{\rm Cs(6p)-Ar} (\bm Q) -  V_{\rm Cs(6s)-Ar} (\bm Q) +  \sum_{k} \frac{1}{2} \omega_k^2 {Q^{\rm n} _k}^2 - E_i
$.
In the above equation, the total quantized ground state energy $E_i = \sum_k \hbar \omega_k (n_k +1/2)$ and $ \sum_{k} \frac{1}{2} \omega_k^2 {Q^{\rm n} _k}^2 $ both increase linearly with the number of modes.
% The reflection approximation's nonphysical effect is observed here, as it treats ground and excited potentials at different approximation levels - the former being quantized and the latter classical. This has been demonstrated quantitatively through specific calculations, where we selected the energy of each individual phonon mode $k$ according to the geometric distribution $P_n(\omega_k)$, and the normal mode positions according to the $n^{\rm th}$ quantum harmonic oscillator wave functions distribution $|\psi_n|^2$. We calculated these using the Inverse transform method, by interpolating the inverse of the cumulative distribution function of $\psi_n$ (the Acceptance-rejection Von-Neumann sampling method is also a viable option) \cite{yao2021sampling}. For any given $E_i$, $ \sum_{k}  \frac{1}{2} \omega_k^2 {Q^{\rm n} _k}^2 $ fluctuates randomly, which leads to the spectrum broadening as the number of modes increases. 
We can bypass this nonphysical divergence by categorizing the modes as either "active" or "passive", depending on whether they impact $V_{\rm Cs(6p)-Ar} (\bm Q) -  V_{\rm Cs(6s)-Ar} (\bm Q)$. This can be done using symmetry-adapted coordinates, in which only the 6 active coordinates significantly affect the Cs interaction with the Ar crystal.

% It's noteworthy that both the $\sum_{k}  \frac{1}{2} \omega_k^2 {Q^{\rm n} _k}^2$ calculation and $E_i$ can be done solely with the normal modes and not with the 6 symmetric interaction coordinates. 

\begin{figure}[ht]
    \centering
    \includegraphics[width=\linewidth]{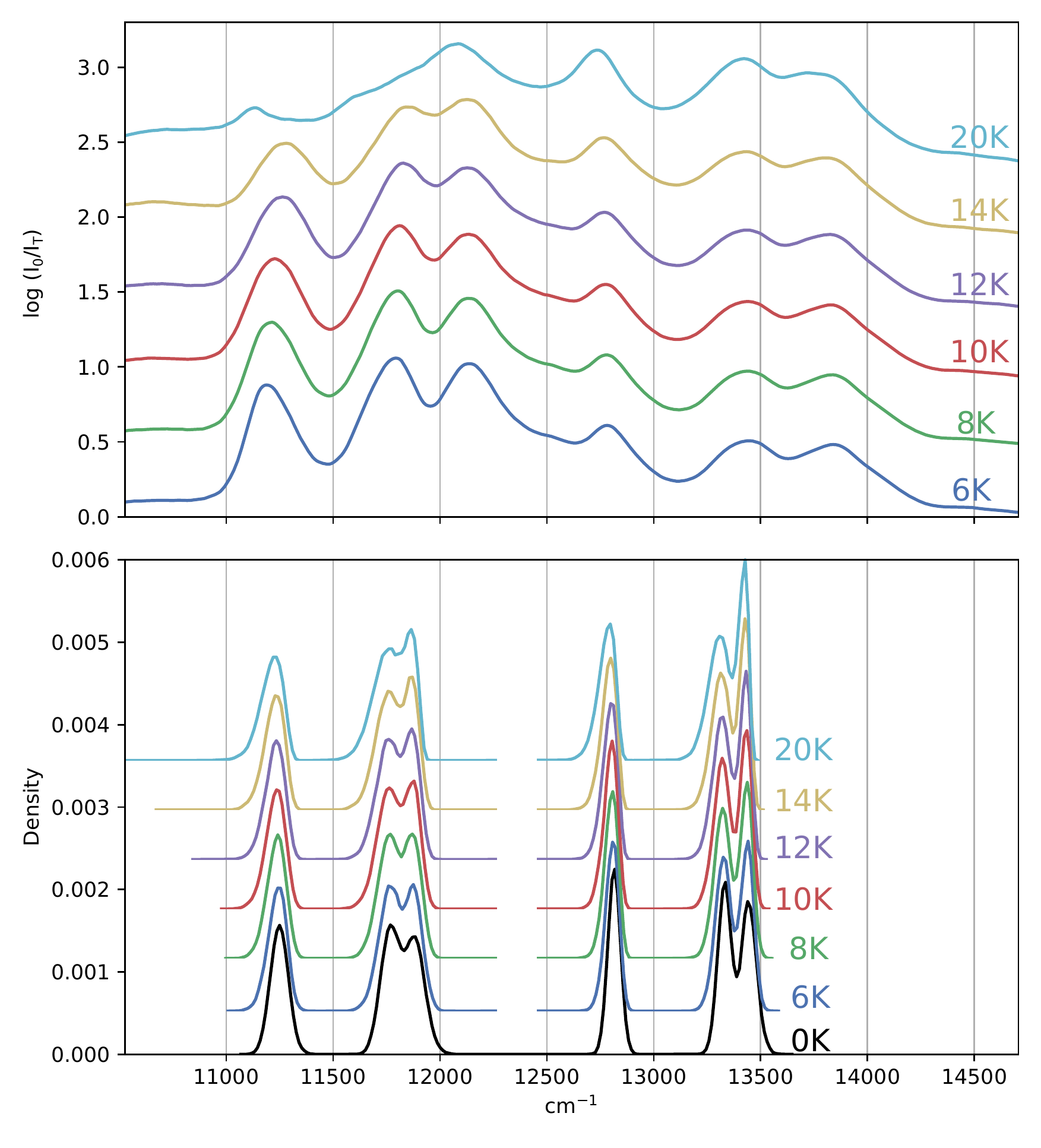}
    \caption{Comparison of observed absorption with the simulated one using the temperature scaled first order crystal fields model. The model incorporates the modified reflection approximation (Eq.~(\ref{Reflection_approximation_inetractin_mode})), using parameters from Table~\ref{table_crystal_field}. The theoretical spectra are generated from a density kernel estimation of the Monte Carlo simulation results. Results for the T$_d$ and O$_h$ sites are shifted in energy by a global offset for easier comparison with the experimental data (that are $ \SI{1650}{cm^{-1}}$ and $ \SI{85}{cm^{-1}}$ respectively for O$_h$ and T$_d$). Vertical offsets have been applied to both experimental and theoretical spectra to enhance visual clarity.}
    \label{fig:Exp_th_Oh_Td_temperature}
\end{figure}
%offsetOh=1650 offsetTd=85

Unfortunately, the ground state Hamiltonian no longer remains diagonal in the interaction modes, making the energy ill-defined. One possible solution to this problem is to use the fact that, in the low temperature regime, the soft normal coordinates phonon modes with low frequency ($\omega \gtrsim k_B T/\hbar$) are populated. This population results in a different spatial distribution for these normal modes and, thus, for their linear combinations present in the six interaction mode coordinates. For each temperature $T$, we scale each normal mode frequency, transforming $\omega$ into $\omega'$, where $\omega'^2/T = \omega^2/T'(\omega)$, to maintain the correct spatial distribution for the normal modes. The resulting orthonormal transformation produces new diagonal $\omega'_{\Gamma \gamma}(T)$ oscillation frequencies for the six interaction mode coordinates. At this temperature $T$, we simulate a spectrum using the classical Boltzmann distribution, $\propto e^{-\frac{1}{2} \omega'_{\Gamma \gamma}(T)^2 Q_{\Gamma \gamma}^2/k_B T}$, for each $Q_{\Gamma \gamma}$ and a classical energy $E_i = \sum_{\Gamma \gamma} \frac{1}{2} \omega'_{\Gamma \gamma}(T)^2 Q_{\Gamma \gamma}^2$. The formula used, with $\bm Q$ representing the interaction mode coordinates, is as follows:

\begin{align}
    A(E)  \propto  \sum_{i } P_i \int  \left|  \Psi_i(\bm Q) \right|^2  \delta[  E -(V_{\rm Cs(6p)-Ar} (\bm Q)\nonumber\\
    - \sum_{k} 
 \frac{1}{2} {\omega'}_k^2 {Q^{\rm n} _k}^2 ) ]  {\rm d} \bm Q \label{Reflection_approximation_inetractin_mode}
\end{align}

The results are presented in Fig.~\ref{fig:Exp_th_Oh_Td_temperature}. This modified reflection approximation restores the triplet shapes, but the splittings do not match closely with the observed ones at the lowest temperature of $T=\qty{6}{K}$.

Furthermore, the simulated spectra show some temperature-dependent evolution that does not fully align with observed data. Experimentally, the red trapping site (attributed to T$_d$), and the blue trapping site (attributed to O$_h$), behave differently with temperature. However, the simulation predicts similar behaviors for O$_h$ or T$_d$ sites, as the Cs$(6p)$ interaction with the Ar matrix is governed by similar couplings (as evidenced by the similar values for the coupling coefficients in Table \ref{table_crystal_field}). Theoretically we observe, for a Gaussian fit, a red shift of all lines (compare to blue shift of some lines in the experiment) of nearly 
$\qty{30}{cm^{-1}}$ for the temperature evolution up to $\qty{20}{K}$. We also see
a quite strong modification (almost 50\%) of the theoretical width, where the experimental evolution is smaller in relative value.
Also care must be taken with experimental temperature calibration, as the temperature of a matrix sample may differ from a temperature sensor nearby \cite{rose1986magnetic}. There are, however, more critical discrepancies that need to be addressed. The most notable is the mismatch in the splitting between the peaks in the experimental and theoretical data. 

It remains unclear whether utilizing more sophisticated approximations for the absorption shape would help alleviate these discrepancies. As an example, the improved crystal field absorption line shape formula proposed in Ref.~\cite{stockmann1984influence} could be considered. Here, S-P transitions are calculated for the most general case, where the coupling of all three vibration modes $A_1,E_g,T_2$, as well as spin-orbit interaction, are taken into account. However, because the $\omega_\Gamma$ frequencies are almost identical for the T$_d$ and O$_h$ case, this formula may also not produce different temperature evolutions for the two sites.

Further, as highlighted in Ref.~\cite{hupper1998uniform}, semiclassical formulas often require correction when the energy involved is comparable to the quantization energy spacing. However, the proposed corrected terms often have a similar order of magnitude, making the convergence unclear.

Hence, it may be more promising to explore other effects to resolve this discrepancy, such as non-Born-Oppenheimer effects like nonadiabatic coupling.

%\subsection{Uncertainty in the short range Cs-Ar excited state potential}\label{sec:Uncertainties-in-the-short-range}

%It can be seen for example in \cite{hewitt2023csar}, that there is no clear consensus on the short-range nature of the $\Sigma_{1/2}$ potential of the Cs-Ar potential yet. This, together with the in section \ref{sec:third-order} shown effect of third order interactions leaves quite a lot of freedom in the shape of the potential. This poses the question if the missing splitting of the P$_{3/2}$ state could be due to the specific potential that we used.\\
%Because of ..., it is clear that the degeneracy of the P$_{3/2}$ state is lifted by the slope of $V_\Sigma$. As previously explained in section \ref{sec:rotation-matrices}, the Cs-Ar distance only takes certain (trapping site dependent) values, which implies that the shape of $V_\Sigma$ only around these specific points has an influence on the simulated spectra while for example the potential below \SI{0.43}{nm} does not matter. \\
%When we only modify the short range of the potential that affects the nearest neighbor interaction as shown in fig. ???,
%we can create line splittings for both trapping sites that matches the experimentally observed one.\\
%The need to make the short range interaction of an alkali-noble gas excited state potential more repulsive to reproduce experimental data has previuously also previously been observed in another study \cite{ryan2010investigations}.

\section{Conclusion}

In conclusion, we have observed Cs absorption spectra in an Ar matrix containing six major resonances. These were assigned to transitions from the $6s_{1/2}$ Cs ground state to the $6p_{1/2}$ and $6p_{3/2}$ excited states, with lifted $|m|$-degeneracy and two different trapping environments. Pair-wise Ar-Ar and Cs-Ar potentials facilitated a stability study, indicating that the two observed triplets are likely due to a 4 vacancy tetrahedral T$_{\mathrm{d}}$ trapping site and a 6 vacancy cubic O$_{\mathrm{h}}$ one in an fcc Ar matrix. 

To qualitatively reproduce the line locations using the pair-wise approach, we derived effective two-body potential curves. These included zero-point energy and third-order terms. We found that zero-point energy has a minor effect, but adding third-order many-body corrections significantly modifies the potentials. Our derivation of the effective two-body potential from the long-range part of the triple dipolar interaction, based on an exact derivation of the long-range part, is not very accurate. More work is needed on effects such as the Pauli repulsion, exchange interaction, and three-body dispersion interactions for the rare gases - alkali atoms, especially in excited states and for the short-range part \cite{anatole2010two}.

Our study reasonably assigns the nearly unshifted (relative to the gas phase) triplet to the 4 vacancy $T_{\mathrm{d}}$ site, and the blue-shifted one to the 6 vacancy cubic $O_{\mathrm{h}}$ trapping site. Using the computationally demanding core polarization pseudopotentials will be an interesting approach to confirm this fact and to achieve better agreement between theory and experiment for the line separations \cite{gross1998pseudopotential,jacquet2011spectroscopic}.

Within our current approach, we demonstrate that a triplet shape can only be reproduced by using quantized energy for the ground state vibrational motions. Indeed, the classical simulation may produce artificial broadening of the lines, on the order of half the Debye oscillation frequency. Therefore, in Ar this would be on the order of $\sim30\,cm^{-1}$. Hence, our semi-classical Mulliken simulation fails to produce a triplet structure, instead producing a broad doublet. A simulation using the reflection approximation, on the other hand, only yields a triplet at \SI{0}{K}, but otherwise diverges with increasing numbers of Ar atoms due to unphysical treatment of the phononic modes.

Finally, a quantized simulation based on a modified reflection approximation reproduces a triplet structure. We show that modifying the population of the soft (low-frequency) phonon modes can also alter the absorption shapes. Using only the matrix gradient (linear, first-order approximation) for the excited state does not alter the quality of the results but leads to significantly faster calculations. Moreover, at low temperatures when only ground state vibrational (zero-phonon) levels are populated, the simplest crystal field model suffices to reproduce the observed line broadening caused by the dynamic Jahn-Teller thermal effect. This requires only three electron-lattice parameters given in Table \ref{table_crystal_field} (together with Eq. (\ref{line_shape}) and diagonalization of spin-orbit and (\ref{matrice_CF}) matrices) \cite{cho1968optical}. The reduced number of modes is advantageous as it leads to faster calculations than the full Monte Carlo method and captures the essential aspects of the electrostatic interaction, providing a powerful starting point for further studies.

However, our work was unable to reproduce the triplet splitting and did not provide accurate red and blue line separations. The reason for this discrepancy is unclear, but could be due to our neglect of the dipole transition dependence on the $Q$ coordinates, lack of coupling with other excited states (like Cs$(5d)$), or an inappropriate inclusion of the third-order non-additive terms in the potentials. The role of nonadiabatic couplings should also be studied.

Following \cite{ryan2010investigations}, modifying the potential curves, such as the B$\Upsigma$ state (recently suspected to be inaccurate \cite{hewitt2023csar}), might lead to better agreement between experiment and theory. However, our preliminary attempts suggest that substantial modification of the position and the slope of the curve in the region near the first nearest neighbors (see Fig. \ref{fig:potentialCurves_diff}) may be necessary. This should be further investigated, along with experimental and theoretical studies of other possibilities such as trapping in other crystal defects of lower symmetry (for instance, 5 vacancy of D$_{3h}$ symmetry in the hcp phase \cite{davis2018investigation,ozerov2021generic}), which may explain the discrepancy between experiment and simulation.

Experimentally, many tools are available such as magnetic circular dichroism, fluorescence, electron (paramagnetic) spin resonance, oscillator strength determination, time-resolved emission spectra, nonradiative branching ratios, optical pumping, or bleaching studies. Comparing with the rubidium case and using other rare gases, or even mixed ones, would be interesting \cite{corbin1994optical}, to confirm that the matrix shift seen in the S-P transitions is approximately linear with the polarizability of the matrix host \cite{laursen1991multiplicity,ryan2010investigations,tao2015heat}.

We hope that our work will be useful for further studies of similar systems used for fundamental physics experiments in rare gas solid crystals \cite{guarise2022particle,lambo2022calculation,koyanagi2022accurate}.
%\newpage

\section{Acknowledgements}

T.B. and S.L. contributed equally to this work.

This research was financed in whole or in part by
Agence Nationale de la Recherche (ANR) under the project ANR-21-CE30 -0028-01. 

 A CC-BY public copyright license has been applied by the authors to the present document and will be applied to all subsequent versions up to the Author Accepted Manuscript arising from this submission, in accordance with the grant's open access conditions.

We acknowledge   W. Chin, B. Darquier,  O. Dulieu, B. Gervais, H. Lignier, Ch. Malbrunot, C. Smai, B. Viaris for fruitful discussions, and B. Vivan and L. Marriaux for the design and mechanical realization.
% With the aim of its publication in the access, the author/rights holder applies an open access license CC- 8 Y to any article/manuscript accepted for publication (AAM) resulting from this submission 

\appendix

% The sample temperature could give indication about the trapping site by comparing the position of the absorption maxima and the  evolution of the linewidths \cite{brewer1968absorption,forstmann1977analysis}. For Cs in Ar\cite{weyhmann1965optical} did not see any modification of the linewidth with temperature. 

% Assuming pure $A_1$ breathing mode, in a trapping site with $n$ nearest neighbors,  give the FWHM linewidth of the transition $W = 2\sqrt{\ln{2}} \frac{\hbar}{n m_{Ar} \omega} n \frac{d V_{Ca(6p),Ar} }{d Q_{A_1}}$ $\omega$ being given by the Debye frequency with a Debye temperature of $\SI{93.3}{K}$ in argon.

\section{Perturbation matrices for different symmetries}
\label{crystal_fiel_matrix}
A given trapping site is defined by the near neighbor environment of the impurity atom and by its overall crystal orientation in the laboratory frame. We detail here the crystal field method that we use to simulate the absorption spectrum.

 Using vector notation for the cartesian coordinate of each position of the Cs and the $(N-n)$ Ar nuclei $\bm R =\{ \bm R_1  , \ldots, \bm R_{3(N-n+1)} \} $) and $\bm r$ as the Cs valence electron coordinates, the electron-lattice interaction 
 can be written as  $ V = V(\bm r, \bm R)$, where for simplicity, we use, in this article, the same notation for operators and their representations in coordinate basis.

\subsection{Crystal field interpretation}
\label{crystal_field}

A given single trapping site, at
the equilibrium  position $\bm R_0$, possesses a certain symmetry group with irreducible representation $\Gamma$ (and its row $\gamma$). 
So it is more convenient to use coordinates, $Q_{\Gamma \gamma}$, that transforms well under such a representation. They are linear combinations of the $  R_i - {R_0}_i$ ones, such that at a linear (first order) approximation, in the nuclear coordinate $\bm R$ or $\bm Q^\Upgamma$, the electron-lattice interaction can be written as $\hat V = V^{(0)}({\bm r}) + \sum_{ \Gamma  \gamma} V_{ \Gamma  \gamma}^{(1)}({\bm r}) Q_{ \Gamma  \gamma} $. We refer to this expression as the "crystal-field approximation".

At zero order,
the electron lattice
$V^{(0)}({\bm r}) =  V(\bm r, \bm R_0)$ 
modifies the Cs electron wavefunction
throughout the 
$\langle n_1   l_1   m_1 | V^{(0)} | n_2   l_2   m_2 \rangle$ couplings  (we use here the non-relativistic wavefunctions, without fine structure that depend on the spherical harmonics $Y_l^m$).
The trapping site equilibrium position $\bm R_0$ determines the symmetry group for the nuclear environment that produces the electron lattice interaction $V_0(\bm r) $. 
For simplicity, we keep the same notation, as for the nuclear coordinate representation $\Gamma $, 
and  name
  the eigenstates of the electronic Hamiltonian that are linear combinations of the $|n l m \rangle$ states, $|  \Gamma  \gamma \rangle $. They are given by the projection operators for the irreducible representation $\Gamma$. 

Once the Cs electronical eigenstates $|  \Gamma  \gamma \rangle $, that are adapted for the ground state equilibrium position, are found, we can use them to
calculate the next order of the electron lattice interaction $ \sum_{ \Gamma  \gamma} V_{ \Gamma  \gamma}^{(1)}({\bm r}) Q_{ \Gamma  \gamma} $.

The key result \cite{tsukerblat2006group,dresselhaus2007group} is that the 
  $\langle \Gamma_1   \gamma_1  | V_{ \Gamma  \gamma}^{(1)}  | \Gamma_2   \gamma_2 \rangle$ coupling terms can be calculated using 
the Wigner-Eckart's theorem 
$\langle \Gamma_1 \gamma_1 | V_{ \Gamma  \gamma}^{(1)} | \Gamma_2 \gamma_2 \rangle  = \langle \Gamma_1 || \hat V_\Gamma^{(1)} || \Gamma_2 \rangle  \langle \Gamma_1 \gamma_1 | \Gamma_2 \gamma_2,\Gamma \gamma \rangle$. Non zero 
Clebch-Gordan coefficient $ \langle \Gamma_1 \gamma_1 | \Gamma_2 \gamma_2,\Gamma \gamma \rangle$   occurs only if the 
$\Gamma $ representation is part of the $\Gamma_1 \times \Gamma_2$ representation \cite{tsukerblat2006group,dresselhaus2007group}.  Since $V$ is real, we only need to consider the symmetric part of the product:
$[\Gamma_1 \times \Gamma_2]$.

Thus, group theory automatically leads to the proper basis $| \Gamma  \gamma \rangle $ and provides directly the useful representations  $\Gamma $ with non-zero matrix elements $V_\Gamma = \langle \Gamma_1 || \hat V_\Gamma^{(1)} || \Gamma_2 \rangle$ and the Clebch-Gordan coupling coefficients. 
 
 We will now illustrate this crystal field procedure for several possible trapping site symmetries that might occur for atoms in matrices \cite{cotterill1966morse,damask1968defects,johnson1973empirical,lam1985multiple,lam1986calculations,sabochick1988atomistic,nemirovich2007vacancies,wang2013defect,zenia2016stability,lounis2016stability,lara2019effects,peng2020solute}. For this we use 
  the free Mathematica group theory package GTPack \cite{gtpack1,gtpack2}. 
 
\subsection{Cubic: O$_h$ or T$_d$ symmetry}

In fcc crystals, the 1 vacancy substitution and the 6 vacancies site  have both O$_{\rm h}$ symmetry
(same symmetry as the SF$_6$ molecule)
while the 4 vacancies site has a T$_{\rm d}$ symmetry
(as CH$_4$).

For such symmetries, the electron-lattice potential at equilibrium $V^{(0)}({\bm r})$ does not lift the 6s nor the 6p degeneracy.
This can be understood in a simple manner by decomposing it
in spherical harmonics: $V^{(0)}({\bm r}) = \sum_{l,m} r^l a_l^m Y_l^m(\theta,\varphi) $.
We now only have to take into account that this interaction potential has necessarily the same
O$_{\rm h}$ or T$_{\rm d}$ symmetry as the nuclear position that creates it. Thus most of the terms are  zero, and the potential becomes
$
V^{(0)}({\bm r})= a_0 Y_0^0 +  r^4 a_4 ( Y_4^{ -4} + \sqrt{14/5} Y_4^0 + Y_4^{ 4} ) + \cdots $ ($a_0$ is not to be confused with the Bohr radius). 
 These leading order terms formed the so-called Devonshire's potential \cite{devonshire1936rotation}.
The triangle inequality $|l_1-l_2|\leq l \leq l_1+l_2$ for the Clebsch-Gordan coefficients appearing when calculating
$\langle n_1   l_1   m_1 | V^{(0)}= \sum_{l,m} r^l a_l^m Y_l^m | n_2   l_2   m_2 \rangle$ indicates that
the $l\geq 4$ terms  cannot couple to $6s$ ($l=0$) nor to $6p$ ($l=1$) states but would lift the degeneracy of the 5d ($l=2$) level for instance.

For completeness, we indicate that the same calculation can be done elegantly using directly the quantum states $| n   l   m \rangle$, without using spherical harmonics, by directly writing the potential 
 $ \hat V(\bm r, \bm R_0)$
 using angular momentum operators. For this, we simply have to replace, in the Devonshire's potential expression, $ Y_0^0 \rightarrow 1$, $ Y_4^{ \pm 4}  \rightarrow L_\pm^4 $ and
$ Y_4^{ 0} \rightarrow 35 L_z^4 -(30 L(L+1)-25)L_z^2+2L^2(L+1)^2 - 6 L (L+1)$ where $\bm L$ is the orbital angular momentum operator
\cite{misra2011spin, ryabov2009operator}.

It would be possible to use the $|6s m=0\rangle $ and $|6p m\rangle = |m\rangle $  basis but for cubic symmetry
the so-called real (or tesseral) spherical harmonics: $ |x\rangle = (|m=-1 \rangle -  |m=+1 \rangle)/\sqrt{2},\;  |y\rangle = i (|m=-1 \rangle +  |m=+1 \rangle)/\sqrt{2},\rangle,\;  |z\rangle = |m=0\rangle $ is better.
Indeed, the
 projection operator technique for the
angular momentum basis  $|6p m\rangle$ under the symmetry group O$_h$ or T$_d$ and their representations $\Gamma$ leads to the following:
for the 6s electronic state
$\Gamma = A_{1_g}$ (or $\Gamma = A_1$ for $T_d$) and $|6s m=0\rangle $ forms a good basis. For the 6p electronics state 
 $\Gamma = T_{\rm 1u}$ (or $\Gamma = T_{\rm 1}$ in the $\rm T_d$ case) and the basis
$|  \Gamma  \gamma \rangle $ is just given by $|x\rangle,\; |y\rangle,\; |z\rangle$.

Now that the zero-order symmetry   $|  \Gamma  \gamma \rangle $ basis adapted to the  nuclear equilibrium configuration is established, we can use it to calculate the first-order correction
  $\sum_{ \Gamma  \gamma} \langle \Gamma_1   \gamma_1  |  V_{ \Gamma  \gamma}^{(1)}({\bm r})   | \Gamma_2   \gamma_2 \rangle Q_{ \Gamma  \gamma}$.
  Only the non zero $\langle 6p = T_{\rm 1u} ||V^{(1)}||6p = T_{\rm 1u} \rangle$ terms are interesting.
Then because 
$[T_{1u} \times T_{1u}] = A_{1g} \oplus E_{g} \oplus T_{2g} $ (or 
$[T_{1} \times T_{1}] = A_{1} \oplus E \oplus T_{2} $ 
in the T$_{\rm d}$ case
), the $\langle \Gamma_1   \gamma_1  | V_{ \Gamma  \gamma}  | \Gamma_2   \gamma_2 \rangle Q_{ \Gamma  \gamma} $ calculation leads
to the  matrix (with the $u,g$ label removed in the $T_d$ case), in the 6p manifold, for the total 
$\langle \Gamma_1   \gamma_1  | \hat V = V^{(0)} + V^{(1)} | \Gamma_2   \gamma_2 \rangle Q_{ \Gamma  \gamma} $ crystal field (CF) interaction:

 \begin{widetext}
\begin{equation}
M_{\rm CF} = \left( \frac{a_0}{\sqrt{4\pi}} + V_{A_{1g}} Q_{A_{1g}} \right)\bm I_3
+
 \bordermatrix{
 & |x\rangle & |y\rangle & |z\rangle \cr
 &  V_{E_{g}} \left( Q_{E_{g},1} -\frac{1}{\sqrt{3}} Q_{E_{g},2} \right) & V_{T_{2g}} Q_{T_{2g},3} & V_{T_{2g}} Q_{T_{2g},2} \cr
 & V_{T_{2g}} Q_{T_{2g},3} & - V_{E_{g}} \left( Q_{E_{g},1} +\frac{1}{\sqrt{3}} Q_{E_{g},2} \right) & V_{T_{2g}} Q_{T_{2g},1} \cr
 & V_{T_{2g}} Q_{T_{2g},2} & V_{T_{2g}} Q_{T_{2g},1} &  V_{E_{g}}  \frac{2}{\sqrt{3}} Q_{E_{g},2} \cr
}
    \label{MCF_Td_Oh}
\end{equation}
\end{widetext}
Where we have used 
 $V_{A_1} = \langle \Gamma_1 || \hat V_{A_1}^{(1)} || \Gamma_2 \rangle/\sqrt{3}$,
  $V_{E_g} = \langle \Gamma_1 || \hat V_{E_g}^{(1)} || \Gamma_2 \rangle/\sqrt{2}$,
   $V_{T_{2g}} = \langle \Gamma_1 || \hat V_{T_{2g}}^{(1)} || \Gamma_2 \rangle/\sqrt{2}$ and performed an orthonormal transformation for the
 $\{  Q_{E_{g},1} ,Q_{E_{g},2} \}$ coordinates
 to obtain the same matrix as the one given in Ref. \cite{cho1968optical}.

This matrix highlights again the fact that, with no thermal effect, so with $Q_{\Gamma  \gamma } =0$, the 6p levels are not lifted by the crystal field, only a ($a_0$) global shift of the free transition $6s\rightarrow 6p$ will occur.

In order to complete the calculation, we add the spin degree of freedom with the spin $1/2$ basis sets $|\pm\rangle$.

The spin-orbit matrix is
\begin{equation}
    H_{\rm SO} = \frac{A^{\rm SO}}{2}  
 \bordermatrix{
 & |x+\rangle & |y+\rangle & |z+\rangle & |x-\rangle & |y-\rangle & |z-\rangle \cr
 & 0 & -i & 0 & 0 & 0 & 1  \cr
 & i & 0  & 0 & 0 & 0 & -i  \cr
 & 0 & 0  & 0 & -1& i & 0   \cr
 & 0 & 0  & -1& 0 & i & 0  \cr
 & 0 & 0 & -i& -i& 0 & 0  \cr
 & 1 & i & 0 & 0 & 0 &0   \cr
}
\label{SO_matrix_cho}
\end{equation}

And the crystal field matrix takes  a simple block form:
$$
H_{\rm CF} =
\begin{pmatrix}
M_{\rm CF} & 0 \\
0 & M_{\rm CF} 
\end{pmatrix} $$ 
% The internal coordinates can be of the following type: (i) Stretching:  a change in bond length (ii) Bending:  a change in bond angle or angle deformation (iii) Rocking:  a change in angle between a group of atoms and rest of the molecule (iv) Twisting:  a change in angle between the plane of two groups of atoms (v) Out of plane deformation: the movement of the central atom in and out of the plane of the rest of the atoms (vi) Wagging:  a change in the angle between the plane of a group of atoms and a plane through the rest of atoms

\subsection{Hexagonal: D$_{\rm 3h}$, C$_{\rm 3v}$}

\subsubsection{Hexagonal: D$_{\rm 3h}$}

The D$_{\rm 3h}$ symmetry 
(as C$_2$H$_6$ Eclipsed Ethane), can arise from 5 vacancy in hcp phase, for instance. Its basis vectors are
 $|E',1\rangle ,\; |E',2\rangle$, and $|A''_2\rangle  = |z\rangle$. 
The crystal field is, at the lowest order,
$
V^{(0)}({\bm r}) = a_0^0 Y_0^0  + r^2 a_2 Y_2^0 +\cdots $
(with $\hat Y_2^0 = 3L_z^2-L(L+1) $).

And with 
$[E' \times E'] = A'_{1} \oplus E' $ we have
 \begin{widetext}
$$
M_{\rm CF} =  \left( \frac{a_0}{\sqrt{4\pi}} \right)\bm I_3 +
 \bordermatrix{
 & |E',1\rangle & |E',2\rangle & |z\rangle  \cr
 & - \frac{\langle r^2 \rangle a_2}{\sqrt{20 \pi}}  + V_{E'}  Q_{E',1}  &  V_{A'_1}  Q_{A'_1} & 0 \cr
  &   V_{A'_1}  Q_{A'_1} &  - \frac{\langle r^2 \rangle a_2}{\sqrt{20 \pi}} + V_{E'}  Q_{E',2}   &  0 \cr
   & 0 & 0  & \frac{2 \langle r^2 \rangle a_2}{\sqrt{20 \pi}} \cr
}
$$
\end{widetext}
where $\langle r^2 \rangle = \langle 6p | r^2|6p \rangle$
is the square of the size of a Cs atom in its 6p level.

Here the $6p$ ($l=1$) degeneracy is partially lifted into two levels by the crystal field 
and only the spin-orbit will create the triplet structure that is then broadened and modified by the 
$A'_1$ and $E'$ terms.
This symmetry would likely produce a  triplet structure.

\subsubsection{Hexagonal: C$_{\rm 3v}$}

The C$_{\rm 3v}$ symmetry (as Ammonia NH$_3$ molecule), can arise in 4 vacancy in an hcp phase. It is very similar to the D$_{\rm 3h}$ symmetry case and can be obtained from it through the substitutions $E'\rightarrow E$, $A'_1\rightarrow A_1$, and $A''_2\rightarrow A_1$. Therefore, the basis becomes
 $|E,1\rangle, |E,2\rangle  , |A_1\rangle  = |z\rangle$ and 
the crystal field that at lowest order is given by:
$
V^{(0)}({\bm r}) = a_0^0 Y_0^0  + r^2 a_2 Y_2^0 +\cdots $. 
With
$[E \times E] = A_{1} \oplus E $ 
 we have
 \begin{widetext}
$$
M_{\rm CF} =  \left( \frac{a_0}{\sqrt{4\pi}} \right)\bm I_3 +
 \bordermatrix{
 & |E,1\rangle & |E,2\rangle & |z\rangle  \cr
 & - \frac{\langle r^2 \rangle a_2}{\sqrt{20 \pi}}  + V_{E}  Q_{E,1}  &  V_{A_1}  Q_{A_1} & 0 \cr
  &   V_{A_1}  Q_{A_1} &  - \frac{\langle r^2 \rangle a_2}{\sqrt{20 \pi}} + V_{E}  Q_{E,2}   &  0 \cr
   & 0 & 0  & \frac{2 \langle r^2 \rangle a_2}{\sqrt{20 \pi}} \cr
}
$$
\end{widetext}
 This is identical to the previous case, so this symmetry would likely produce a  triplet structure. 

\subsection{Tetragonal: C$_{\rm 4v}$}
\label{C4v} 

The C$_{\rm 4v}$ symmetry
( as SF$_5$Cl  or IF$_5$),  can arise in the 10 vacancy case. Its basis is
 $|E,1\rangle = |x\rangle$, $|E,2\rangle  = |y\rangle$, and $|A_1\rangle  = |z\rangle$. 
The crystal field is at the lowest order:
$
V^{(0)}({\bm r}) = a_0^0 Y_0^0  + r^2 a_2 Y_2^0 +\cdots $.
And with 
$[E \times E] = A_{1} \oplus B_2 \oplus B_1 $ 
we get
 \begin{widetext}
$$
M_{\rm CF} =  \left( \frac{a_0}{\sqrt{4\pi}} + V_{A_{1g}} Q_{A_{1g}} \right)\bm I_3 +
 \bordermatrix{
 & |x\rangle & |y\rangle & |z\rangle  \cr
 & - \frac{\langle r^2 \rangle a_2}{\sqrt{20 \pi}}  -  V_{B_2}  Q_{B_2}  &  V_{B_1}  Q_{B_1} & 0 \cr
  &  V_{B_1}  Q_{B_1} &  - \frac{\langle r^2 \rangle a_2}{\sqrt{20 \pi}} + V_{B_2}  Q_{B_2}  &  0 \cr
   & 0 & 0  & \frac{2 \langle r^2 \rangle a_2}{\sqrt{20 \pi}} \cr
}
$$
\end{widetext}

 This is very similar to the previous cases, and this symmetry would likely produce a  triplet structure. 

\subsection{Lower symmetry: C$_{\rm 2v}$}
\label{C2v} 

The C$_{\rm 2v}$ symmetry (as H$_2$O water molecule), can arise in the 8 vacancy case and has the basis
 $|B_2\rangle = |x\rangle, |B_1\rangle  = |y\rangle, |A_1\rangle  = |z\rangle$. 
The crystal field is at the lowest order:
 $
V^{(0)}({\bm r}) = a_0^0 Y_0^0  + r^2 [a_2^0 Y_2^0 + a_2^2 (Y_2^{-2} + Y_2^2)] + \cdots $.
(with $\hat Y_2^{\pm 2} = L_\pm^2 $)
and its matrix takes a diagonal form:
 \begin{widetext}
$$
M_{\rm CF} = 
 \left( \frac{a_0}{\sqrt{4\pi}} + V_{A_{1g}} Q_{A_{1g}} \right)\bm I_3
 +\frac{ \langle r^2 \rangle}{\sqrt{20 \pi}}
 \bordermatrix{
 & |x\rangle & |y\rangle & |z\rangle  \cr
 & - a_2^0 +\sqrt{6} a_2^2 & 0 & 0 \cr
  & 0 & -  a_2^0 - \sqrt{6} a_2^2   &  0 \cr
   & 0 & 0  & 2 a_2^0    \cr
}
$$
 \end{widetext}

Here the $6p$ ($l=1$) degeneracy is fully lifted in the three levels by the crystal field. The spin-orbit will also  shift  the levels and the triplet structure will be broadened in a symmetric way by the 
$A_{1g}$ terms.

\section{Potential curves calculation}
\label{Energy_calculation}

We discuss here the Cs-Ar potential for Cs in its 6p excited level, the origin of the energy is taken at the 6p level in gas phase.

\subsection{Hund's case (a) and (c) curves}

Without spin-orbit interaction
the Cs$(6p)$-Ar interaction is diagonal
in the basis $ |  \Uppi,\pm \rangle = |L=1 , M= \pm 1 \rangle $,
 $ | \Upsigma \rangle = |L=1 , M=0\rangle $
with $V_\Uppi(R) , V_\Upsigma(R)$ Hund's case (a) potential curves, where $R$ is the Cs-Ar separation.

%Using the primitive cell and lattice	vectors	we sum the interaction 
% $\langle L' M' | \hat V_{Cs,Ar} (\bm R_{Cs,Ar}) | L M \rangle   $ 
%between one Cs and all Ar atoms. Where $| L M \rangle =| L M \rangle_z $ 
%is quantized along the fix $z$ axis. 

When adding the spin-orbit $A^{\rm SO}(R) \bm L.\bm S$, the Hamiltonian becomes
$$
 \bordermatrix{
 & | \Uppi, +  \rangle | m_s=  - 1/2\rangle  & |\Upsigma \rangle|m_s=  1/2\rangle & |\Uppi,+ \rangle |m_s= 1/2\rangle \cr
 & V_{\Uppi} -\frac{A^{\rm SO}}{2} & \frac{A^{\rm SO}}{\sqrt{2}}  & 0 \cr
&  \frac{A^{\rm SO}}{\sqrt{2}}  & V_{\Upsigma} &0  \cr
& 0 & 0 &  V_{\Uppi} + \frac{A^{\rm SO}}{2} \cr
}
$$
and the matrix will be identical for the basis with opposite signs for all values of the projection of angular momenta: $  | \Uppi, - \rangle | m_s=   1/2\rangle  , |\Upsigma \rangle|m_s=  -1/2\rangle, |\Uppi,- \rangle |m_s= - 1/2\rangle $.
The
spin-orbit parameter 
$A^{\rm SO}(R)$ 
is not in general constant and deviates from the asymptotic atomic value as shown in Fig. \ref{fig:potentialCurves} \cite{lewis2017theoretical}.
Taking the eigenvalues leads to the potential curves with spin-orbit $ V_{\Uppi_{\frac{1}{2}} }, V_{\Uppi_{\frac{3}{2}} }, V_{\Upsigma_{\frac{1}{2}} }$ with the correspondence:
\begin{widetext}
\begin{eqnarray}
V_\Upsigma &= &
 \frac{1}{3}  (2 V_{\Uppi_{\frac{1}{2}} }- V_{\Uppi_{\frac{3}{2}} }+ 2 V_{\Upsigma_{\frac{1}{2}} }+ 
     \sqrt{V_{\Uppi_{\frac{1}{2}}}^2 + 2 V_{\Uppi_{\frac{1}{2}} }V_{\Uppi_{\frac{3}{2}} }- 
      2 V_{\Uppi_{\frac{3}{2}}}^2 - 4 V_{\Uppi_{\frac{1}{2}} }V_{\Upsigma_{\frac{1}{2}} }+ 
      2 V_{\Uppi_{\frac{3}{2}} }V_{\Upsigma_{\frac{1}{2}} }+ V_{\Upsigma_{\frac{1}{2}}}^2}) \label{Hunda_c} \\
V_\Uppi & = &
 \frac{1}{6} (V_{\Uppi_{\frac{1}{2}} }+ 4 V_{\Uppi_{\frac{3}{2}} }+ V_{\Upsigma_{\frac{1}{2}} }- 
     \sqrt{V_{\Uppi_{\frac{1}{2}}}^2 + 2 V_{\Uppi_{\frac{1}{2}} }V_{\Uppi_{\frac{3}{2}} }- 
      2 V_{\Uppi_{\frac{3}{2}}}^2 - 4 V_{\Uppi_{\frac{1}{2}} }V_{\Upsigma_{\frac{1}{2}} }+ 
      2 V_{\Uppi_{\frac{3}{2}} }V_{\Upsigma_{\frac{1}{2}} }+ V_{\Upsigma_{\frac{1}{2}}}^2}) \nonumber  \\
A^{\rm SO} &= &  \frac{1}{3}  (-V_{\Uppi_{\frac{1}{2}} }+ 2 V_{\Uppi_{\frac{3}{2}} }- V_{\Upsigma_{\frac{1}{2}} }+ 
       \sqrt{V_{\Uppi_{\frac{1}{2}}}^2 + 2 V_{\Uppi_{\frac{1}{2}} }V_{\Uppi_{\frac{3}{2}} }- 
      2 V_{\Uppi_{\frac{3}{2}}}^2 - 4 V_{\Uppi_{\frac{1}{2}} }V_{\Upsigma_{\frac{1}{2}} }+ 
      2 V_{\Uppi_{\frac{3}{2}} }V_{\Upsigma_{\frac{1}{2}} }+ V_{\Upsigma_{\frac{1}{2}}}^2}) \nonumber 
 \end{eqnarray}
\end{widetext}

From this  correspondence we have extracted the $V_\Upsigma(R)$ and $V_\Uppi(R)$ (and $A^{\rm SO}(R)$) potential curves from the $ V_{\Uppi_{\frac{1}{2}} }, V_{\Uppi_{\frac{3}{2}} }, V_{\Upsigma_{\frac{1}{2}} }$ ones.
The calculation is done using the most recent $V_{\rm X\, ^2\Sigma_{1/2}^+}$, $ V_{\rm A\, \Uppi_{1/2}}$, $V_{\rm A \,\Uppi_{3/2}}$, $V_{\rm B\, ^2\Sigma_{1/2}^+}$ potentials (in Hund's case (a) notation, but calculated with fine structure included) with spin-orbit interactions from \cite{kobayashi2016ab} and with a cubic interpolation between points and above $\SI{2}{nm}$ the long-range part is taken from
\cite{blank2012m+}. 

\subsection{Cs-Ar sum}
\label{Cs_Ar_sum}
Using the primitive cell and lattice vectors, we sum the interaction 
$\langle L' M' | \hat V_{Cs,Ar} (\bm R_{Cs,Ar}) | L M \rangle   $ between a Cs atom and all Ar ones. Where $| L M \rangle =| L M \rangle_z $ is quantized along the fixed $z$ axis. 

We then use the fixed right-handed frame, right-hand screw counterclockwise rule active interpretation (extrinsic) rotation z y z convention  $ {\cal R}(\alpha,\beta,\gamma) = e^{- i \alpha L_z}  e^{- i \beta L_y} e^{- i \gamma L_z}$ \cite{varshalovich1988quantum,man2016wigner} such that $ {\cal R}$ rotates the $| L M \rangle_z $ state to the $|L M\rangle_{ {\cal R}(z)}$ where the new axis $z'= {\cal R}(z)$ is along $\bm R_{Cs,Ar}$. Thus 
$\bm L_{ {\cal R}(z)} = {\cal R} \bm L_z {\cal R}^\dag$. 
The calculation is done using active rotation of the state vector and Wigner D-matrix with the convention: 
 $_z\langle L M' |  {\cal R}(\alpha,\beta,\gamma)| L M \rangle_z  = D^L_{M' M} (\alpha,\beta,\gamma) 
 =
 e^{- i \alpha m'} {}_z\langle L' M' | e^{- i \beta L_y} | L M \rangle_z e^{- i \gamma m}
= {\rm WignerD}[\{L,M',M\},-\alpha,- \beta, -\gamma]$ (the last notation being adapted for the Mathematica software).
Using the spherical coordinate, polar angle $\theta$ and azimuthal angle $\varphi$, for the vector $\bm R_{Cs,Ar}$ (from Cs to Ar) we have $\alpha=\varphi,\beta = \theta,\gamma = 0$.

% Furthermore, using for the Cs-Ar state the convention of a  vertical plane of symmetry $\sigma_v = \sigma_{yz}$ we have $\sigma_{yz} |lm\rangle = |l -m\rangle  $ but  $\sigma_{xz} |lm\rangle = (-1)^m |l -m\rangle  $.

% This simplifies greatly because no phase $\varphi$ appears.

Finally, we find the following matrix for the $\langle L' M' | \hat V_{Cs,Ar} (\bm R_{Cs,Ar}) | L M \rangle  $ interaction \cite{boatz1994monte,kenney2005theory}:

\begin{widetext}
\begin{equation}
 \bordermatrix{
 & | \Uppi, - \rangle  & |\Upsigma \rangle & |\Uppi,+ \rangle \cr
 & \frac{ 3 V_\Uppi + V_\Upsigma+ ( V_\Uppi - V_\Upsigma) \cos (2\theta)  }{4} &   \frac{e^{i \varphi} ( V_\Upsigma - V_\Uppi) \cos \theta  \sin \theta }{\sqrt{2}} &
\frac{e^{2 i \varphi} ( V_\Uppi - V_\Upsigma )  \sin^2 \theta }{2} \cr
  & 
 \frac{e^{-i \varphi} ( V_\Upsigma - V_\Uppi) \cos \theta  \sin \theta }{\sqrt{2}} & 
 V_\Upsigma \cos^2 \theta +  V_\Uppi  \sin^2 \theta  
    &  \frac{e^{i \varphi} (  V_\Uppi - V_\Upsigma ) \cos \theta  \sin \theta }{\sqrt{2}}  \cr
 & \frac{e^{-2 i \varphi} (   V_\Uppi - V_\Upsigma)  \sin^2 \theta }{2} & \frac{e^{-i \varphi} (  V_\Uppi - V_\Upsigma ) \cos \theta  \sin \theta }{\sqrt{2}}  &
\frac{ 3 V_\Uppi + V_\Upsigma+ ( V_\Uppi - V_\Upsigma) \cos (2\theta)  }{4}  \cr
   }
   \label{rot_matrix_spherical}
\end{equation}
\end{widetext}

for a vector $\bm R$ between the Cs atoms and the Ar atoms with $X,Y,Z$ cartesian coordinate where $\cos(\theta) = Z/R,
e^{\pm i\varphi} \sin(\theta) = (X \pm i Y)/R$.
Its  cartesian form is given by Eq. \ref{rot_matrix}.

We then sum over all Ar atoms, using  $V_\Upsigma$ and $ V_\Uppi$ potentials without spin-orbit interactions in Eq. \ref{rot_matrix}, and after  we add a constant spin-orbit interaction that is

\begin{widetext}

\begin{equation}
H_{\rm SO} = A^{\rm SO} 
 \bordermatrix{
 &  |\Uppi_-   \downarrow \rangle   & |\Upsigma  \downarrow \rangle  & |\Uppi_+  \downarrow \rangle  &   | \Uppi_- \uparrow \rangle  & |\Upsigma  \uparrow \rangle  & |\Uppi_+ \uparrow \rangle   \cr
 & \frac{1}{2}  & 0 & 0 & 0 & 0 & 0  \cr
 & 0 & 0  & 0 & \frac{1}{\sqrt{2} }  & 0 & 0  \cr
 & 0 & 0  & - \frac{1}{2} & 0& \frac{1}{\sqrt{2} }  & 0   \cr
 & 0 & \frac{1}{\sqrt{2} }  & 0 & -\frac{1}{2}  & 0 & 0  \cr
 & 0 & 0 & \frac{1}{\sqrt{2} }  & 0 & 0 & 0  \cr
 & 0 & 0 & 0 & 0 & 0 & \frac{1}{2}   \cr
}
   \label{SO_matrix}
\end{equation}
\end{widetext}

\section{Vibrational modes}

\label{JahnTeller}

In order to study the vibronic interactions \cite{hergenhahn2004vibrational} and the Jahn-Teller effect 
 \cite{STURGE1968a,bersuker2006jahn,bersuker2013jahn} we  use several different mode coordinates. We will detail here all coordinates we used:  
 \begin{itemize}
     \item cartesian $\bm R = \{ R_j \}$.
     \item 
 mass-weighted   $\bm Q^{\rm m} = \{ Q^{\rm m}_j \}  = {\bm m}^{1/2} (\bm R -\bm R _0)$
 \item 
 normal  modes $\bm Q^{\rm n} = \{ Q^{\rm n}_j \} = {\bm O_{\rm m}^{\rm n} }^\dag \bm Q^{\rm m} $ with oscillations $\bm \omega_{\rm n}$ simply noted $\bm \omega$.
  \item interaction modes  $ \bm Q^{\rm int} = \{ Q^{\rm int}_j \}=  \bm V_r^\dag  {\bm Q^{\rm n}} $
   \item crystal field symmetry adapted interaction modes  $\bm Q^{\Upgamma} = \{Q_{\Upgamma \gamma} \} = {\bm O^{\Upgamma}_{\rm n}}^\dag  {\bm Q^{\rm n}}  $ with oscillations $\bm \omega^\Upgamma$.
    \item frequency scaled  coordinates
$ {\tilde {\bm Q}}^{\rm n} =\frac{1}{\sqrt{2}} \bm \omega \bm Q^{\rm n} $ or $ {\tilde {\bm Q}}^\Upgamma =\frac{1}{\sqrt{2}} \bm \omega^\Upgamma \bm Q^{\rm \Upgamma} $ with unit oscillations frequencies.
 \end{itemize}
and their corresponding momenta conjugate respectively  $\hat{\bm P}, {\hat {\bm P}}_{Q^{\rm m}} , {\hat {\bm P}}_{Q^{\rm n}} , {\hat {\bm P}}_{Q^{\rm int}} , {\hat {\bm P}}_{Q^{\Upgamma}},
{\hat {\bm P}}_{\tilde {Q^{\rm n}}} $, ${\hat {\bm P}}_{\tilde {Q^{\Upgamma}}} $  .

\subsection{Quantization in the normal modes}

We consider the matrix elements 
 $V(\bm R)=\langle n'l'm' |\hat V | nlm\rangle$ for the  Cs(6s) and 6p levels.

Neglecting the anharmonic terms, the electron-lattice crystal field
$V = V_0 +  {\bm \nabla}_{\! \! \bm R} V^\dag (\bm R-\bm R_0)  + \dfrac{1}{2} (\bm R-\bm R_0)^\dag  {\bm \nabla}_{\! \! \bm R}^2 V  (\bm R-\bm R_0) $ 
is numerically evaluated
by finite difference: ${\bm \nabla}_{\! \! \bm R} V_k = \frac{\partial V}{\partial R_k} (\bm R^0) \approx \frac{V( \bm R^0+\delta R_k) - V(\bm R^0-\delta R_k)}{2 \delta R_k}  
  $ and
${\bm \nabla}_{\! \! \bm R}^2 V _{kl} = \frac{\partial^2 V}{\partial R_k \partial R_l} (\bm R^0) \approx   \frac{\frac{\partial V}{\partial R_l} ( \bm R^0+\delta R_k) - \frac{\partial V}{\partial R_l} (\bm R^0-\delta R_k)}{2 \delta R_k}  $
% \approx \frac{ V( \bm R^0 +\delta R_k + \delta R_l) - V(\bm R^0+\delta R_k-\delta R_l ) -( V( \bm R^0 -\delta R_k + \delta R_l) - V(\bm R^0-\delta R_k-\delta R_l ) ) }{4 \delta R_k \delta R_l }  $.
Where we have used standard row vector 
$\bm R^\dag = (  R_1 ,  R_2  ,\cdots), {\bm \nabla}_{\! \! \bm R} V^\dag = ( \frac{\partial V }{\partial R_1}  , \frac{\partial V }{\partial R_2}  ,\cdots) $ ($^\dag$ being the  Hermitian adjoint, that is the conjugate transpose) and matrix $ {\bm \nabla}_{\! \! \bm R}^2 V$ notations.

In order to treat all kinetic energy terms $\frac{{\hat P}_k^2}{2 m_k}$, ($ {\hat P}_k  = - i\hbar \frac{\partial }{\partial R_k } $ 
being the  quantized  momentum conjugate coordinate)
in a similar manner,
 we use the $3(N-n+1)$
mass-weighted Cartesian coordinates $Q^{\rm m}_j = (R_j-{R_0}_j) \sqrt{m_j} $ 
where $ m_j$  is the mass of the atom on which the $j^{\rm th}$  coordinate resides.
We define the
diagonal mass matrix $\bm m$ with $m_{jj} = m_j$
and
${\bm Q^{\rm m}}^\dag = (  Q^{\rm m}_1 , Q^{\rm m}_2  ,\cdots),
{{\bm \nabla}_{\! \! {\bm Q}^{\rm m}}}^\dag = 
( \frac{\partial V }{\partial Q^{\rm m}_1}  , \frac{\partial V }{\partial Q^{\rm m}_2}  ,\cdots) $.

So
the nuclear Hamiltonian in the Born-Oppenheimer approximation 
$\hat H = 
\dfrac{1}{2}  {\hat {\bm P}}^\dag {\bm m}^{-1} {\hat {\bm P}} +
V_0 +  {\bm \nabla}_{\! \! \bm R} V^\dag (\bm R-\bm R_0)  + \dfrac{1}{2} (\bm R-\bm R_0)^\dag {\bm \nabla}_{\! \! \bm R}^2 V (\bm R-\bm R_0)$
becomes simply 
$\hat H = 
\dfrac{1}{2}  {\hat {\bm P}_{\rm Q^{\rm m}}}^\dag {\hat {\bm P}_{\rm Q^{\rm m}}} +
V_0 +  {\bm \nabla}_{\! \! \bm Q^{\rm m}} V^\dag \bm Q^{\rm m}  + \dfrac{1}{2} {\bm Q^{\rm m}}^\dag {\bm \nabla}_{\! \! \bm Q^{\rm m}}^2 V \bm Q^{\rm m}$,
where we have used ${\hat {\bm P}_{\rm Q^{\rm m}}} = {\bm m}^{-1/2} \hat {\bm P} $ vector notation for the momentum conjugate with the $\bm Q^{\rm m}$ coordinates, that  verifies the
canonical commutation relation $[\bm Q^{\rm m}, {\hat {\bm P}_{\rm Q^{\rm m}}}] = [\bm R, \hat {\bm P}] = \bm R {\hat {\bm P}}^\dag -  {\hat {\bm P}} {\bm R}^\dag   =  i \hbar \bm 1$.

The eigendecomposition of the Hessian matrix
$ {\bm \nabla}_{\! \! \bm Q^{\rm m}}^2 V  = \bm O_{\rm m}^{\rm n}  {\bm \omega}^2  {\bm O_{\rm m}^{\rm n} }^\dag$  creates the
 the normal vibrational modes $\bm Q^{\rm n} = {\bm O_{\rm m}^{\rm n} }^\dag \bm Q^{\rm m} $ with   ${\bm O_{\rm m}^{\rm n} }$ the  change-of-basis   orthonormal matrix  and
 ${\bm \omega}$ a diagonal matrix with the k-th element being the (mass scaled) oscillation angular frequency $\omega_k$. The Hamiltonian is thus written as
\begin{eqnarray}
\hat H &= &
\dfrac{1}{2}  {\hat {\bm P}_{\rm  Q^{\rm n}}}^\dag {\hat {\bm P}_{\rm  Q^{\rm n}}} +
V_0 +  {\bm \nabla}_{\! \! \bm Q^{\rm n}} V^\dag \bm Q^{\rm n} + \dfrac{1}{2} {\bm Q^{\rm n}}^\dag \bm \omega^2 \bm Q^{\rm n} \label{correct_pos} \\
  &=&   \sum_{k} 
   - \hbar^2 \frac{\partial^2 }{\partial^2 Q^{\rm n}_k } + V_0 +
    (   {\bm \nabla}_{\! \! \bm Q^{\rm n}} V ^*)_k Q^{\rm n} _k +\frac{1}{2} \omega_k^2 {Q^{\rm n} _k}^2  \nonumber
\end{eqnarray}
with
 $   {\bm \nabla}_{\! \! \bm Q^{\rm n}} V   =   {\bm O_{\rm m}^{\rm n} } ^\dag {\bm \nabla}_{\! \! \bm Q^{\rm m}} V $ and
${\hat {\bm P_{\rm  Q^{\rm n}}}}=  {\bm O_{\rm m}^{\rm n} }^\dag  {\hat {\bm P}_{\rm Q^{\rm m}}} $
is the conjugate momentum with $[\bm Q^{\rm n}, {\hat {\bm P}_{\rm  Q^{\rm n}}}]  =  i \hbar \bm 1$.

\subsection{First order: crystal field coefficient}

We will now use the decomposition (\ref{correct_pos}) for
 $V(\bm R)=\langle 6p m' |\hat V | 6p m\rangle$.

\subsubsection{Crystal field coefficients}

In order to  determine the (real) crystal fields coefficients $V_\Gamma$, that are the
$({\bm \nabla}_{\! \! \bm Q^{\Upgamma}} V^\dag )_{m' m} = \langle m' |  \frac{\partial V}{\partial Q_{ \Gamma  \gamma} } |m \rangle  $ coefficients present in the 3$\times$3 crystal field matrix $ M_{\rm CF}$,
we restrict ourselves to only the first order series in nuclear coordinates. Thus $V=\langle 6p m'  |\hat V | 6pm  \rangle$ contains only $ (V_0)_{m' m}+ ({\bm \nabla}_{\! \! \bm Q^{\rm n}} V^\dag)_{m' m} \bm Q^{\rm n} $ type of terms. 
We will put the $({\bm \nabla}_{\! \! \bm Q^{\rm n}} V^\dag)_{m' m}$  values  in a 3$\times$3=9 rows matrix, with  each row  linked to a given $m',m$. 

The first order potential can be calculated using the normal mode basis $ \bm Q^{\rm n}$  or expressed in the expected $ M_{\rm CF}$ crystal field ones. Equalizing the two expressions
leads to the equation
$ {\bm \nabla}_{\! \! \bm Q^{\Upgamma}} V^\dag    \bm Q^{\Upgamma}  =  {\bm \nabla}_{\! \! \bm Q^{\rm n}} V^\dag \bm Q^{\rm n}  $.

Without  knowing the orthonormal transformation,
$\bm Q^{\Upgamma} = {\bm O^{\Upgamma}_{\rm n}}^\dag  {\bm Q^{\rm n}}  $
that links the crystal field symmetry adapted interaction mode coordinates $Q_{ \Gamma  \gamma}$ to the normal mode coordinates $Q^{\rm n}_k$, the use of the fact that the transformation is orthonormal
leads to a 9$\times$9 matrix equation
\begin{equation}
 {\bm \nabla}_{\! \! \bm Q^{\Upgamma}} V^\dag    {\bm \nabla}_{\! \! \bm Q^{\Upgamma}} V  = {\bm \nabla}_{\! \! \bm Q^{\rm n}} V^\dag   {\bm \nabla}_{\! \! \bm Q^{\rm n}} V
\label{crystal_field_coeff}
\end{equation} 
that, when solved, gives the crystal field coefficients ${\bm \nabla}_{\! \! \bm Q_{\Gamma}} V $.

\subsubsection{Interaction mode coordinates}

In order to perform Monte Carlo simulations, it might be important to know how the crystal field symmetry adapted interaction mode coordinates 
$\bm Q^{\Upgamma}  $ can be calculated.

They
 can be found using the singular value decomposition of the 9 rows matrix ${\bm \nabla}_{\! \! \bm Q^{\rm n}} V^\dag = \bm U \bm W  \bm V^\dag $ of rank $r$ (the matrix $\bm V$ not being confused with the potential $V$). 
Equation
$ {\bm \nabla}_{\! \! \bm Q^{\Upgamma}} V^\dag    \bm Q^{\Upgamma}  =  {\bm \nabla}_{\! \! \bm Q^{\rm n}} V^\dag \bm Q^{\rm n}  $
 leads to 
${\bm \nabla}_{\! \! \bm Q_{\Gamma}} V^\dag     \bm Q^{\Upgamma}    = (\bm U \bm W )  ( \bm V^\dag {\bm Q^{\rm n}} ) $. We cannot simply  equalize 
${\bm \nabla}_{\! \! \bm Q_{\Gamma}} V^\dag  $ with $\bm U \bm W$ and  $ \bm Q^{\Upgamma}$ with $\bm V^\dag {\bm Q^{\rm n}}$ because $ \bm Q^{\rm int} = ( \bm V^\dag  {\bm Q^{\rm n}}) $ 
 is only one choice for the interaction mode coordinates, that is not necessarily the same basis (for instance not in the same order, same sign, ...) as the $\bm Q^{\Upgamma} = \{Q_{\Gamma \gamma} \}$  used to determine from symmetry consideration the crystal field matrix $M_{\rm CF}$. 
However,  using the only useful first $r$ rows of the
 $r$ blocks matrix restriction (noted with index $_{\rm r}$) of 
 $(\bm W_r)^{-1} \bm U_r^\dag \left({\bm \nabla}_{\! \! \bm Q^{\Upgamma}} V^\dag  \cdot \bm Q^{\Upgamma}\right)    =  \bm Q^{\rm int}
 =  ( \bm V_r^\dag  {\bm Q^{\rm n}})
 $ leads  to the  transformation   $\bm Q^{\Upgamma} = {\bm O^{\Upgamma}_{\rm n}}^\dag  {\bm Q^{\rm n}}  $, where,
for simplicity, we have kept the same notation $\bm Q^{\Upgamma}$ or $\bm Q^{\rm int}$ for the first $r$ interaction modes coordinates than for the $3N$ ones.
 
 The important fact is that the diagonal matrix $\bm W $ has only $r$ non-zero singular values on the diagonal (so $\bm W_r$ is square diagonal of size $r$), and thus we will often be interested only in the $r$ relevant ($r$ first) interaction modes coordinates. 
 
With these new coordinates it turns out that the $r \times r$ matrix $ (\bm 
 \omega_{\Upgamma})^2=  {\bm O^{\Upgamma}_{\rm n}}^\dag \bm \omega^2 {\bm O^{\Upgamma}_{\rm n}}$ matrix is diagonal. We thus
have
 \begin{equation}
  \langle 6s   |\hat V | 6s  \rangle  = V_0 +  \dfrac{1}{2} {\bm Q^{\Upgamma}}^\dag  (\bm  \omega^{\Upgamma})^2 \bm Q^{\Upgamma}
     = V_0 +  \sum_{ \Gamma  \gamma=1}^r \frac{1}{2}  \omega_\Gamma^2 Q_{ \Gamma  \gamma}^2
  \label{ground_state_pot_int_coord}
\end{equation}

Once again, we see the advantage of using the interaction mode coordinates 
 to reduce drastically the number of modes to be calculated. The Monte Carlo simulation becomes very simple because we have to calculate only few ($r\leq 9$) interaction mode coordinates,
such as $\bm Q^{\Upgamma} = \{Q_{A_1},Q_{E,1}, Q_{E,2},Q_{T,1},Q_{T,2},Q_{T,3} \}$
 to be compared to $\sim 3N$ modes in standard molecular dynamic simulation (almost 10000 in our case for the $\sim 3000$ movable atoms).

\subsection{Frequency scaled coordinates}

\label{frequency_scaled}

Following \cite{toyozawa1966dynamical,cho1968optical}, it might be interesting to
use the 
oscillation frequency scaled  coordinates
$ {\tilde {\bm Q}}^n =\frac{1}{\sqrt{2}} \bm \omega \bm Q^{\rm n} $ and 
$ {\tilde {\bm Q}}^\Upgamma =\frac{1}{\sqrt{2}} \bm \omega^\Upgamma \bm Q^{\rm \Upgamma} $
such that  Eq. (\ref{ground_state_pot_int_coord}) become
\begin{eqnarray*}
      \langle 6s   |\hat V | 6s  \rangle &= & V_0  + 
      { {\tilde {\bm Q}}^{\rm n} } {}^\dag 
      {\tilde {\bm Q}}^{\rm n}
 =  V_0+ \sum_{k} 
 { {\tilde Q}_k^{\rm n} }{}^2  \cr
  &=& V_0 +  \sum_{ \Gamma  \gamma=1}^r  {{\tilde Q}_{ \Gamma  \gamma}}^2
\end{eqnarray*}

This form is interesting, because
 in a pure classical molecular dynamical simulation, the Boltzmann statistics indicates that 
${\tilde  {\bm Q}}^{\rm n} $  follows a multivariate normal distribution 
$ \propto e^{- \frac{1}{2} 
{\tilde  {\bm Q}}^{\rm n} {}^\dag 
{\tilde {\bm \Sigma}}^{-1} 
{\tilde{\bm Q}}^{\rm n}  }$
with a  variance ${\tilde {\bm \Sigma}}^{\rm n} = \frac{1}{2} k_B T \bm I $. That way the interaction mode coordinate vector   ${\tilde {\bm Q}}^\Upgamma = {\bm O^{\Upgamma}_{\rm n}}^\dag   {\tilde {\bm Q}}^n $ follows 
a multivariate normal distribution with a variance that is also diagonal: ${\tilde {\bm \Sigma}}^\Upgamma = {\bm O^{\Upgamma}_{\rm n}}^\dag {\tilde {\bm \Sigma}}^{\rm n} {\bm O^{\Upgamma}_{\rm n}}= \frac{1}{2} k_B T \bm I $.
Thus,
in a standard classical Monte Carlo simulation at temperature $T$, each
$\frac{  {{\tilde Q}_{ \Gamma  \gamma}} }{ \sqrt{k_B T/2} }$
 coordinate has the same distribution: a  standard unit normal (Gaussian).

% The first order excited state interaction can then be calculated by $V=   {\bm \nabla}_{\! \!  {\tilde {\bm Q}}^{\rm n}} V ^\dag  {\tilde {\bm Q}}^{\rm n} $ where $   {\bm \nabla}_{\! \!  {\tilde {\bm Q}}^{\rm n}} V   =  \frac{1}{\sqrt{2}} \bm \omega {\bm \nabla}_{\! \! \bm Q^{\rm n}} V   $

It is even possible to keep the normalized coordinates
in a classical simulation that includes the temperature scales $T'$ by having 
${\tilde  {\bm Q}}^{\rm n}$  following a
multivariate normal distribution with  a new diagonal variance ${\tilde {\bm \Sigma}} {}^{\rm n} {}'$ with diagonal element given by $  \frac{1}{2} k_B T'(\omega_k)  $ according to Eq. (\ref{scaled_temp}), namely   
$\frac{{\tilde Q}^{\rm n}_k}{ \sqrt{k_B T'(\omega_k)/2} }$.
In this case, $ {\tilde {\bm Q}}^\Upgamma $ has simply to be chosen with  a Gaussian variance ${\tilde {\bm \Sigma}}^\Upgamma {}' 
= {\bm O^{\Upgamma}_{\rm n}}^\dag {\tilde {\bm \Sigma}}^{\rm n} {}' 
{\bm O^{\Upgamma}_{\rm n}} $. Another strategy, that leads to the same final result,
is to keep the unit normal distribution for $\frac{\tilde Q_k^{\rm n}}{ \sqrt{k_B T/2} }$ but change each $\omega_k$ into $\omega_k'$ such that 
$\omega_k'^2/T = \omega_k^2/T'(\omega_k) $. % Then Eq. \ref{crystal_field_coeff} gives the interaction coefficients  ${\bm \nabla}_{\! \! \bm Q_{\Gamma}} V $, that depend now on the temperature $T$ chosen, this is the drawback of this method. Still, the advantage is that it does not require the knowledge of the orthonormal transformation ${\bm O^{\Upgamma}_{\rm n}}$.

So, as used in Ref. \cite{toyozawa1966dynamical,cho1968optical},  such frequency scaled coordinates are useful in classical simulations because the oscillations frequencies $\omega^\Upgamma$ (or the scaled ones $\omega'^\Upgamma$) of the modes are not needed, if we include them in the definition of new crystal field parameters $\tilde V_\Upgamma = \sqrt{2}  V_\Upgamma / \omega_\Gamma $ . Then all terms in the interaction matrix can be written using
$ V_\Upgamma Q_{\Upgamma \gamma}  = \tilde V_\Upgamma  \tilde Q_{ \Gamma  \gamma} $
with now
Eq. (\ref{line_shape}) becoming Eq. (\ref{line_shape_normalized}) that is
\begin{equation}
A(E) \propto \int \sum_{i=1}^6 \delta[E-X_i ({\tilde{\bm Q}}^{\Upgamma})]  e^{- \sum_{ \Gamma  \gamma} 
 {\tilde Q}_{ \Gamma  \gamma}^2/  k_B T}  {\rm d} {\tilde{\bm Q}}^{\Upgamma}
 \label{line_shape_normalized_Appendix}
\end{equation}
with the $\frac{  {{\tilde Q}_{ \Gamma  \gamma}} }{ \sqrt{k_B T/2} }$
 coordinates having a  standard unit normal (Gaussian) distribution.

To illustrate this, we mention that the values given in Ref. \cite{Kanda1971} for the crystal field in Table \ref{table_crystal_field} are

$A = \frac{\sqrt{k_BT/2}}{A^{SO}} \tilde V_{A_1} = \frac{\sqrt{k_BT/2}}{A^{SO}}\frac{\sqrt{2}}{\omega_{A_1}} V_{A_1} = 0.294$,

$B = \frac{\sqrt{k_BT/3}}{A^{SO}} \tilde V_{E} = \frac{\sqrt{k_BT/3}}{A^{SO}}\frac{\sqrt{2}}{\omega_{E}} V_{E} = 0.6$,

$C = \frac{\sqrt{k_BT/4}}{A^{SO}} \tilde V_{T_2} = \frac{\sqrt{k_BT/4}}{A^{SO}}\frac{\sqrt{2}}{\omega_{T_2}} V_{T_2} = 0.2$.\\

Indeed the linear part of the potential is

$V_{\Gamma \gamma} Q_{\Gamma \gamma} = A^{SO}  \left( \frac{\sqrt{k_BT/x_\Gamma}}{A^{SO}} \frac{\sqrt{2}}{\omega_\Gamma} V_{\Gamma_\gamma} \right) \left( \frac{1}{\sqrt{k_BT/x_\Gamma}} \frac{\omega_\Gamma}{\sqrt{2}} Q_{\Gamma_\gamma} \right)$

($x_\Gamma$ is 2 for the case of $\Gamma = A_1$, 3 for $E_2$ and 4 for $T_2$\cite{cho1968optical}).

In the above notation, the ground state potential reduces to $\sum_{\Gamma \gamma} \frac{1}{2} \omega_\Gamma^2 Q_{ \Gamma  \gamma}^2 =   \sum_{\Gamma \gamma} {{\tilde Q}_{ \Gamma  \gamma}}^2 $ .

For the quantized simulation, this frequency scaled coordinates procedure does not work anymore because the quantized energy levels are intrinsically linked to the frequencies, that thus cannot be simply scaled. We could think of using the frequency scaled coordinates with unity frequencies such that all modes look similar and the number of different modes would appear to matter less. However
this approach is futile because
 the kinetic energy of the lattice 
(from
Eq. 
(\ref{correct_pos}))
 would not be diagonal in the momenta conjugate to the scaled interaction mode anymore, as already  noticed in Ref. \cite{toyozawa1966dynamical}. This is not important in the classical Franck-Condon Mulliken approximation where the kinetic energy plays no role because there, we only need to find the positions that are determined by the Boltzmann distribution involving only the potential energy. It, however, becomes crucial for quantized energies, such as in the reflection approximation.

\subsection{Coordinate  distribution}

Let us mention that at $T=\qty{0}{K}$  each normal mode $ |  \Psi_i(\bm Q_k) |^2$ distribution is gaussian (ground state of the harmonic oscillator)  and so
 the multivariate normal distribution  $\propto e^{- \frac{1}{2}  {  {\bm Q}}^{\rm n} {}^\dag  { {\bm \Sigma}}^{-1}  {{\bm Q}}^{\rm n}  }$ of the $\bm Q^{\rm n}$ normal modes is transformed in a multivariate normal distribution with a  variance ${ {\bm \Sigma}}^\Upgamma = {\bm O^{\Upgamma}_{\rm n}}^\dag { {\bm \Sigma}}^{\rm n} {\bm O^{\Upgamma}_{\rm n}}$ for the interaction mode. It turns out that the marginal distribution for the 6 mode coordinates is indeed a variance that is diagonal. This is a consequence of the fact that  the projection on the  6 mode coordinates on the ground state Hessian matrix is diagonal with the frequencies $\omega^\Upgamma$ given in Table \ref{table_crystal_field}. Together with the fact that the first order approximation is quite accurate for the excited state interaction this leads to the fact that the $T=\qty{0}{K}$ spectra can be calculated in a pure crystal field model with only 6 coordinates simply using  frequencies $\omega^\Upgamma$ to determine the gaussian $Q^\Upgamma$ coordinates using temperature scaling or the ground state wavefunction of the $\frac{1}{2} \omega_{\Gamma \gamma}^2 Q_{\Gamma \gamma}^2 $ harmonic oscillator. The result for $T=\qty{0}{K}$ gives the same spectra as the one in Fig. \ref{fig:Exp_th_Oh_Td_temperature}. 
Indeed at $T=\qty{0}{K}$  each $E_i$ is constant and is  the zero point energy of the $i$-th modes. By offsetting to this value, we have at $T=\qty{0}{K}$, $E_i=0$ and so the reflection approximation is  $A(E)  \propto  \sum_{i=1 }^6 P_i \int  \left|  \Psi_i(\bm Q) \right|^2  \delta[  E -(V_{\rm Cs(6p)-Ar} (\bm Q)  - E_i) ]  {\rm d} \bm Q$ which can be compared to the formula (\ref{Reflection_approximation_inetractin_mode}).

\section{Semi-classical approximations}
\label{Semi_classical_approximations}

To study laser excitation of the $(6s)$ cesium atom toward the $6p$ manifold we start (thanks to the Beer-Lambert-Bouguer's law)  by using the fact that
the spectral density optical absorption coefficient $A(E)$ for a photon of energy $E$, is given by the sum over all initial vibronic levels $|i\rangle$, populated with probability $P_i$
with wavefunction $\Psi_i(\bm Q)$ of energy $E_i$, 
towards all possible final ones $\Psi_f(\bm Q)$ of energy $E_f$:
\begin{equation}
A(E) \propto \sum_{i f} P_i \delta[ E -(E_f-E_i ) ]  \left| \int \Psi_f^\dag(\bm Q) d_{i f}(\bm Q) \Psi_i(\bm Q) {\rm d} \bm Q\right|^2
\label{transition_rate_full}
\end{equation}
Thus $A(E) dE$ is the absorption coefficient for a photon in the energy band $E, E+d E$.

\subsection{Semi-classical Franck-Condon  approximation}

In our case, the variation of the dipole transition strength from X to $\Upsigma$ or $\Uppi$ states is on the order of 10\% \cite{blank2012m+}. So we can reasonably 
 consider 
 variation of the dipole $d_{i f}(\bm Q)$ with the internuclear distances  $\bm Q$ as negligible (this is the "Condon" approximation \cite{lax1952franck}).
 If needed, a better approximation would be the Herzberg-Teller centroid one with $d_{i f}$ taken at the point $\bar {\bm Q}$, which cancels the first order evolution in $\bm Q$.
 In the following, for simplicity we will assume  $d_{i f}(\bm Q) \approx d_{i f}(\bar {\bm Q}) $  to be constant (it is simply the 6s to 6p dipole transition and can thus be left out of the equation).
Furthermore, we can also
justify an isotropic nature of the dipole in our case because of the poly-crystalline structure of our sample leads to random orientations of the crystal axes.

In our case, after laser excitation of the Cs(6s) atom from its equilibrium position,  the Cs(6p) atom is far from being in its equilibrium position  and 
the excited electron potentials are mostly linear and not quadratic. Thus
the  motional excited states will be  states with large vibrational quantum
numbers or even quasi continuum one,  which, according to the Bohr correspondence
principle, can be treated as approximately classical. Thus, following \cite{lax1952franck} we can assume 
transitions near the classical turning points and for the sum over $f$ 
 we can replace $E_f$ in the delta-function by a mean value
independent of $f$ that is the electron excited (6p) potential energy curve $V_{\rm e}(\bm Q)$ so:
\begin{equation}
A(E) \propto \sum_{i } P_i \int  \delta[  E -(V_{\rm e}(\bm Q) - E_i) ]  \left|  \Psi_i(\bm Q) \right|^2 {\rm d} \bm Q
\label{Reflection_approximation_appendix}
\end{equation}
 This expression is also called  the reflection approximation
 \cite{heller1978quantum}.

If many initial states (vibrational for instance) $i$ are involved, it is convenient to
use a  mean-value approximation using for the ground state energies, the potential energy $V_{\rm g}(\bm Q)$ leading to the so-called
 semiclassical  Franck-Condon  
formula 
\begin{equation}
A(E) \propto\int  \delta[  E -(V_{\rm e}(\bm Q) - V_{\rm g}(\bm Q)) ]  P_{\rm g}(\bm Q) {\rm d} \bm Q
\label{eq_Mulliken_1} 
\end{equation}
with $  P_{\rm g}(\bm Q)  =  \sum_{i } P_i \left|  \Psi_i(\bm Q) \right|^2 $ is the
quantum-statistical mechanical probability distribution. In the case where 
$  P_{\rm g}(\bm Q)$ is given by  classical statistics,  the expression becomes the  standard classical Franck-Condon  
 formula \cite{lax1952franck}.

\subsection{Semiclassical transition in energy or in phase space}

It is  useful to present the spectral density transition probability $P(E_0)$, for a transition energy $E_0$ given by Eq. (\ref{transition_rate_full}), in a state picture
\begin{equation}
P(E_0) =  \sum_{f,i} P_i  \left| \langle \Psi_f | \hat d | \Psi_i \rangle \right|^2
\delta( E_0 -(E_f-E_i) ) \label{state_transfert_Fermi}
\end{equation}
because $2\pi P(E_0)/\hbar$   can now be interpreted as a rate given by Fermi's Golden rule   \cite{siebrand1967radiationless,freed1975nuclear,stockmann1984influence}.

 This can be written simply as (in the Condon or centroid approximation where we forget about $d^2$ from now on), for a pure state ($P_i=1$ to simplify the notation) as
\begin{eqnarray*}
P(E_0) & = & \sum_{f,i}  \langle \psi_f | \psi_i \rangle \langle \psi_i|     \psi_f \rangle    \langle \psi_f | \delta(E_0-(\hat H_{\rm f}-E_i))   | \psi_f \rangle   \\
& = & \sum_{f,i}  \langle \psi_f | \psi_i \rangle \langle \psi_i|  \sum_{f'}  | \psi_{f'} \rangle    \langle \psi_{f'} | \delta(E_0-(\hat H_{\rm f}-E_i))   | \psi_f \rangle   \\
& = & \sum_{f} \langle \psi_f | \Big( \sum_i  |\psi_i \rangle \langle \psi_i|   \delta(E_0-(\hat H_{\rm f}-E_i))  \Big) | \psi_f \rangle   \\
 \\
&= & Tr \Big[ \sum_i  |\psi_i \rangle \langle \psi_i|   \delta(E_0-(\hat H_{\rm f}-E_i)) \Big] \\
&= & Tr [\hat \rho_{\rm i} \delta(E_0-(\hat H_{\rm f}-E_i) )  ]  \\ 
&= & \langle \delta(E_0-(\hat H_{\rm f}-E_i) )  \rangle \\
&= & \langle \delta(E_0-(\hat V_{\rm f} - \hat V_{\rm i}) -(\hat H_{\rm i} - E_i) )  \rangle 
\end{eqnarray*}
 where
$\hat H_{\rm i} = \sum_k {\hat P}_k^2/(2m_k) + V_{\rm i} (\hat{\bm Q})  $ and
$\hat H_{\rm f} = \sum_k {\hat P}_k^2/(2m_k) + V_{\rm f} (\hat{\bm Q})  $ are the Hamilton operators respectively for the initial and final electronic state
and $\hat \rho_{\rm i} =  \sum_{i}  | \Psi_i \rangle \langle \Psi_i | $  is the density matrix operators for  the initial  state.
The formula can also be written, in the case of a pure state as
$P(E_0) =   \frac{1}{2\pi \hbar } \int d t e^{i(E_0-\hat E_{\rm i}) t/\hbar } \langle \Psi_i  | \Psi_i (t)  \rangle  $where $ |\Psi_i (t) \rangle=  e^{-i \hat H_{\rm f} t/\hbar }  |\Psi_i \rangle$.
This time (and Fourier transform) picture opens another way to treat the transition using the (wave packet) time evolution of the system  to find $|\psi(t) \rangle $ 
\cite{heller1976wigner,mukamel1982semiclassical,shao1998quantum,hupper1998uniform,ansari2022instanton}.

One of the best approximations for the formula is clearly to use $\hat H_{\rm i} - E_i$ as a perturbation, especially when projected on $ | \psi_i   \rangle $ because 
$
 \langle \psi_i | \hat H_{\rm i} - E_i | \psi_i   \rangle 
=0$.
Therefore the simplest natural approximation is the zero-order one for a $d$ dimensional space:
\begin{eqnarray}
   P(E_0) &=&  \langle \delta(E_0-(\hat V_{\rm f} - \hat V_{\rm i}) )  \rangle \label{classical_approx} \\
       & = &  (2\pi \hbar)^{-d} \sum_i  \int   d \bm Q  | \psi_i (\bm Q) |^2 \delta((E-{V_{\rm f}}(\bm Q)  + {V_{\rm i}}(\bm Q)  )   )  \nonumber
\end{eqnarray}
This is exactly the same as the formula  (\ref{eq_Mulliken_1}). But,
note that this derivation
differs
considerably from the previous traditional derivation despite the final results being the same.
For instance, it
gives a very clear physical understanding that
 the reflection
approximation of Eq. (\ref{Reflection_approximation}) is reproduced under the extra approximation of $ \delta(E-({H_{\rm f}}_{\rm cl} -{H_{\rm i}}_{\rm cl} )  $  replaced
by $ \delta(E-(V_{\rm f}( \bm Q) -E_i) ) $. 

 A very similar derivation has been done by moving to the   Wigner representation of quantum mechanics \cite{1932Wigner,heller1976wigner,tatarskiui1983wigner,hillery1984distribution,case2008wigner,ferry2018wigner}
 where the exact formula for $P(E_0)$ becomes, in $d$ dimension, simply
$P(E_0) = (2\pi \hbar)^{-d}  \int  d \bm P d \bm Q {\rho_{\rm i}}_W \delta(E_0-(\hat H_{\rm f}-E_i) )_W    $ where $_W$ designing the Wigner transform. 
The leading order in $\hbar$ is simply the semi-classical limit with 
$\hat H_{\rm f,i}$ replaced by their classical counterpart 
$H_{\rm f,i} = \sum_k {\hat P}_k/(2m_k) + V_{\rm f,i} (\bm Q) $
and ${\rho_{\rm i}}_W $ replaced by the classical phase space density $\rho_{\rm cl} \propto e^{-H_{\rm cl}/k_B T}$. It is obviously slightly better to keep the true Wigner function ${\rho_{\rm i}}_W$ (also because it is known analytically  for the harmonic potential case) to have  \cite{segev2003fermi}:
$
    P(E) 
    =  (2\pi \hbar)^{-d}  \int  d \bm P d \bm Q  {\rho_{\rm i}}_W \delta(E-({H_{\rm f}}_{\rm cl} -{H_{\rm i}}_{\rm cl} )   ) $. This is again the same formula  (\ref{eq_Mulliken_1})
if using the fact that integration of the Wigner function over $\bm P$ gives exactly the wavefunction probability distribution $ | \psi_i (\bm Q) |^2$ \cite{tatarskiui1983wigner}.
 This derivation has the advantage of justifying the use of the true quantum (or Wigner) phase-space initial distribution and not the classical one. Thus the probability $P_i$ has to be chosen accordingly to the actual quantum distribution.  

Another advantage of this semi-classical phase space picture is that
it
 allows for systematic series expansion, typically in power of $\hbar$ order corrections \cite{1932Wigner,heller1976wigner,heller1978quantum,shao1998quantum,japha2002semiclassical,segev2003fermi,sergeev2003semiclassical,pollak2023thermal}. Finally, the method
 can  be   generalized for 
non-radiative transitions, such as surface hoping non-adiabatic effects or electron transfer with
instanton theory or other non Born--Oppenheimer effects
\cite{kallush2002surface,segev2003fermi,segev2003dominant,heller2020semiclassical,agostini2022chemistry}.

\section{Third order many-body term}
\label{third_order_appendix}

We will detail some calculations to determine the effective two-body potential from the sum of the third-order terms. For this, we will first derive the general expression for the third-order terms and calculate it for the long-range dipolar case and in a simple two-level approximation. This will allow us to create a mean field of long-range third-order terms that we compare to the long-range part of the two-body case. We finally extend this comparison for all internuclear distances to produce an effective two-body potential that includes the third-order effect.

\subsection{Perturbation theory up to third order}

The energy shift $\Delta E$ of the energy $E$ of the full system of a given atom A (Cs in our case) in a given state $|l m \rangle$ within the crystal formed by many other atoms (Ar in our case) can be estimated by the perturbation theory from the full hamiltonian energy
$H=\sum_i H_0(i) + \frac{1}{2} \sum_{i\neq j} V_{ij} $ where $H_0(i)$ is the single $i^{\rm th}$ atom hamiltonian and $V_{ij}(\bm r_i,\bm r_j,\bm R_i,\bm R_j) $ contains all electrostatic interactions between the  $i^{\rm th}$ atom and the  $j^{\rm th}$ with electron coordinates  $\bm r_i,\bm r_j$ and nuclear ones $\bm R_i,\bm R_j$.

 We note the eigenstates of the single atom Hamiltonian $H_0(i)$ as $|0_i\rangle$ for the initial (not necessarily the ground state in the Cs case) state of atom $i$ of energy $E_{0_i}$ and $|m_i\rangle$ the other states of energy $E_{m_i}$.

 We calculate the energy shift $\Delta E(\bm R)$ that only depends on the nuclear coordinates $\bm R$ by perturbation theory.
 Up to the third-order, the full crystal energy is 
$\frac{1}{2} \sum_{AB} E_{AB}^{(2)} + \frac{1}{6} \sum_{ABC} E_{ABC}^{(3)} =
 \sum_{A<B} E_{AB}^{(2)} +  \sum_{A<B<C} E_{ABC}^{(3)}
$. 
The first order is  zero and
the second order leads to  
 a shift depending on atom A given by
$ \sum_B E_{AB}^{(2)}$ where $E_{12}^{(2)} = \sum_{m_1,m_2} - \frac{ |\langle 0_{1}  0_{2} |  V_{12} | m_{1}  m_{2} \rangle|^2 }{
\Delta_{m_1}(1)  + \Delta_{m_2}(2)  
} $ with 
  $\Delta_{m_i}(i)= E_{m_i} - E_{0_i} $.
 % The terms dependent on atom A in third-order perturbation theory are given by
% $E_{ABC}^{(3)} = E_{BAC}^{(3)} = E_{BCA}^{(3)} $$
% $E_{A}^{(3)} =   \frac{1}{2} \sum_{B\neq C}  E_{A B C}^{(3)} $ with 

The third-order perturbation theory 
 terms depend on three atoms A, B, and C. 
 Using the symmetry group $S_3$ of permutations of the 3 atoms A=1, B=2, C=3, and using $V_{ij} = V_{ji}$,
$E_{A B C}^{(3)} = E_{123}^{(3)}$ 
is the sum of six terms (see \cite{bell1970multipolar} (7b)):
 \begin{widetext}
\begin{equation}
E_{123}^{(3)} =  \sum_{\sigma \in S_3,m_1,m_2,m_3}
\frac{\langle 0_{\sigma(1)}  0_{\sigma(2)} |  V_{\sigma(1) \sigma(2)} | m_{\sigma(1)}  m_{\sigma(2)} \rangle \langle   m_{\sigma(2)} 0_{\sigma(3)} |  V_{\sigma(2) \sigma(3)} |  0_{\sigma(2)} m_{\sigma(3)}  \rangle \langle m_{\sigma(3)}  m_{\sigma(1)} |   V_{\sigma(3) \sigma(1) } | 0_{\sigma(3)}  0_{\sigma(1)} \rangle }{
(\Delta_{m_{\sigma(1)}}(\sigma(1))  + \Delta_{m_{\sigma(2)}}(\sigma(2)) ) 
(\Delta_{m_{\sigma(1)}}(\sigma(1))+ \Delta_{m_{\sigma(3)}}(\sigma(3)) ) 
}
\label{sum_third_order}
\end{equation}
 \end{widetext}

\subsection{dipole dipole dipole long-range interaction}
We will calculate these terms in the dipolar long-range interaction case.

We thus
assume a dipolar type of interaction (with a fixed quantization axis named z).
For two atoms labeled 1 and 2 with internuclear separation $\bm R_{12} = R_{12} \bm e_{12}$ where $\bm e_{12}$ is a unit vector, the dipole-dipole interaction is
$V_{12} = \frac{e^2}{4\pi \epsilon_0 R_{12}^3} [ \bm r(1). \bm r(2)-3 (\bm r(1) . \bm e_{12})  (\bm r(2) . \bm  e_{12})]  = \frac{e^2}{4\pi \epsilon_0 R_{12}^3}  \bm r(1). (1-3 \bm e_{12} \bm  e_{12} ).\bm r(2) 
$. 
% This last writing restores the real physical insight of the dipole interaction between  of one atom (1) interacting with the (dipolar-shaped) electric field produced by the other dipole (2).

That can be written using the irreducible tensors notation,
$r_q = r \sqrt{\frac{4\pi }{3}}  Y_{1q}(\bm r)$, as 

$V_{12} = - \frac{e^2}{4\pi \epsilon_0 R^3} \sqrt{6}  C(\bm e_{12}). \{ \bm r(1) \otimes \bm r(2) \}_{2}  $
% that is $V_{12}= - \frac{e^2}{4\pi \epsilon_0 R^3} \sqrt{6} \sum_M (-1)^M C_{-M}(\bm e_{12}) \{ \bm r(1) \otimes \bm r(2) \}_{2M}  $
where $C_{M}(\bm e_{12}) = \sqrt{\frac{4\pi }{5}} Y_{2M}(\bm e_{12})$ and $\{ \bm r(1) \otimes \bm r(2) \}_{2M} = 
\sum_{q,q'}  C_{1 q 1 q'}^{2 M}  r_q (1)  r_{q'} (2) $.
So
$$V_{12} = - \frac{\sqrt{6} e^2 }{ 4\pi \epsilon_0 R_{12}^3}  \sum_{M,q_1,q_2} (-1)^M C_{-M}(\bm e_{12}) C_{1 q_1 1 q_2}^{2 M}  r_{q_1}(1)  r_{q_2} (2)  $$

We will often use atomic units where
$\frac{e^2}{4\pi \epsilon_0} =1 $. This can be simply realized by  changing $r(i) $ by $r(i) \sqrt{\frac{e^2}{4\pi \epsilon_0} } $.

\subsubsection{Two level approximation}
\label{2_level_app_dip}

One of the simplest approximations is to consider a single effective atomic energy transition, in the pure case, a $S\leftrightarrow P$ transition for each atom (energy difference $E_{SP}(Cs)$ for Cs and $E_{SP}(Ar)$ for Ar). 
In our case, all Ar atoms are in the ground state $S$, but the Cs can be either in ground $S$ or in excited state $P$. Therefore, $S$ or $P$ degeneracy means that, in Eq. (\ref{sum_third_order}), all energy terms are independent of $m_i$. We write
$\Delta_{m_i}(\sigma(i)) $ as $ \Delta(\sigma(i)) $ that can be factorized out of the sum in Eq. (\ref{sum_third_order}).

Using Cs as atom 1 and Ar for 2 and 3, we have for $i=2,3$ $|0_i\rangle = |l=0 m=0 \rangle$ and $|m_i\rangle = |l=1 m=m_i \rangle$ states.

We then follow the elegant derivation done in ref. \cite{midzuno1956non}. 
We first look on terms depending on $m_2$ so for $i$ label for which $\sigma(i)=2$, in Eq. (\ref{sum_third_order}) we have
 terms 
$\langle  0_2 |  V(j 2) | m_2   \rangle$ and $ \langle  m_2 |  V(2 k) | 0_2 \rangle $. So 
using 
the Wigner-Eckart theorem 
$ \langle l' m'  |  r_2^{(q)} | l m  \rangle  = \frac{C_{l m 1 q}^{l' m'}}{\sqrt{2l'+1}}
\langle l'   ||  r_2 || l   \rangle
$ the 
$ \sum_{m_2} \langle  0_2 |  V(j 2) | m_2   \rangle .... \langle  m_2 |  V(2 k) | 0_2 \rangle$ sum (the $ ....$ are simply here to indicate that the terms are not necessarily neighbors in  Eq. (\ref{sum_third_order}))
contains
$ \sum_{m_2} \langle  0_2 |  r_{q_2}(2) | m_2   \rangle .... \langle  m_2 |  r_{q_2'}(2) | 0_2 \rangle = - \sum_{m_2}  C_{1 m_2 1 q_2}^{0 0}  ...     \frac{C_{0 0 1 q_2'}^{1 m_2}}{\sqrt{3}}  \langle 1   ||  r(2) || 0   \rangle^2   $, where we have used $\langle 0   ||  r(2) || 1   \rangle = - \langle 1   ||  r(2) || 0   \rangle$.
Then, $\sum_{m_2=-1}^{m_2=1} $  leads to
$
\frac{  \langle 1   ||  r(2) || 0   \rangle^2}{3} \delta_{q_2, -q_2'} (-1)^{q_2} 
$.
Similar results arise for atoms 3, for $j$ label for which $\sigma(j)=3$.

\subsubsection{Interaction between ground state atoms}

If the Cs atom is in its ground state, the levels are $|0_1 \rangle =  |l=0 \, m=0 \rangle$ and $|m_1\rangle =  | l=1 \, m_1 \rangle$.

The second order leads to
\begin{eqnarray}
E_{12}^{(2)} & =  - \frac{C_6}{R_{12}^6}  \label{C6_ground}\\
C_6 &= & % \frac{2}{3} \frac{\langle 1   ||  r(1) || 0   \rangle^2 \langle 1   ||  r(2) || 0   \rangle^2    }{\Delta(1) +  \Delta(2)}  =
\frac{2}{3} \frac{ r_{SP}^2(Cs) r_{SP}^2(Ar)  }{ E_{SP}(Cs) +  E_{SP}(Ar) }  \nonumber
\end{eqnarray}
with $ r_{SP}(Cs) = \langle 1   ||  r(1) || 0   \rangle$ and $ r_{SP}(Ar) = \langle 1   ||  r(2) || 0   \rangle$

For the third order calculation, the  results we just derived, in section \ref{2_level_app_dip}, indicate that, in the sum of Eq. (\ref{sum_third_order}),  one numerator becomes

\begin{eqnarray*}
E &=& - \frac{\sqrt{6} }{ R_{12}^3} \frac{\sqrt{6} }{ R_{13}^3} \frac{\sqrt{6} }{ R_{23}^3} \frac{  \langle 1   ||  r(1) || 0   \rangle^2}{3}
\frac{  \langle 1   ||  r(2) || 0   \rangle^2}{3}
\frac{  \langle 1   ||  r(3) || 0   \rangle^2}{3} \times   \\
& &
\sum_{q_1,q_2,q_3,q'_1,q'_2,q'_3}
 (-1)^{q_1+q_2+q_3}   \delta_{q_1, -q_1'}
 \delta_{q_2, -q_2'} 
 \delta_{q_3, -q_3'}\times  
 \\
 & & \sum_{M,M',M''}
 (-1)^{M+M'+M''}  C_{-M}(\bm e_{12}) 
 C_{-M'}(\bm e_{13}) 
 C_{-M''}(\bm e_{23})  \times\\
 & & 
 C_{1 q_1 1 q_2}^{2 M}  
 C_{1 q'_1 1 q_3}^{2 M'} 
 C_{1 q'_2 1 q'_3}^{2 M''}  
\end{eqnarray*} 
but using (8.4(10) and 8.7(15) from \cite{varshalovich1988quantum}) we have
$ \sum_{q_1,q_2,q_3}  (-1)^{q_1+q_2+q_3}  C_{1 q_1 1 q_2}^{2 M}  
 C_{1 -q_1 1 q_3}^{2 M'} 
 C_{1 -q_2 1 -q_3}^{2 M''} = - (-1)^{-M''} 5
  C_{2 M 2 M'}^{2 -M''} 
   \begin{Bmatrix}
  1 & 1 & 2 \\
  2 & 2 & 1 
 \end{Bmatrix} =  - (-1)^{-M''} 
  C_{2 M 2 M'}^{2 -M''}    \sqrt{7/12} $.
So
\begin{eqnarray*}
E &= &  \frac{\sqrt{14}}{9} \frac{  \langle 1   ||  r(1) || 0   \rangle^2}{R_{12}^3}
\frac{  \langle 1   ||  r(2) || 0   \rangle^2}{R_{13}^3}
\frac{  \langle 1   ||  r(3) || 0   \rangle^2}{R_{23}^3} \times\\
& & \sum_{M,M',M''}  
(-1)^{M+M'} C_{-M}(\bm e_{12}) 
 C_{-M'}(\bm e_{13}) 
 C_{-M''}(\bm e_{23})  
  C_{2 M 2 M'}^{2 -M''}   
\end{eqnarray*}
 and finally using $z$ axis along $\bm e_{12}$ (so $M=0$ in the previous sum) and $\theta_i$ being the inner angles of the 123 triangle with $\cos \theta_1= \bm e_{12}.\bm e_{13}$ and $\cos (\pi -\theta_2)= \bm e_{12}.\bm e_{23}$. We get
\begin{eqnarray*}
E & = &  \frac{  \langle 1   ||  r(1) || 0   \rangle^2  \langle 1   ||  r(2) || 0   \rangle^2   \langle 1   ||  r(3) || 0   \rangle^2 }{36 R_{12}^3 R_{13}^3 R_{23}^3} \times \\
& &  (1-3(\cos(2\theta_1) +  \cos(2\theta_2) +  \cos(2\theta_3)))
\end{eqnarray*}
 So when summing the 6 terms, this finally leads to
\begin{eqnarray*}
 E_{123}^{(3)} &= &\frac{ 4 \langle 1   ||  r(1) || 0   \rangle^2  \langle 1   ||  r(2) || 0   \rangle^2   \langle 1   ||  r(3) || 0   \rangle^2 }{36 R_{12}^3 R_{13}^3 R_{23}^3} \times \\
 & & ( 1-3(\cos(2\theta_1) +  \cos(2\theta_2) +  \cos(2\theta_3))  )  \times \\
 & &
\frac{ \Delta(1)  + \Delta(2) + \Delta(3) }{
(\Delta(1)  + \Delta(2) ) 
(\Delta(2)  + \Delta(3) ) 
(\Delta(1)+ \Delta(3) ) 
}
\end{eqnarray*}
So in summary
\begin{eqnarray}
E_{123}^{(3)} &= &  \frac{C_9 }{R_{12}^3  R_{23}^3 R_{31}^3 }  \frac{1-3(\cos(2\theta_1) +  \cos(2\theta_2) +  \cos(2\theta_3))}{4} \label{C9_ground} \\
C_9 & = & \frac{ 4}{9}  r_{SP}^2(Cs) r_{SP}^4(Ar)   
\frac{ E_{SP}(Cs)  + 2 E_{SP}(Ar)  }{
2 (E_{SP}(Cs)  +  E_{SP}(Ar)  )^2 
 E_{SP}(Ar)  } \nonumber
\end{eqnarray} 
We wrote the $C_9$ coefficient such that we can restore  the historical (Axilrod-Tenner-Mut\^{o} \cite{axilrod1943interaction,muto1943force} form  using
$ 1-3(\cos(2\theta_1) +  \cos(2\theta_2) +  \cos(2\theta_3))= 4( 1+3\cos(\theta_1)   \cos(\theta_2)   \cos(\theta_3) ) $.

These calculations are technical but similar in a way to the case of two atoms under the effect of a static external electric field \cite{lepers2011long}, the physical picture
 of the dipole of one atom (1 for instance) interacting with the  vector electric fields produced by the other dipoles (2) and (3)
helps to understand that the final result depends on the angles and distance between atoms.

\subsubsection{Interaction with Cs in an excited state $| l=1 \, m \rangle $}

To our knowledge, no general simple formula has been derived for the dipole-dipole-dipole interaction with one atom in the excited state (see however Ref. \cite{yan2021long} and reference therein). We derive one formula here. For this, we start by the fact that
in this case, $|0_1\rangle =  | l=1 \, m \rangle$ and $|m_1 \rangle =  |l=0 \, m=0 \rangle$. So
\begin{eqnarray}
\sum_{m_1} \langle  0_1 |  r_{q_1}(1) | m_1   \rangle .... \langle  m_1 |  r_{q_1'}(1) | 0_1 \rangle\nonumber\\
=\langle  1 m  |  r_{q_1}(1) | 00   \rangle .... \langle  00 |  r_{q_1'}(1) | 1 m \rangle\nonumber\\
 =-   C_{1 m 1 q'_1}^{0 0}  \frac{C_{0 0 1 q_1}^{1 m}}{\sqrt{3}}  \langle 1   ||  r(1) || 0   \rangle^2\nonumber\\
= \delta_{m q_1}   \delta_{q_1, -q_1'} (-1)^{q_1}  \frac{\langle 1   ||  r(1) || 0   \rangle^2 }{3}\nonumber
\end{eqnarray}

The second order leads to  $C_6^* = \frac{2}{3} \frac{\langle 1   ||  r(1) || 0   \rangle^2 \langle 1   ||  r(2) || 0   \rangle^2    }{\Delta(1) +  \Delta(2)}$ with
\begin{eqnarray}
 E_{12}^{(2)} &=& - \frac{2 C_6^*}{3 R_{12}^6}  = - \frac{C_6^*(\Uppi)}{ R_{12}^6}   \mbox{ for } m=0\label{C6_excited} \\ 
E_{12}^{(2)} &= & - \frac{C_6^*}{6 R_{12}^6} = - \frac{C_6^*(\Sigma)}{ R_{12}^6}  \mbox{ for }  m=\pm 1 \nonumber \\
C_6^* & = & \frac{2}{3} \frac{ r_{SP}^2(Cs) r_{SP}^2(Ar)  }{E_{SP}(Ar)  -  E_{SP}(Cs) } \nonumber
\end{eqnarray}
so as expected $C_6^*(\Upsigma)+2C_6^*(\Pi) =  C_6^*$.

For the third order, we have for a given numerator term:
\begin{eqnarray*}
E &=& - \frac{\sqrt{6} }{ R_{12}^3} \frac{\sqrt{6} }{ R_{13}^3} \frac{\sqrt{6} }{ R_{23}^3} \frac{  \langle 1   ||  r(1) || 0    \rangle^2}{3}
\frac{  \langle 1   ||  r(2) || 0   \rangle^2}{3}
\frac{  \langle 1   ||  r(3) || 0   \rangle^2}{3} \times  \\
& &
\sum_{q_1,q_2,q'_1,q'_2,q_3,q'_3} 
\delta_{m q_1} \delta_{q_1, -q_1'} (-1)^{q_1+q_2+q_3}
 \delta_{q_2, -q_2'} 
 \delta_{q_3, -q_3'}   \times \\
 & & 
  \sum_{M,M',M''}  
(-1)^{M+M'+M'} C_{-M}(\bm e_{12}) 
  C_{-M'}(\bm e_{13})
 C_{-M''}(\bm e_{23}) \times \\
 &  &  C_{1 q_1 1 q_2}^{2 M}  
  C_{1 q'_1 1 q_3}^{2 M'} 
   C_{1 q'_2 1 q'_3}^{2 M''}
\end{eqnarray*} 
but 
using $z$ axis along $\bm e_{12}$ (so $M=0$) we have
\begin{eqnarray*}
E &= & - \sqrt{\frac{8}{27}} \frac{  \langle 1   ||  r(1) || 0   \rangle^2}{R_{12}^3}
\frac{  \langle 1   ||  r(2) || 0   \rangle^2}{R_{13}^3}
\frac{  \langle 1   ||  r(3) || 0   \rangle^2}{R_{23}^3} \times\\
& & \sum_{M',M''}  
(-1)^{M'+M''} C_{0}(\bm e_{12}) 
 C_{-M'}(\bm e_{13}) 
 C_{-M''}(\bm e_{23})  
 \times \\
 &  &
  C_{1 m 1 -m}^{2 0} 
  \sum_{q_3}  (-1)^{q_3}   
 C_{1 -m 1 q_3}^{2 M'} 
 C_{1 m 1 -q_3}^{2 M''}
\end{eqnarray*}
So,
% and finally with  $\theta_i$ being the inner angles of the  123 triangle with  $\cos \theta_1= \bm e_{12}.\bm e_{13}$ and  $\cos (\pi -\theta_2)= \bm e_{12}.\bm e_{23}$ and $\theta_3=\pi-\theta_1-\theta_2$, we have
  $
E =   \frac{  \langle 1   ||  r(1) || 0   \rangle^2  \langle 1   ||  r(2) || 0   \rangle^2   \langle 1   ||  r(3) || 0   \rangle^2 }{54 R_{12}^3 R_{13}^3 R_{23}^3}  f(\theta_1,\theta_2,\theta_3,m)
 $
  with 
  \begin{eqnarray}
 f(\theta_1,\theta_2,\theta_3,m) & = & -(1 + 3 \cos(2\theta_1)  +  3 \cos(2\theta_2) +  9 \cos(2\theta_3) ) \nonumber \\ & &  \mbox{for} \  m=0  \\  
  f(\theta_1,\theta_2,\theta_3,m) & =  & (5 - 3 \cos(2\theta_1)  -  3 \cos(2\theta_2) +  9 \cos(2\theta_3)) /4 \nonumber \\ & &
   \mbox{for} \  m=\pm 1
  \end{eqnarray}
Then, summing the 6 terms (see \cite{midzuno1956non}) finally leads to 
\begin{eqnarray*}
 E_{123}^{(3)} &=& \frac{ 2 \langle 1   ||  r(1) || 0   \rangle^2  \langle 1   ||  r(2) || 0   \rangle^2   \langle 1   ||  r(3) || 0   \rangle^2 }{27 R_{12}^3 R_{13}^3 R_{23}^3} \times \\
 &  & f(\theta_1,\theta_2,\theta_3,m) \times \\
 & &
\frac{ \Delta(1)  + \Delta(2) + \Delta(3) }{
(\Delta(1)  + \Delta(2) ) 
(\Delta(2)  + \Delta(3) ) 
(\Delta(1)+ \Delta(3) ) 
}
\end{eqnarray*}
so,
\begin{eqnarray}
 E_{123}^{(3)} &=&  \frac{C_9^* }{R_{12}^3 R_{13}^3 R_{23}^3} \frac{f(\theta_1,\theta_2,\theta_3,m)}{6}   \label{C9_excited}  \\
C_9^*  &=&   \frac{4}{9}  r_{SP}^2(Cs) r_{SP}^4(Ar)  
\frac{  2 E_{SP}(Ar) - E_{SP}(Cs)  }{
2 (E_{SP}(Ar) -E_{SP}(Cs) )^2 
E_{SP}(Ar)
} \nonumber
\end{eqnarray}

\subsection{Mean field effect and effective two-body potentials}

The most accurate way to include this third-order correction for the energy of atoms A is to sum over all B,C pairs of atoms. But due to the lack of information on the short-range part and due to the crude (two-level) estimation made up to now, we can simplify the problem further and use the mean-field approach done in Ref. \cite{stenschke1994effective} (see also \cite{dridi2022development,muser2022interatomic}) by creating a mean-field potential $\bar E_{AB}^{(3)}=\sum_C E_{ABC}^{(3)}$ such that
the full crystal energy 
$\frac{1}{2}  \left( \sum_{AB} E_{AB}^{(2)} + \frac{1}{3} \sum_{ABC} E_{ABC}^{(3)} \right) $ can be written as sum of two body terms $E_{AB}^{\rm eff} =  (E_{AB}^{(2)} + \frac{1}{3} \bar E_{AB}^{(3)} $.

 We can thus stay at a two-body level simply by modifying the
 two-body interaction between atoms by adding $ \frac{1}{3} \bar E_{AB}^{(3)}$ to the $ E_{AB}^{(2)} $ potential interaction curves we already have. This mean-field approach
 has been proven to be quite accurate for pure crystals, as shown by the so-called Marcelli-Wang-Sadus potential \cite{grimme2016dispersion,xu2020many,muser2022interatomic,deiters2021interatomic,stroker2022thermodynamic,dridi2022development} and as we demonstrate in Fig. \ref{fig:potentialCurves} in the Ar-Ar case.

\subsubsection{Mean field homogeneous assumption}

Strangely enough, almost no mathematical derivation of the
mean field approach exists though it
has been performed  as back as in the original paper by Mut\^{o}
\cite{muto1943force} (whose journal reference is almost always wrong) and by Stenschke\cite{stenschke1994effective}. Following these works, we replace the sum $\bar E_{AB}^{(3)}=\sum_C E_{ABC}^{(3)}$ by an integral assuming Ar atoms uniformly distributed with density $\rho =N/V$ (for solid argon the density $\rho = \SI{280}{a_0^{-3}}$ where $a_0$ is the Bohr radius). We thus sum over the C atoms (the number 3 above); that is, we average the potential $V=\sum_j E_{12j}^{(3)}$ as
$\bar V = \rho \int V r^2 \sin \theta    {\rm d} r  {\rm d}\theta  {\rm d}\varphi =  2\pi \rho \int V r^2 \sin \theta     {\rm d} r  {\rm d}\theta   $ in spherical coordinate with $\theta=\theta_1$ and $r=R=R_{12}$. We found simpler to use $r_1 = R_{13} ,r_2 = R_{23}$ coordinates,
as
 in Ref. \cite{stenschke1994effective}, such that
$\bar V = 2 \pi \rho \int V \frac{r_1 r_2}{R  } {\rm d} r_1  {\rm d} r_2 $, with an exclusion sphere of radius $\sigma$ around the atoms A=1 and B=2.
Equations (\ref{C9_ground}) and (\ref{C9_excited}) lead to
\begin{eqnarray}
 \bar E_{12}^{(3)} & =& 2\pi \rho \frac{ C_9}{R^6} \frac{4}{3}  \label{averaged_pot} \\
  \bar {E^*_{12}}^{(3)} & =& 2\pi \rho \frac{ C_9^*}{R^6} 
  \frac{4}{9}
(4 - 3 \log [(R^2 - \sigma^2)/ \sigma^2]) \mbox{ for } m=0 \nonumber \\
\bar {E^*_{12}}^{(3)} & =&  2\pi \rho \frac{ C_9^*}{R^6} 
  \frac{2}{9}
(-1 + 3 \log [(R^2 - \sigma^2)/ \sigma^2]) \mbox{ for } m=\pm 1 \nonumber
\end{eqnarray}

Therefore, the triple dipole interaction does not depend on $\sigma$ only when the state is isotropic, such as the S ground state, or when summing over $m=-1,0,1$ for the P state.

\subsubsection{Effective potentials}
All results are summarized by Eqs (\ref{C6_ground}-\ref{averaged_pot}) from where we can extract
 the effective potentials for the internuclear distance $R$:
\begin{eqnarray}
E_{AB}^{\rm eff} (R)  & =& E_{AB}^{(2)} (R)  \left(1- \frac{8\pi \rho}{9}  \frac{ C_9}{C_6}  \right)  \label{correction_potential} \\ 
 E_{AB}^{*,\rm eff}(\Upsigma) & =& E_{AB}^{(2)}(\Upsigma) \left(1-  \frac{4\pi \rho}{9}  \frac{ C_9^*}{C_6^*}  
\left( 4 - 3 \log \left[ \frac{R^2 - \sigma^2}{ \sigma^2} \right] \right)  \right)   \nonumber \\
 E_{AB}^{*,\rm eff}(\Uppi) & =& E_{AB}^{(2)}(\Uppi) \left(1-  \frac{8\pi \rho}{9}  \frac{ C_9^*}{C_6^*}  
\left( -1 + 3 \log \left[ \frac{R^2 - \sigma^2}{ \sigma^2} \right] \right)  \right)   \nonumber
\end{eqnarray}
This form, derived from the long-range triple dipole interaction, can however naturally be extended to the full range of the potential by simply keeping the same formula for all $R$. The only requirement is the choice of the cut-off because $R>2\sigma$. Because the short-range part of the potential is clearly not a dipole-dipole one, we smoothly reduce the $C_9^*$ part for $R<2\sigma$ by multiplying it in Eq. (\ref{correction_potential}), for all $R$, by a sharp cut-off function $1-e^{-(R/2\sigma)^{10}}$.
 A reasonable value for $\sigma$ might be  the
nearest neighbor distance in the fcc crystal, the lattice size or the first nearest neighbor Cs-Ar atoms in a given trapping site (see Fig. \ref{fig:potentialCurves_diff}). This may depend on the chosen trapping site to be studied. Another cut-off, in principle independent, should be of the order of the LeRoy radius that corresponds to the change in the multipole expansion at small $R$. In that case the theoretical curves are too different from the long-range part, which we used to derive the formula \cite{le1974long,kielkopf1974semiempirical}. 
From all this, we found that
a  reasonable value can be $\sigma \sim \SI{8}{a_0}$. However,
 all these parameters (cut-offs and the power 10 in the sharp function) are arbitrary choices that can be optimized. We have modified them a bit (typically by a factor of 2) to produce the uncertainty in the line position shown in Fig. \ref{fig:abs_spectra_th}.

\subsubsection{Approximate calculation of $C_6$ and $C_9$ coefficients}

Our previous formulae for second Eq. (\ref{C6_ground}) and (\ref{C6_excited}), and third-order perturbations Eq. (\ref{C9_ground}) and (\ref{C9_excited}), contain the  van der Waals interaction coefficients $C_6,C_6^*,C_9,C_9^*$, which we now need to evaluate.
For this, we will use an effective two-level approximation, known to be 25\% accurate for ground state interactions between rare-gas and alkali atoms \cite{tang1969dynamic,diaz1980combination}, and that we are going to use for the excited states as well. We will also compare the value found by fitting them to the long-range part of our potential curves to existing results to assess the method accuracy.

From now on, we will use atomic units for this calculation.

The atomic unit value for the $C_6$ coefficient for Ar-Ar, found by fitting the theoretical Ar-Ar potential curve by $-C_6/R^6$, is 67; and 380 for Ar-Cs. 
Using Eq. (\ref{C6_ground}) with the well-known experimental values,
 $E_{SP}(Cs) \approx 0.0524 $ and
 $E_{SP}(Ar) \approx 0.43  $ for the first excited $P$ levels,
leads to    $ r_{SP}(Cs) =5.4 $ and $ r_{SP}(Ar) =3.0 $.

These values are acceptable because 
they are reasonable: the $C_6$ Cs-Cs coefficient is 5500 (compare to the real value of nearly 6700 \cite{gould2016c}) ;  $C_9(Ar-Ar-Ar) \approx 650$ (compared to $525 $ in Refs.
\cite{bell1966van,tang1969dynamic}); and $C_9(Cs-Cs-Cs) \approx 1 500 000 $ (compared to $C_9(Cs-Cs-Cs) \approx 2 200 000 $ in Ref. \cite{diaz1980combination}).
However, these values for
 $ r_{SP}(Cs), r_{SP}(Ar), E_{SP}(Cs) $ and $E_{SP}(Ar)$
lead to quite wrong values for the excited Cs$^*$-Ar coefficient of 
$C_6^*(\Uppi) = \frac{2}{3} C_6^* = 310$ compared to 600 in our potential curves and $C_6^*(\Upsigma) = \frac{1}{6} C_6^* = 80$ compared to 270. 
 Unfortunately, it is not possible to solve this discrepancy for all excited states at once simply because 
 Eq. (\ref{C6_excited})  gives a ratio 4 between $C_6^*(\Pi)$ and $C_6^*(\Sigma)$ while in our case, the theoretical potential curves give only a factor $\sim 2$. This factor 2 is indeed found in many other cases \cite{bussery1986calculated,zhang2007long}. 

 This discrepancy probably comes from the fact that
we have neglected other excited states (like Cs$(7s)$) that are quite near the Cs excited (6p) level. But,
because our intention is to simulate the Cs spectrum in argon, we will focus more on the Cs-Ar and Cs$^*$-Ar interaction curves than on the Ar-Ar one in order 
 to try to reduce the discrepancy. 
For this, we choose an intermediate strategy and allow modification of the dipole transitions $r_{SP}(Cs)$, $r_{SP}(Ar)$ and the energy level $E_{SP}(Ar)$, that become effective values. This allows us to better reproduce the long-range part of the theoretical potential curve for Cs-Ar and Cs$^*$-Ar.
For this,
we start with fitting the theoretical potential curve that leads to $\frac{C_6^*}{C_6}  \approx 3 $. Then Eq. (\ref{C6_ground}) and (\ref{C6_excited})   give $ \frac{C_6^*}{C_6} = \frac{E_{SP}(Ar)  +  E_{SP}(Cs)}{E_{SP}(Ar)  -  E_{SP}(Cs) }$. This leads to 
 $E_{SP}(Ar) = 2 E_{SP}(Cs) = 0.10  $. Then, using expression Eq. (\ref{C6_ground}) for the $C_6$ coefficient for Ar-Ar (67) and Ar-Cs (380) leads  to (in atomic units):
 \begin{eqnarray}
 E_{SP}(Cs) & = & 0.052 \label{coeffrE} \\
 E_{SP}(Ar) & \approx & 0.10 \nonumber \\
 r_{SP}(Cs) & \approx & 4.4 \nonumber \\
 r_{SP}(Ar) & \approx & 2.1  \nonumber \\
C_6 & \approx & 370  \label{coeff_C6_C9}  \\
C_6^* & \approx & 1200 \nonumber \\
 C_9 & \approx & 9100 \nonumber \\
C_9^* & \approx & 54000  \nonumber 
 \end{eqnarray}

 \begin{figure}[ht]
    \centering  
    \includegraphics[width=1\linewidth]{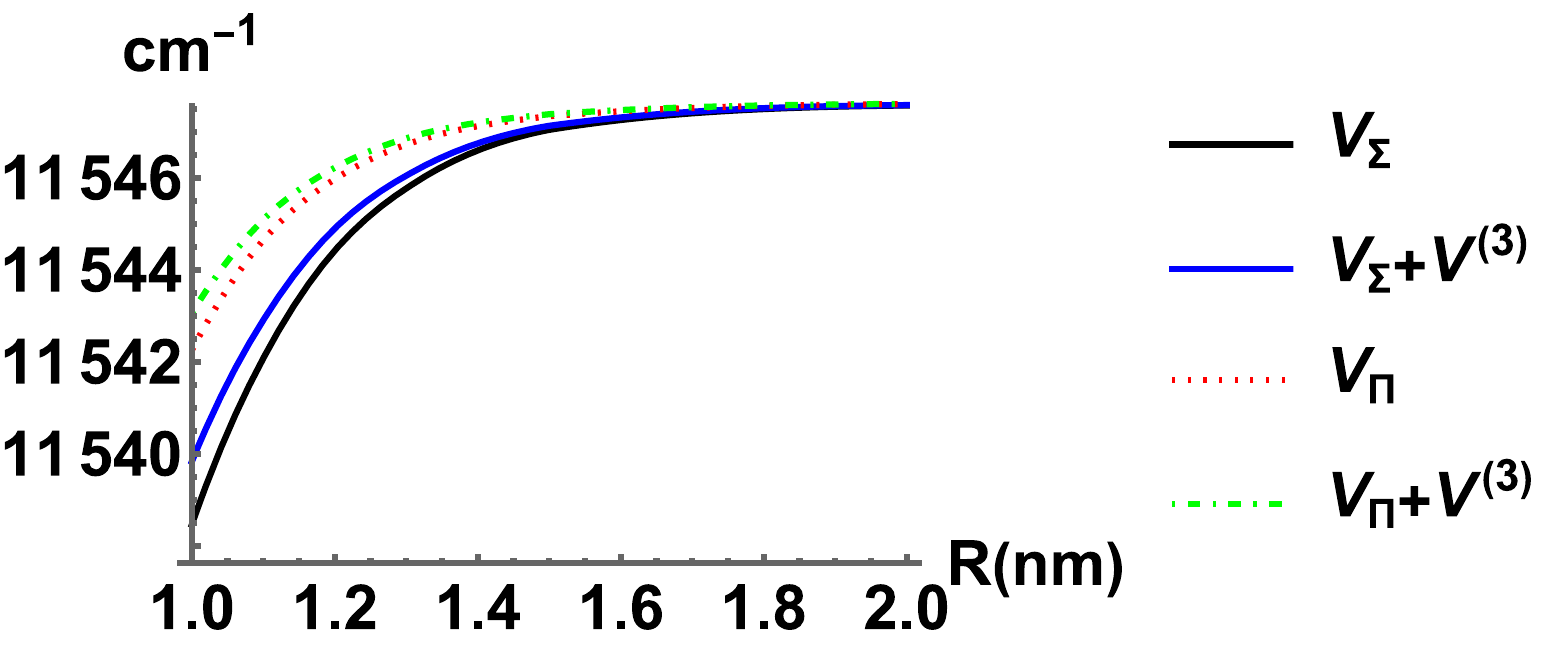}
    \caption{Cs(6p)-Ar potential curves compared to the one with the effective third-order effect included.    }
    \label{fig:third_order_Curves_long_range}
\end{figure}

% \cite{diaz1980combination2} two body

The long-range parts of some potential curves are shown in Fig. \ref{fig:third_order_Curves_long_range}. These parameter values, and the modification by a factor $ \sim 2$ for the cut-off and the power of the sharp function, have been used in order to produce the uncertainty in the line positions shown in Fig. \ref{fig:abs_spectra_th}.

%   $u_3(R)= \frac{-\lambda \nu  \rho}{\epsilon \sigma^6 } u_2(r)$ where $\epsilon$ is the depth of the 2B potential well; $\sigma$ is the distance at which $u_2(R) = 0$; and $\lambda = 0.85$ is a parameter obtained from the comparison of 2B and 3B molecular simulation data. $\rho =N/V$ is the density,  $\nu = C_9 = 7.35x10^{108} J m^9$ is the non-additive coefficient (for argon \cite{solanathermodynamic}) which can be determined experimentally from dipole oscillator strength \cite{deiters2021interatomic,stroker2022thermodynamic}.

%\bibliographystyle{unsrt}
\bibliographystyle{h-physrev}
\bibliography{biblio}

\end{document}